\UseRawInputEncoding
\pdfoutput=1
\documentclass[preprint,review,11pt]{elsarticle}
\usepackage[a4paper,margin=0.75in]{geometry}
\usepackage[numbers]{natbib}
\usepackage{mathptmx,amsmath,amsfonts,amsthm,bm,algorithm,algpseudocode,float,gensymb,hyperref} % Math packages
\usepackage[thinc]{esdiff}
\usepackage{array}
\usepackage{booktabs}
\usepackage[table,xcdraw]{xcolor}

\usepackage{amssymb}
\journal{MSSP}

\begin{document}

\begin{frontmatter}



\title{Virtual sensing of subsoil strain response in monopile-based offshore wind turbines via Gaussian process latent force models}


\author[inst1]{Joanna Zou}
\author[inst1]{Eliz-Mari Lourens}
\author[inst1]{Alice Cicirello}

\affiliation[inst1]{organization={Delft University of Technology},%Department and Organization
            addressline={Faculty of Civil Engineering and Geosciences, Stevinweg 1}, 
            postcode={2628 CN},
            city={Delft},
            country={Netherlands}}

\begin{abstract}
Virtual sensing techniques have gained traction in applications to the structural health monitoring of monopile-based offshore wind turbines, as the strain response below the mudline, which is a primary indicator of fatigue damage accumulation, is impractical to measure directly with physical instrumentation. The Gaussian process latent force model (GPLFM) is a generalized Bayesian virtual sensing technique which combines a physics-driven model of the structure with a data-driven model of latent variables of the system to extrapolate unmeasured strain states. In the GPLFM, \textcolor{black}{unknown sources of excitation are modeled as a Gaussian process (GP) and endowed with a structured covariance relationship with response states, using properties of the GP covariance kernel as well as correlation information supplied by the mechanical model}. It is shown that posterior inference of the latent inputs and states is performed by Gaussian process regression of measured accelerations, computed efficiently using Kalman filtering and Rauch-Tung-Striebel smoothing in an augmented state-space model. While the GPLFM has been previously demonstrated in numerical studies to improve upon other virtual sensing techniques in terms of accuracy, robustness, and numerical stability, this work provides one of the first cases of in-situ validation of the GPLFM. The predicted strain response by the GPLFM is compared to subsoil strain data collected from an operating offshore wind turbine in the Westermeerwind Park in the Netherlands. A number of test cases are conducted, where the performance of the GPLFM is evaluated for its sensitivity to varying operational and environmental conditions, to the instrumentation scheme of the turbine, and to the fidelity of the mechanical model. In particular, this paper discusses the capacity of the GPLFM to achieve relatively robust strain predictions under high model uncertainty in the soil-foundation system of the offshore wind turbine by attributing sources of model error to the estimated stochastic input.

\end{abstract}




\end{frontmatter}

% \linenumbers

\section{Introduction}
\label{sec:intro}

With many existing monopile-based offshore wind farms approaching the end of their service life and the frequency of extreme climate events on the rise in recent years, offshore wind turbines (OWTs) face increasingly greater risk of fatigue-induced structural failure. The fatigue life of an offshore wind turbine (OWT) is driven by its dynamic response to the rotor motion of the nacelle in addition to wind and wave loading, which fluctuates over the lifetime of the structure \cite{Maes2016}. For newly constructed offshore wind farms, where OWTs are larger and more slender in design in order to maximize power generation capacity, there is a greater risk of resonance effects from the rotor frequency coinciding with the fundamental frequency of the structure, leading to accelerated fatigue damage \cite{Devriendt2014}.
\\
\indent Structural health monitoring (SHM) techniques offer a means to obtain information necessary for fatigue life evaluation of existing OWTs and design validation of recently installed OWTs. In practice, the strain response at critical locations can be measured over time to predict the accumulation of fatigue damage. However, current approaches to SHM which rely on physical measurements of strains are not well adapted for OWTs, because fatigue damage in monopile-based turbines concentrates in the foundation below the mudline where accurate strain readings are impractical to obtain. The installation and maintenance of sensors at subsoil fatigue hotspots is cost-prohibitive, and for monopiles instrumented prior to installation, the sensors are often damaged during the pile-driving process \cite{Henkel2019}.
\\
\indent To overcome these challenges, \textit{virtual sensing} approaches to SHM have received considerable attention in OWT monitoring applications, since they provide a full-field estimation of an unobserved quantity of interest (e.g. strains) using measurements of a proxy response variable (e.g. accelerations) from physical sensors. These methods can be efficiently implemented using a small number of standard sensors, such as accelerometers, placed at accessible locations of an operating structure. The response at unmeasured locations of the structure is then reconstructed from the measurements through a dynamics model of the system or transfer function. This approach to SHM not only reduces sensor deployment costs significantly, but also improves the quality of data collected, as accelerometers are more reliable, less sensitive to noise, and capable of recording at higher sampling rates compared to strain gauges \cite{Noppe2016, Smyth2007}. Virtual sensing approaches may be categorized into two broad classes: 1) deterministic model-based extrapolation, where the estimation of response states beyond measured locations is obtained through a well-calibrated mechanical model, such as a finite element model, and 2) probabilistic state estimation, where strain response states are treated as random variables whose probability distributions are learned from measured data.
\\
\indent Among the deterministic techniques, the predominant method in practice is modal decomposition and expansion (MD\&E) by Iliopoulos et al. \cite{Iliopoulos2016}, which decomposes the dynamical system into mode shape components from which the response between measured points may be inferred. The accuracy of MD\&E depends on the parameters of the finite element model, which are calibrated using system identification techniques such as operational modal analysis. In applications to operational data from OWTs, MD\&E produces a good estimate of the strain response in the time domain but faces difficulty with estimating higher frequency regions of the response spectrum \cite{Maes2016}. Moreover, because MD\&E derives displacements from the double integration of measured accelerations, the strain estimate is susceptible to low-frequency noise amplification, a numerical issue termed the ``drift effect" which is detailed in \cite{Smyth2007}. To address the difficulty in the frequency domain strain estimate, Iliopoulos et al. \cite{Iliopoulos2017} introduce a multi-band method where the dynamic frequency range is separated into two sections, such that a variable number of sensors and modes may be chosen for each frequency band. Studies by Noppe et al. \cite{Noppe2016, Noppe2018} show that the multi-band method may lead to discontinuities in the estimate at the defined frequency band limits. Henkel et al. \cite{Henkel2018, Henkel2019} extends the studies in \cite{Maes2016, Noppe2016, Noppe2018} to the estimation of subsoil strains, where multi-band MD\&E is observed to result in a phase lag in the time domain and an underestimation of the first mode spectral density for periods of high wind speeds. Generally, ill performance of MD\&E can be attributed to its fully deterministic nature, such that it remains highly sensitive to discrepancies between the physical system and idealized model. In particular, MD\&E does not explicitly account for stochastic measurement noise and model error resulting from uncertainty in soil-foundation interaction properties. Moreover, since calibration of the model must be performed offline, MD\&E cannot account for shifts in the modal parameters of the system (e.g. natural frequencies, damping ratios) across operational states and loading conditions, which limits the robustness of its use. 
\\
\indent The other class of techniques takes a Bayesian inference approach to virtual sensing and has a number of notable differences from deterministic methods. First, in estimating the trajectory of the unknown state, Bayesian inference methods produce a probability distribution of values of the state over time, reflecting additional information on the level of uncertainty in the predicted state rather than a single estimate of the time series. Secondly, sources of uncertainty from stochastic measurement noise and model error are explicitly incorporated in the construction of the inference problem. Finally, these virtual sensing methods offer a statistical framework to account for the direct dependence of the dynamic response on qualities of the force, or ``input", acting on the system, such as its magnitude, variability, and correlation structure. As no practical means exist for direct measurement of forces, the unknown input is either implicitly modeled as a form of process noise or treated as an additional random variable whose time evolution may be derived simultaneously with that of the strain response. The latter of these approaches, in which the joint distribution of unknown inputs and states is inferred from measurement data, is referred to as ``joint input-state estimation" and constitutes the focus of this paper. 
\\
\indent Bayesian inference approaches to virtual sensing are commonly implemented using the Kalman filter \cite{Kalman1960}, whose sequential structure to inferring states from measurements is suitable for long time series-type data. Among the Kalman filter-based techniques, one version comprises of minimum variance unbiased filters derived from optimal control theory which do not assume prior knowledge of the time-varying behavior of the input. While early methods addressed the state estimation problem exclusively \cite{Kitandis1987, Hou1998}, Hsieh et al. \cite{Hsieh2000} recognized the value of performing input estimation to improve the state estimation. The methods of \cite{Kitandis1987} and \cite{Hsieh2000} were extended in Darouach et al. \cite{Darouach2003} to include direct feedthrough between the inputs and states. These contributions lead to the development of the joint input-state estimator by Gillijns \& de Moor \cite{Gillijns2007a, Gillijns2007b}, abbreviated as GDF or JIS in literature, which formalized a filter for simultaneously estimating inputs and states in linear systems satisfying minimum variance and unbiasedness conditions. Since then, the method has been extended for cases with a time delay using Rauch-Tung-Striebel (RTS) smoothing \cite{RTS1965, Maes2018a}, for cases with a reduced order model and equivalent forces \cite{Lourens2012}, and with a method to account for estimation uncertainty and correlation between inputs and states \cite{Maes2016b}. The GDF is evaluated extensively in literature, using numerical tests in \cite{Azam2015, Tatsis2017, Nayek2019}, experimental tests in \cite{Lourens2012, Azam2017}, and operational data in \cite{Maes2016, Maes2016c, Lourens2012, Petersen2017}. Notably, the GDF is compared to other state-of-the-art virtual sensing techniques using in-situ data from an offshore wind turbine in \cite{Maes2016a}, where it is concluded that methods are competitive and interchangeable. However, other work has noted difficulties with the robustness of the GDF in virtual sensing applications. In particular, the GDF is observed to be susceptible to numerical instability \cite{Tatsis2017, Lourens2012, Petersen2017} arising from rank deficiency of system matrices, which occurs in overdetermined systems where the number of sensors exceeds the number of modes in the model \cite{Lourens2012} and where the sensor configuration results in near linear dependence between the inputs and states \cite{Petersen2017}. A comparison test in \cite{Azam2017} demonstrates that the GDF is more susceptible to the drift effect from the accumulation of integration errors than other state-of-the-art joint input-state estimators. While the theoretical performance of the GDF produces minimum variance unbiased estimates for systems without model error, Maes et al. \cite{Maes2016b, Maes2016c} note that a primary limitation of the GDF is that input and state estimates are no longer minimum variance in applications where loading, noise, or structural conditions deviate from those assumed in the model. Moreover, the estimated error covariance matrices of the inputs and states no longer correspond to the true error. Optimality of  estimates by GDF are with respect to the noise covariance matrices on the inputs and states, which must be known a priori or tuned appropriately for a given application.
\\
\indent In contrast to the GDF, which makes no prior assumption on the input, more recent developments in Kalman filter-based estimators assume that the input takes the form of a random walk model, where the time evolution of the input is driven by zero-mean white Gaussian noise. The augmented Kalman filter, or AKF \cite{Lourens2012b},  performs simultaneous estimation of the input and state variables in augmented state-space; in comparison to Maes et al. \cite{Maes2016, Maes2016a}, which implements a classical Kalman filter for pure state estimation by assuming the process noise term encompasses contributions from the input, the AKF defines the covariance structure of the unknown input separate from that of process noise. To address issues of unobservability of the augmented state matrix when acceleration-only measurements are used, Naets et al. \cite{Naets2015} proposes the AKF-dm, supplementing the AKF with artificial ``dummy displacements" as additional measurements in order to stabilize the state estimate. The dual Kalman filter, or DKF \cite{Azam2015}, has also been proposed to overcome unobservability issues by introducing a successive structure to the filter, where the input estimation is followed by the state estimation. The performance of the Kalman filter-based estimators is compared using numerical data in \cite{Tatsis2017, Nayek2019}, experimental data in \cite{Azam2017}, and operational data in \cite{Noppe2018}. A well-known shortcoming of the AKF, AKF-dm, and DKF is that their accuracy largely depends on an appropriate choice of noise covariance matrices, including process noise on dynamic states, measurement noise of observed states, and modeled noise in the unknown input \cite{Azam2015, Lourens2012b}. While tuning procedures for covariance matrices have been proposed, including heuristic rules of thumb, a maximum likelihood estimation approach \cite{Yuen2007}, and exhaustive search by the L-curve criterion \cite{Azam2015, Lourens2012b}, manual adjustment of the covariance matrix is often implemented to estimate an appropriate order of magnitude for the covariances. In several instances, the problem of estimating the covariance matrix is simplified by constraining the matrix to have a diagonal structure with constant values \cite{Azam2015, Lourens2012b}, placing the strict assumption that all inputs and states are uncorrelated and have equal magnitude variance. All these methods face the additional important limitation that the white noise assumption on the input may be too restrictive for real life applications. In the case of OWTs, which are assumed as linear time-periodic (LTP) systems, cyclic load sources are expected to contain harmonic components which violate the broadband assumptions of white noise input.  
\\
\indent Recently, latent force models have been proposed as a generic case of Bayesian estimators which allow for greater flexibility in the treatment of the unknown input. Latent force models mark a connection between the control theory-based perspective and machine learning-based perspective of the state estimation problem; whereas purely data-driven approaches in statistical modeling traditionally suffer in cases of extrapolation and sparse datasets, the hybrid approach of a latent force model supplements the statistical model with a mechanical model of the underlying system dynamics to guide regression in the latent variable space \cite{Alvarez2009}. In virtual sensing applications, the latent force model relates the measured accelerations, or observed variables, to the inputs and states, which comprise the unobserved or ``latent" variables of the structural system \cite{Nayek2019}. Alvarez et al. \cite{Alvarez2009} introduced the concept of the Gaussian process latent force model, or GPLFM, where a linear system is driven by an unknown Gaussian process whose covariance matrix is characterized by a kernel function. The GPLFM exploits properties of Gaussian processes to construct covariance relationships between the inputs and states, using parameters of the assumed kernel as well as correlation information supplied by the mechanical model. Posterior inference of latent variables may then be reduced to a problem of Gaussian process regression of the observed variables with joint inference of the unobserved variables. Whereas \cite{Alvarez2009} provides a convolution-based solution to Gaussian process regression, Hartikainen \& Sarkka \cite{Hartikainen2010} demonstrate that Gauss-Markov processes may be reformulated in state-space such that regression can be more efficiently performed using Kalman filtering \cite{Kalman1960} and RTS smoothing \cite{RTS1965}. Their study is extended to the joint input-state estimation problem in \cite{Hartikainen2011}, which provides the state-space framework for the GPLFM.
\\
\indent A  numerical comparison of the GPLFM with competing joint input-state estimation algorithms for linear structural systems is carried out in Nayek et al. \cite{Nayek2019}. This investigation demonstrated that the GPLFM consistently outperforms or is on par with the performance of the AKF, AFK-dm, and DKF for joint input-state estimation across various loading scenarios. While Bayesian joint input-state estimators share many core characteristics, the superior performance of the GPLFM can be attributed to the nature of the statistical model parameters which require tuning \cite{Nayek2019}. \textcolor{black}{The GDF, AKF, AKF-dm, and DKF all rely on individual definition and inference of full covariance matrices, leading to a complex covariance matrix estimation problem. To simplify the problem,} Yuen et al. \cite{Yuen2007} propose one approach of taking a maximum likelihood estimate of the matrix diagonal using the negative log likelihood expression recovered from the Kalman filter; however, in this approach, the covariance matrix is parameterized using constraining assumptions, such as independent variance in each state dimension, and requires the solution to an optimization problem whose complexity scales with the dimension of the model. In \cite{Azam2015, Lourens2012}, a modified L-curve criterion is adopted, which exhaustively searches for a constant variance parameter defining the noise covariance matrix which minimizes estimation error. \textcolor{black}{Both of these procedures for covariance matrix estimation correspond to model fitting based on variance minimization}, but either do not explicitly account for frequency content of the input, as in the case of the GDF, or fix an assumption of broadband frequency in the input using a white noise model, as in the case of the AKF, AKF-dm, and DKF. In contrast, \textcolor{black}{the GPLFM characterizes the relationship between inputs and states in terms of a GP covariance kernel. The kernel function is parameterized by a relatively small set of hyperparameters, the number of which are independent of the state dimension. Consequently, tuning hyperparameters by parameter estimation techniques such as maximum likelihood estimation \cite{Nayek2019} or Bayesian inference \cite{Rogers2020} leads to a much more efficient exploration of the tunable parameter space compared to full covariance matrix estimation.} Moreover, these hyperparameters often correspond to interpretable properties of the input and response, such as the amplitude, frequency content, and smoothness or differentiability of the signal, in contrast to parameters describing individual covariance terms.
\\
\indent The GPLFM is first applied to structural health monitoring applications in Nayek et al. \cite{Nayek2019}, where it is shown that the GPLFM is a generalization of the AKF which offers greater flexibility in the definition of the input in augmented state-space. Rogers et al. \cite{Rogers2020} extends the study to address system parameter estimation in addition to the joint input-state estimation problem. While the GPLFM has been evaluated in numerical studies, literature on the experimental or operational validation of the GPLFM in SHM is sparse. Petersen et al. \cite{Petersen2022} demonstrates the potential of combining analytical wind load models with the latent force model for long-span suspension bridges, but without the availability of validation data. Using data collected from an operating onshore wind turbine, Bilbao et al. \cite{Bilbao2022} evaluates the performance of the GPLFM for strain estimation across the height of the tower above the foundation. 
\\
\indent This contribution provides a critical study of the in-situ validation of the GPLFM and associated practical considerations for its use in fatigue monitoring. The fatigue assessment of offshore wind turbines serves as a suitable case study, as the GPLFM is expected to offer particular advantages in state estimation for systems subject to high model uncertainty and variable excitation sources which are not well characterized by the white noise assumption grounding other virtual sensing techniques. Moreover, this work presents the rare opportunity for validation of strain estimates in the OWT foundation below the mudline, as measurements from strain gauges embedded within the monopile are available in the considered case study. The paper is organized as follows: Section \ref{sec:theory} presents the formulation of both the mechanical model of the system and the stochastic model of the unknown input in state-space, as well as the theoretical framework of the GPLFM. Section \ref{sec:casestudy} presents the case study for in-situ validation, using subsoil data obtained in a monitoring campaign of an OWT in the Westermeerwind Park in the Netherlands. Section \ref{sec:results} presents the results of strain estimation for fatigue assessment and an evaluation of the sensitivity of the GPLFM to both measurement data fidelity and model fidelity. Section \ref{sec:conclusions} concludes and provides recommendations for future research directions.  

\section{Mathematical formulation}
\label{sec:theory}

\subsection{Mechanical model of structural system}
\label{sec:mechmodel}

In structural dynamics, the response states of a physical system under forced vibration are described by a mechanical model which takes the form of a linear second-order differential equation:  

\begin{equation}
    \label{eq:eom}
    \mathbf{M\ddot{u}}(t)
    +\mathbf{C\dot{u}}(t)
    +\mathbf{Ku}(t)
    =\mathbf{S}_{\mathrm{p}}\mathbf{p}(t)
\end{equation}

where $\mathbf{u}(t) \in \mathbb{R}^{n_{\mathrm{u}}}$ are the displacement states at $n_{\mathrm{u}}$ degrees of freedom of a system with mass, damping, and stiffness properties defined by $\mathbf{M}, \mathbf{C}, \mathbf{K} \in \mathbb{R}^{n_{\mathrm{u}} \times n_{\mathrm{u}}}$ respectively. The system is externally excited by $n_{\mathrm{p}}$ forces $\mathbf{p}(t) \in \mathbb{R}^{n_{\mathrm{p}}}$, with spatial distribution indicated by the selection matrix $\mathbf{S}_{\mathrm{p}} \in \mathbb{R}^{n_{\mathrm{u}} \times n_{\mathrm{p}}}$ \cite{Lourens2012}. In this presentation, the system is assumed to be time-invariant, i.e. the model parameters and force locations are constant in time. 
\\ 
\indent For complex models with several degrees of freedom, the solution for the response states may be more efficiently computed using model order reduction as in \cite{Lourens2012, Maes2015}. The eigensolution to $\mathbf{K}\Phi = \mathbf{M}\Phi\bm{\Omega}^2$ returns the mass-normalized mode shapes $\phi_j$, given by the columns of matrix $\Phi \in \mathbb{R}^{n_{\mathrm{u}} \times n_{\mathrm{u}}}$, and the natural frequencies $\omega_j^2$ in rad/s, which are collected in the diagonal matrix $\bm{\Omega}^2 \in \mathbb{R}^{n_{\mathrm{u}} \times n_{\mathrm{u}}}$. In the modally reduced order formulation, the matrices are truncated into $\bar{\Phi} \in \mathbb{R}^{n_{\mathrm{u}} \times n_{\mathrm{m}}}$ and $\bar{\bm{\Omega}}^2 \in \mathbb{R}^{n_{\mathrm{m}} \times n_{\mathrm{m}}}$  to retain a subset of modes $n_{\mathrm{m}}$ which dominate the system response. By casting into modal states $\mathbf{r}(t) \in \mathbb{R}^{n_{\mathrm{m}}}$ with the transformation $\mathbf{u}(t) = \bar{\Phi} \mathbf{r}(t)$ and premultiplying by $\bar{\Phi}^\textup{T}$, Eq. \eqref{eq:eom} becomes: 

\begin{equation}
    \label{eq:eom2}
    \bar{\Phi}^\textup{T} \mathbf{M} \bar{\Phi} \mathbf{\ddot{r}}(t)
    +\bar{\Phi}^\textup{T} \mathbf{C} \bar{\Phi} \mathbf{\dot{r}}(t)
    +\bar{\Phi}^\textup{T} \mathbf{K} \bar{\Phi} \mathbf{r}(t)
    =\bar{\Phi}^\textup{T} \mathbf{S}_{\mathrm{p}}\mathbf{p}(t)
\end{equation}

With mass-normalized mode shapes, the orthogonality conditions $\bar{\Phi}^\textup{T} \mathbf{M} \bar{\Phi} = \mathbf{I}$ and $\bar{\Phi}^\textup{T} \mathbf{K} \bar{\Phi} = \bar{\bm{\Omega}}^2$ are satisfied. Assuming proportional damping, the modal damping ratios $\xi_j$ are given by $\bar{\Phi}^\textup{T} \mathbf{C} \bar{\Phi} = \bar{\bm{\Gamma}}$, where $\bar{\bm{\Gamma}} = {\rm diag}(2 \xi_j \omega_j)$ for $j=1,...,n_{\mathrm{m}}$. The forces are transformed into their modal contributions by $\mathbf{f}(t)=\bar{\Phi}^\textup{T} \mathbf{S}_{\mathrm{p}}\mathbf{p}(t)$, where $\mathbf{f}(t) \in \mathbb{R}^{n_{\mathrm{m}}}$ is the vector of modal force time histories. The modally reduced order model then becomes: 
\begin{equation}
    \label{eq:eom3}
    \mathbf{\ddot{r}}(t)
    +\bar{\bm{\Gamma}} \mathbf{\dot{r}}(t)
    +\bar{\bm{\Omega}}^2 \mathbf{r}(t)
    =\mathbf{f}(t)
\end{equation}

To formulate the model for filtering and inference, the continuous-time linear differential equation is converted to state-space form. By defining the state vector $ \mathbf{x}(t) = \begin{bmatrix}
\mathbf{r}(t) & 
\mathbf{\dot{r}}(t)
\end{bmatrix}^\textup{T}$, Eq. \eqref{eq:eom3} can be expressed as:

\begin{equation}
    \label{eq:dynamicsmodelfull}
    \begin{bmatrix} \mathbf{\dot{r}}(t) \\ \mathbf{\ddot{r}}(t) \end{bmatrix}
    = \underbrace{ \begin{bmatrix} \mathbf{0} & \mathbf{I} \\ -\bar{\bm{\Omega}}^2 & -\bar{\bm{\Gamma}} \end{bmatrix} }_\text{$\mathbf{A}_{\mathrm{c}}$}
    \begin{bmatrix} \mathbf{r}(t) \\ \mathbf{\dot{r}}(t) \end{bmatrix}
    + \underbrace{ \begin{bmatrix} \mathbf{0} \\ \mathbf{I} \end{bmatrix} }_\text{$\mathbf{B}_{\mathrm{c}}$}
    \mathbf{f}(t)
\end{equation}

which is succinctly written as: 
\begin{equation}
    \label{eq:dynamicsmodelshort}
    \mathbf{\dot{x}}(t) = \mathbf{A}_{\mathrm{c}}\mathbf{x}(t) + \mathbf{B}_{\mathrm{c}}\mathbf{f}(t)
\end{equation}

Eq. \eqref{eq:dynamicsmodelshort} constitutes the dynamical model of the state-space system, which describes the evolution of response states over time as a function of the system parameters, given by $\mathbf{A}_{\mathrm{c}}$, and of the input forces, whose influence is given by $\mathbf{B}_{\mathrm{c}}$. 
\\
\indent To facilitate numerical implementation, the dynamical model is converted from continuous-time to discrete-time with time step size $\Delta t$ using the zero-order hold assumption. The discretized dynamical model is as follows: 

\begin{equation}
    \label{eq:dynamicsmodeldiscr}
    \begin{gathered}
        \mathbf{x}_k = \mathbf{A}\mathbf{x}_{k-1} + \mathbf{B}\mathbf{f}_{k-1}
        \\
         \mathbf{A} = \textup{exp}(\mathbf{A}_{\mathrm{c}} \Delta t), \quad
         \mathbf{B} = [\mathbf{A} - \mathbf{I}] \mathbf{A}_{\mathrm{c}}^{-1} \mathbf{B}_{\mathrm{c}}
    \end{gathered}
\end{equation}

\indent In addition to the dynamical model, a discrete-time measurement model is defined to describe the relation between state variables and observed variables of a distinct sampling frequency. While accelerometers are predominantly used in SHM applications, axial strain readings from strain gauges may be additionally available with certain sensor configurations. Heterogeneous data fusion with displacement measurements is presented in \cite{Smyth2007, Ferrari2015}; for the case where strain measurements are available at an equal sampling rate as acceleration measurements, the measured quantities $\mathbf{y}_k \in \mathbb{R}^{n_\mathrm{y}}$ for $k=1,...,N$ are related to the original displacement, velocity, and acceleration states by: 

\begin{equation}
    \label{eq:measuremodel}
    \mathbf{y}_k = 
    \begin{bmatrix} \bm{\epsilon}_{k} \\ \mathbf{\ddot{u}}_{k} \end{bmatrix}
    = \begin{bmatrix} \mathbf{S}_{\mathrm{s}}\mathbf{T}_{\mathrm{s}} & \mathbf{0} & \mathbf{0} \\
    \mathbf{0} & \mathbf{0} & \mathbf{S}_{\mathrm{a}} \end{bmatrix}
    \begin{bmatrix} \mathbf{u}_{k} \\ \mathbf{\dot{u}}_{k} \\ \mathbf{\ddot{u}}_{k} \end{bmatrix}
\end{equation}

where $\mathbf{S}_{\mathrm{a}} \in \mathbb{R}^{n_{\mathrm{acc}} \times n_{\mathrm{u}}}$ is the selection matrix of degrees of freedom where $n_{\mathrm{acc}}$ accelerations are recorded and $\mathbf{S}_{\mathrm{s}} \in \mathbb{R}^{n_{\mathrm{str}} \times n_{\mathrm{e}}}$ is the selection matrix of elements where $n_{\mathrm{str}}$ strains are recorded. $\mathbf{T}_{\mathrm{s}} \in \mathbb{R}^{n_{\mathrm{e}} \times n_{\mathrm{u}}}$ is the transformation matrix which computes axial strain deformation $\bm{\epsilon} \in \mathbb{R}^{n_{\mathrm{e}}}$ from lateral displacements of a finite element model with $n_{\mathrm{e}}$ elements based on shape functions from beam theory, provided in \ref{sec:app_disptransf}.  The total dimension of the observation vector is therefore $n_\mathrm{y} = n_{\mathrm{acc}} + n_{\mathrm{str}}$.
\\
\indent The observation equation with respect to the modal states $\mathbf{x}$ is then given by:
\begin{equation}
    \label{eq:measuremodelfull}
    \begin{bmatrix} \bm{\epsilon}_{k} \\ \mathbf{\ddot{u}}_{k} \end{bmatrix}
    = \underbrace{ \begin{bmatrix} \mathbf{S}_{\mathrm{s}}\mathbf{T}_{\mathrm{s}}\Phi & \mathbf{0} \\
    -\mathbf{S}_{\mathrm{a}}\Phi\bm{\Omega}^2 & -\mathbf{S}_{\mathrm{a}}\Phi\bm{\Gamma} \end{bmatrix} }_\text{$\mathbf{G}$}
    \begin{bmatrix} \mathbf{r}_{k} \\ \mathbf{\dot{r}}_{k} \end{bmatrix}
    + \underbrace{ \begin{bmatrix} \mathbf{0} \\ \mathbf{S}_{\mathrm{a}}\Phi \end{bmatrix} }_\text{$\mathbf{J}$}
    \mathbf{f}_k
\end{equation}

which can be reduced to the following equation, with output influence matrix $\mathbf{G}$ and direct feedthrough matrix $\mathbf{J}$: 

\begin{equation}
    \label{eq:measuremodelshort}
    \mathbf{y}_k = \mathbf{G}\mathbf{x}_k + \mathbf{J}\mathbf{f}_k
\end{equation}

Together, Equations \eqref{eq:dynamicsmodeldiscr} and \eqref{eq:measuremodelshort} give the discrete-time state-space model of the physical system: 

\begin{subequations}
    \label{eq:ssm}
    \begin{gather}
    \mathbf{x}_k = \mathbf{A}\mathbf{x}_{k-1} + \mathbf{B}\mathbf{f}_{k-1} + \mathbf{\eta}_{k-1} \\
    \mathbf{y}_k = \mathbf{G}\mathbf{x}_k + \mathbf{J}\mathbf{f}_k + \mathbf{v}_k
    \end{gather} 
\end{subequations}

The process noise term $\mathbf{\eta}_{k}$ with covariance $\mathbb{E}[\mathbf{\eta}_k\mathbf{\eta}_l^\textup{T}] = \mathbf{Q}^{\mathrm{x}}$ and measurement noise term $\mathbf{v}_k$ with covariance $\mathbb{E}[\mathbf{v}_k\mathbf{v}_l^\textup{T}] =\mathbf{R}$ are added to account for model uncertainty and measurement error, respectively. The noise terms are mutually uncorrelated white Gaussian noise processes, i.e.\ $\mathbb{E}[\mathbf{\eta}_k\mathbf{v}_l^\textup{T}] = \mathbf{0}$.
\\
\indent In its current form, the linear state-space model in Eq. \eqref{eq:ssm} cannot be used to directly solve for the response states because the forcing process $\mathbf{f}_{k}$ is unknown. Whereas Maes et al.\ \cite{Maes2016a} simplifies the problem by grouping the unknown input with the white noise term $\mathbf{w}_k$, this approach has limited robustness in capturing the frequency range of dynamic strains for events with non-white input.

\subsection{Gaussian process models}
\label{sec:inputmodel}

This section provides a brief overview of properties of Gaussian processes (GPs) and GP regression, which form the basis of the GPLFM methodology. In the case of a single dimensional input $t \in \mathbb{R}$, a Gaussian process defines a probability density over possible functions $h(t)$ of a random variable $t$. A GP has the property of being completely characterized by the first two moments of the process, namely its mean and covariance functions, expressed as follows: 

\begin{equation}
    \label{eq:gp}
    h(t) \sim GP(\mu(t), \kappa(t,t';\theta))
\end{equation}

 where $\mu(t)$ denotes the mean function and $\kappa(t,t';\theta)$ denotes the covariance function, also referred to as the kernel, representing the covariance as a function of distance between any two points $t$ and $t'$ of the process. Several options for forms of the kernel and its corresponding hyperparameters $\theta$ exist to control the smoothness and differentiability properties of $h(t)$, discussed further in Section \ref{subsec:GPkernels}. 
\\
\indent Gaussian processes are commonly employed in regression problems to define a non-parametric mapping between inputs and outputs, e. g. $h \, : X \rightarrow Y$. Consider a training dataset $D = \{ X, Y \}$ with inputs $X = \{t_1,...,t_N\}$ and corresponding outputs $Y = \{y_1,...,y_N\}$. The observed $Y$ are assumed to be noisy measurements of the underlying process, modeled as $h(X)$:

\begin{equation}
    \label{eq:gpreal}
    Y = h(X) + \bm{\nu}, \quad \bm{\nu} \sim \mathcal{N}(\mathbf{0},\sigma_\nu^2\mathbb{I})
\end{equation}

Discrepancy between measurements $Y$ and the model $h(X)$ is represented with additive white Gaussian noise $\bm{\nu}$ with variance $\sigma_\nu^2$. In regression, $h$ provides an approximation of the unobserved output corresponding to a test set of values $X^*$ which are separate (or held out) from the observed dataset $D$. In GP regression, moments of the Gaussian posterior distribution $p(h(X^*)|Y)$ may be computed with closed form expressions, as provided in \cite{Hartikainen2010, Rasmussen2006}. The main drawback of this solution to GP regression is the high order of computational complexity, scaling with respect to the number of data points $N$ by $\mathcal{O}(N^3)$, due to the inversion of a $N \times N$ matrix. Rasmussen \& Williams \cite{Rasmussen2006} provide an algorithm for the practical implementation of the closed form solution, by addressing matrix inversion with Cholesky decomposition. However, it can be shown that for stationary Gauss-Markov processes, a recursive solution to GP regression by means of Kalman filtering and RTS smoothing may be implemented with a computational complexity scaling linearly with the dataset size, e. g. $\mathcal{O}(N)$ \cite{Hartikainen2010}. For particularly large datasets, as in cases with time series-type data, the recursive solution has substantial computational advantage.

\subsubsection{Special case of stationary Gaussian processes}
\label{subsec:stationaryGPs} 
 
A stationary GP which satisfies Markov properties may be formulated in state-space such that Kalman filtering and RTS smoothing can be implemented to solve for its posterior distribution \cite{Hartikainen2010}. For the zero-mean stationary case of the GP, the mean function evaluates to zero, e. g. $\mu(t) \rightarrow 0$, and the covariance function does not change over time shifts $\tau = t - t'$, becoming: 
 
 \begin{equation}
    \label{eq:gpstationary}
    h(t) \sim GP(0, \kappa(\tau; \theta))
\end{equation}

The temporal GP $h(t)$ of Eq. \eqref{eq:gpstationary} may be equivalently expressed as the output of a linear time-invariant (LTI) system driven by a white noise process $w(t)$, taking the form of a stochastic differential equation (SDE) of order $\beta$: 

\begin{equation}
    \label{eq:gpSDE}
    \frac{d^{\beta}h(t)}{dt^\beta} + a_{\beta-1}\frac{d^{\beta-1}h(t)}{dt^{\beta-1}} + \hdots + a_1\frac{dh(t)}{dt} + a_0h(t) = w(t) \ , \quad S_w(\omega) = q_w
\end{equation}

in terms of coefficients $a_0, ..., a_{\beta-1}$ and the spectral density of the white noise, $q_w$. Eq. \eqref{eq:gpSDE} may be converted to state-space through its companion form, in which the $\beta$-order linear SDE is represented as a system of $\beta$ first-order linear SDEs. Defining a vector of $h(t)$ and its derivatives up to order $\beta-1$ as $\mathbf{z}(t) = \begin{bmatrix} h(t) \quad \frac{dh(t)}{dt} \quad \hdots \quad \frac{d^{\beta-1}h(t)}{dt^{\beta-1}} \end{bmatrix}^\textup{T}$, the companion form follows: 

\begin{equation}
    \label{eq:singleforcemodel}
    \begin{bmatrix} \frac{dh(t)}{dt} \\ \frac{d^2h(t)}{dt^2} \\ \vdots \\ \frac{d^{\beta}h(t)}{dt^\beta} \end{bmatrix}
    = \underbrace{ \begin{bmatrix} 0  & 1 & 0 & \hdots & 0 \\
    0 & 0 & 1 & \hdots & 0 \\
    \vdots & \vdots & \vdots & \ddots & \vdots \\
    0 & 0 & 0 & \hdots & 1 \\
    -a_0    & -a_1   & -a_2 & \hdots & -a_{\beta-1} \\ \end{bmatrix} }_\text{$\mathbf{F}_{\mathrm{c}}$}
    \begin{bmatrix} h(t) \\ \frac{dh(t)}{dt} \\ \vdots \\ \frac{d^{\beta-1}h(t)}{dt^{\beta-1}} \end{bmatrix}
    + \underbrace{ \begin{bmatrix} 0 \\ 0 \\ \vdots \\ 0 \\ 1 \end{bmatrix} }_\text{$\mathbf{L}_{\mathrm{c}}$}
    w(t)
\end{equation}

or simply as:

\begin{equation}
    \label{eq:singleforcemodelshort}
    \mathbf{\dot{z}}(t) = \mathbf{F}_{\mathrm{c}}\mathbf{z}(t) + \mathbf{L}_{\mathrm{c}}w(t)
\end{equation}

The SDE in Eq. \eqref{eq:singleforcemodelshort} constitutes the state-space model of the GP. It must be highlighted that properties of the SDE - namely, the spectral density $q_w$, the order of the model $m$, and the value of coefficients $a_0, ..., a_{\beta-1}$ - are a function of the kernel chosen for the GP. Certain conditions must be met in order to construct a state-space model of the GP with the intended stationary kernel $\kappa(\tau)$: the kernel must satisfy positive semi-definiteness, and the kernel produces a GP $h(t)$ whose spectral density $S_h(\omega)$ is in rational form. For GPs which meet these criteria, the equivalent state-space model ($\mathbf{F}_{\mathrm{c}}, \mathbf{L}_{\mathrm{c}},$ and $q_w$) may be computed by spectral factorization, the steps of which are detailed in \cite{Nayek2019, Hartikainen2010}. 
\\
\indent From Eq. \eqref{eq:singleforcemodelshort}, our interest is to estimate the probability density of the state vector, $p(\mathbf{z}(t))$. The Fokker-Planck-Kolmogorov (FPK) equation provides the complete probability density of the solution of a generic SDE, as detailed in \cite{Sarkka2012}. However, due to properties of GPs,  computation of the FPK equation can be bypassed and the first two moments of the distribution can be directly solved to provide a complete probabilistic description of the state vector, $p(\mathbf{z}(t)) \sim \mathcal{N}(\hat{\mathbf{z}}(t), \mathbf{P}(t))$. For the case of linear time invariant (LTI) SDEs with the initial condition $\mathbf{z}(t_0) \sim \mathcal{N}(\hat{\mathbf{z}}_0, \mathbf{P}_0)$, the differential equations of the mean and covariance of $\mathbf{z}(t)$ reduce to:

\begin{subequations}
    \begin{gather}
    \label{eq:meanDE}
    \frac{d\hat{\mathbf{z}}(t)}{dt} = \mathbf{F}_{\mathrm{c}}\hat{\mathbf{z}}(t) \\
    \label{eq:covDE}
    \frac{d\mathbf{P}(t)}{dt} = \mathbf{F}_{\mathrm{c}}\mathbf{P}(t) + \mathbf{P}(t)\mathbf{F}_{\mathrm{c}}^\textup{T} + \mathbf{Q}_{\mathrm{c}}
    \end{gather}
\end{subequations}

where the spectral density of the stochastic term is transformed as $\mathbf{Q}_{\mathrm{c}} = \mathbf{L}_{\mathrm{c}}q_w\mathbf{L}_{\mathrm{c}}^\textup{T}$. The solutions are then given by:

\begin{subequations}
    \label{eq:meancovsoln}
    \begin{gather}
    \hat{\mathbf{z}}(t) = \textup{exp}(\mathbf{F}_{\mathrm{c}}(t-t_0)) \ \hat{\mathbf{z}}_0 \\
    \mathbf{P}(t) = \textup{exp}(\mathbf{F}_{\mathrm{c}}(t-t_0)) \ \mathbf{P}_0 \ \textup{exp}(\mathbf{F}_{\mathrm{c}}(t-t_0))^\textup{T} +
    \int_{t_0}^{t} \textup{exp}(\mathbf{F}_{\mathrm{c}}(t-\tau)) \ \mathbf{Q}_{\mathrm{c}} \ \textup{exp}(\mathbf{F}_{\mathrm{c}}(t-\tau))^\textup{T}\,d\tau   \label{subeq:covsoln}
    \end{gather}
\end{subequations}

Rather than computing the integral in Eq. \eqref{subeq:covsoln} directly, the covariance $\mathbf{P}(t)$ can be solved for from the matrix Riccati differential equation in Eq. \eqref{eq:covDE} using the method of matrix fraction decomposition, detailed in \cite{Sarkka2012, Grewal2001}. However, the solution is further simplified in the case of zero-mean stationary processes, where it may be deduced that the SDE has a \textit{steady-state} solution in which the state vector has a constant distribution at all time points, $\mathbf{z}_k \sim \mathcal{N}(\mathbf{0}, \mathbf{P}_\infty)$. Considering the asymptotic behavior $\frac{d\mathbf{P}(t)}{dt} \rightarrow \mathbf{0}$ and $\mathbf{P}(t) \rightarrow \mathbf{P}_\infty$ in the steady-state condition, Eq. \eqref{eq:covDE} becomes the Lyapunov equation, a special case of algebraic Riccati equations \cite{Sarkka2012}:

\begin{equation}
    \label{eq:lyapunov}
    \mathbf{0} = \mathbf{F}_{\mathrm{c}}\mathbf{P}_\infty + \mathbf{P}_\infty\mathbf{F}_{\mathrm{c}}^\textup{T} + \mathbf{Q}_{\mathrm{c}}
\end{equation}

The Lyapunov equation is solvable with numerical algorithms \cite{Bangert1986} to obtain the steady-state covariance matrix $\mathbf{P}_\infty$. To recover the kernel function $\kappa(\tau)$ of the process, we can define the following relationship to extract $h(t)$ from $\mathbf{z}(t)$: 

\begin{equation}
    \label{eq:singleforcemodelshort2}
    \begin{gathered}
    h(t) = \mathbf{H}_{c}\mathbf{z}(t) \\
    \mathbf{H}_{c} = \begin{bmatrix} 1 & 0 & \hdots & 0 \end{bmatrix}
    \end{gathered}
\end{equation}

Then, the kernel may be computed from the solution to the LTI SDE of Eq. \eqref{eq:singleforcemodelshort}, following the algebra in \cite{Sarkka2012}, to obtain: 

\begin{equation}
    \label{eq:covariancesoln}
    \kappa(\tau) = 
    \begin{cases}
    \mathbf{H}_{\mathrm{c}} \, \mathbf{P}_\infty \, \textup{exp}(\mathbf{F}_{\mathrm{c}}\tau)^\textup{T} \, \mathbf{H}_{\mathrm{c}}^\textup{T} \ , \quad \quad \ \ \
    \tau \geq 0 \\
    \mathbf{H}_{\mathrm{c}} \, \textup{exp}(-\mathbf{F}_{\mathrm{c}}\tau)^\textup{T} \, \mathbf{P}_\infty \,  \mathbf{H}_{\mathrm{c}}^\textup{T} \ , \quad \quad 
    \tau < 0
    \end{cases}
\end{equation}

\subsubsection{Recursive solution to Gaussian process regression}
\label{subsec:gpr_recursive}

For the special class of Gauss-Markov processes, GP regression may be more efficiently performed using recursive Bayesian filtering. Unlike the batch solution to GP regression, which considers all data at once by constructing the joint distribution across points, the recursive solution processes the data sequentially by exploiting the Markov property, where upon observation of the present state of the process $h(t)$, future instances of the state are independent of past observations \cite{Sarkka2012}. Let $y_{i:j} = \{y_i,...,y_j\}$ denote the subset of data $Y$ between time steps $t_i$ and $t_j$. First, the steady-state solution is placed as the prior, $p(\mathbf{z}_0) = \mathcal{N}(\mathbf{0}, \mathbf{P_\infty})$. Then, the independence condition between past and future observations allows for the construction of the recursive Bayesian filter as follows: 

\begin{equation}
    \label{eq:bayesrecursive}
    p(\mathbf{z}_k|y_{1:k}) = \frac{p(y_{k} |\mathbf{z}_k) \ p(\mathbf{z}_k | y_{1:k-1}))}
    {\int_{x} \ p(y_{k} |\mathbf{z}_k) \ p(\mathbf{z}_k | y_{1:k-1})) \ dx}
    \quad \forall \ k=1,...,N
\end{equation}

in which the filtering posterior distribution $p(\mathbf{z}_k|y_{1:k}) \ \forall \ k=1,...,N$ is obtained from a sequential "forward pass" through the data via Kalman filtering. What is worthy to note is that the prior distribution at time step $k$ is taken to be the posterior distribution from the previous time step $k-1$. Hence, Bayes' rule can be applied recursively as the probability distribution of $\mathbf{z}_k$ is updated incrementally with each data point in $y_{1:k}$.
\\
\indent Kalman filtering and RTS smoothing is a practical algorithm for performing recursive Bayesian estimation for linear Gaussian systems. First, the state-space equations of the GP are discretized to suit numerical implementation: 

\begin{subequations}
    \begin{gather}
        \mathbf{z}_k = \mathbf{F}\mathbf{z}_{k-1} + \mathbf{L}w_{k-1} \label{subeq:discreteGPstate1}
        \\
        h(t_k) = \mathbf{H}\mathbf{z}_k \label{subeq:discreteGPstate2}
    \end{gather}
\end{subequations}

with discrete matrices $\mathbf{F} = \textup{exp}(\mathbf{F}_{\mathrm{c}} \Delta t)$, $\mathbf{L} = \mathbf{L}_{\mathrm{c}}$, $\mathbf{H} = \mathbf{H}_{\mathrm{c}}$. In this sense, Eq. \eqref{subeq:discreteGPstate1} constitutes the dynamics model representing the time evolution of the states, and Eq. \eqref{subeq:discreteGPstate2} constitutes the measurement model relating samples of $h(t)$ to the states. As the process is a linear SDE with normally distributed states, its posterior distribution can be recovered through the Kalman filter, which provides the prediction $p(\mathbf{z}_k|y_{1:k})$ using past observations in sequence, followed by the RTS smoother, which provides the  prediction $p(\mathbf{z}_k|y_{1:N})$ using a backward pass through the full observation dataset. The combination of filtering and smoothing gives the exact solution to GP regression for a set of noisy samples of the process \cite{Hartikainen2010}. The filtering and smoothing algorithms are detailed in \eqref{sec:app_KFKS}. 
\\
\indent In order to implement Kalman filtering and RTS smoothing, the covariance matrix $\mathbf{Q}$ of the discrete white noise $\mathbf{L}w_k$ is required. $\mathbf{Q}$ may be calculated from the spectral density of the continuous case of $w(t)$, $\mathbf{Q_c}$, as:

\begin{equation}
    \mathbf{Q} = \int_{0}^{\Delta t} \textup{exp}(\mathbf{F}_{\mathrm{c}}(\Delta t - \tau)) \ \mathbf{Q}_{\mathrm{c}} \ \textup{exp}(\mathbf{F}_{\mathrm{c}}(\Delta t - \tau))^\textup{T} \ d\tau
\end{equation}

In the steady-state condition, a practical method of computing $\mathbf{Q}$ is by enforcing $\mathbf{P}_k \rightarrow \mathbf{P}_\infty$ in the Kalman filter prediction equation (see \ref{sec:app_KFKS}): 

\begin{equation}
    \mathbf{Q} = \mathbf{P}_\infty - \mathbf{F}\mathbf{P}_\infty\mathbf{F}^\textup{T}
\end{equation}

\subsection{Gaussian process latent force model}
\label{sec:gplfm}

In the GPLFM, modal components of the unknown input driving the dynamic response of a physical system are modeled probabilistically as a GP, and a posterior estimate of their evolution over time may be computed through GP regression. However, in the context of \textit{latent} forces, no direct measurements of the process are available; rather, regression of the unknown input is performed through its relation with observed dynamics of the system captured in a \textit{hybrid model} consisting of a mechanical component and a stochastic component. The hybrid model provides a stochastic representation of the system dynamics, therefore allowing for the joint posterior inference of all latent variables, including unobserved dynamics such as the strain states. 
\\
\indent Using the reduced-order formulation of the mechanical model from Section \ref{sec:mechmodel}, unique sources of inputs $\mathbf{p}(t)$ may be identically decomposed into modal components $\mathbf{f}(t) = \begin{bmatrix} f_1(t) & ... & f_{n_{\mathrm{m}}}(t) \end{bmatrix}^\textup{T}$. \textcolor{black}{It is assumed that modal forces can be expressed as mutually uncorrelated GPs}, leading to the following linear block-diagonal state-space model for multiple modal forces:

\begin{subequations}
    \label{eq:blockforcemodel}
    \begin{gather}
    \label{eq:blockforcemodel1}
    \begin{bmatrix} \mathbf{\dot{z}}_1(t) \\ \vdots \\\mathbf{\dot{z}}_{n_{\mathrm{m}}}(t) \end{bmatrix} =
    \underbrace{ \begin{bmatrix} \mathbf{F}_{\mathrm{c},1}   &    &  \\
                        &   \ddots & \\
                        &   &   \mathbf{F}_{\mathrm{c},n_{\mathrm{m}}} \\ \end{bmatrix} }_\text{$\mathbf{\tilde{F}}_{\mathrm{c}}$}
     \begin{bmatrix} \mathbf{z}_1(t) \\ \vdots \\ \mathbf{z}_{n_{\mathrm{m}}}(t) \end{bmatrix} +
     \underbrace{ \begin{bmatrix} \mathbf{L}_{\mathrm{c},1}   &    & \\
                        &   \ddots & \\
                        &   &   \mathbf{L}_{\mathrm{c},n_{\mathrm{m}}} \\ \end{bmatrix} }_\text{$\mathbf{\tilde{L}}_{\mathrm{c}}$}
     \underbrace{ \begin{bmatrix} w_1(t) \\ \vdots \\ w_{n_{\mathrm{m}}}(t) \end{bmatrix} }_\text{$\mathbf{\tilde{w}}(t)$}
    \\
    \label{eq:blockforcemodel2}
    \begin{bmatrix} f_1(t) \\ \vdots \\ f_{n_{\mathrm{m}}}(t) \end{bmatrix} =
    \underbrace{ \begin{bmatrix} \mathbf{H}_{\mathrm{c},1}   &    &  \\
                        &   \ddots & \\
                        &   &   \mathbf{H}_{\mathrm{c},n_{\mathrm{m}}} \\ \end{bmatrix} }_\text{$\mathbf{\tilde{H}}_{\mathrm{c}}$}
     \begin{bmatrix} \mathbf{z}_1(t) \\ \vdots \\ \mathbf{z}_{n_{\mathrm{m}}}(t) \end{bmatrix} 
     \end{gather}
\end{subequations}

where for the $j$th modal force, $\mathbf{z}_j(t) = \begin{bmatrix} f_j(t) \quad \frac{df_j(t)}{dt} \quad \hdots \quad \frac{d^{\beta-1}f_j(t)}{dt^{\beta-1}} \end{bmatrix}^\textup{T}$ is the vector of derivatives corresponding to the companion form of the $\beta$-order SDE and $\mathbf{F}_{\mathrm{c},j}, \, \mathbf{L}_{\mathrm{c},j}, \, \mathbf{H}_{\mathrm{c},j}$, the white noise forcing $w_j(t)$ and its spectral density $q_{wj}$ are defined by the GP covariance kernel and associated hyperparameters from Section \ref{sec:inputmodel}. Equations \eqref{eq:blockforcemodel1}, \eqref{eq:blockforcemodel2} may be written in shorthand using the block state vector $\mathbf{s}(t) = \begin{bmatrix} \mathbf{z}_1(t) & ... & \mathbf{z}_{n_{\mathrm{m}}}(t) \end{bmatrix}^\textup{T}$ as: 

\begin{subequations}
    \begin{gather}
    \label{eq:blockforcemodelshort1}
    \mathbf{\dot{s}}(t) = \mathbf{\tilde{F}}_{\mathrm{c}}\mathbf{s}(t) + \mathbf{\tilde{L}}_{\mathrm{c}}\mathbf{\tilde{w}}(t)
    \\
    \label{eq:blockforcemodelshort2}
    \mathbf{f}(t) = \mathbf{\tilde{H}}_{\mathrm{c}}\mathbf{s}(t)
    \end{gather}
\end{subequations}

 The spectral density $\mathbf{\tilde{Q}}_{\mathrm{c}}$ of the term $\mathbf{L}_{\mathrm{c}}\mathbf{\tilde{w}}(t)$ is a block matrix with $\mathbf{L}_{\mathrm{c},j}q_{wj}\mathbf{L}_{\mathrm{c},j}^\textup{T}$ along the diagonal.
\\
\indent Because the mechanical model and the block GP model of the latent modal forces both take the form of a linear differential equation, it is possible to linearly combine the two into an augmented state-space model. We define the augmented state vector $\mathbf{z}^{\mathrm{a}}(t) = \begin{bmatrix} \mathbf{x}(t) & \mathbf{s}(t)\end{bmatrix}^\textup{T}$, where $\mathbf{x}(t) = \begin{bmatrix} \mathbf{r}(t) & \mathbf{\dot{r}}(t) \end{bmatrix}^\textup{T} \in \mathbb{R}^{2n_\mathrm{m}}$ denotes the modal displacement and velocity states. The GPLFM is given as: 

\begin{subequations}
\label{eq:gplfmfull}
    \begin{gather}
    \label{eq:gplfmfull1}
    \begin{bmatrix} \mathbf{\dot{x}}(t) \\ \mathbf{\dot{s}}(t)\end{bmatrix}
    = \underbrace{ \begin{bmatrix} \mathbf{A}_{\mathrm{c}} & \mathbf{B}_{\mathrm{c}} \\ \mathbf{0} & \mathbf{\tilde{F}}_{\mathrm{c}} \end{bmatrix} }_\text{$\mathbf{F}_{\mathrm{c}}^{\mathrm{a}}$}
    \begin{bmatrix} \mathbf{x}(t) \\ \mathbf{s}(t) \end{bmatrix}
    + \underbrace{ \begin{bmatrix} \mathbf{0} \\ \mathbf{\tilde{L}}_{\mathrm{c}}\mathbf{\tilde{w}}(t) \end{bmatrix} }_\text{$\mathbf{\tilde{w}}^{\mathrm{a}}(t)$}
    \\
    \label{eq:gplfmfull2}
    \mathbf{y}_k
    = \underbrace{ \begin{bmatrix} \mathbf{G} & \mathbf{J}\mathbf{\tilde{H}}_{\mathrm{c}} \end{bmatrix} }_\text{$\mathbf{H}^{\mathrm{a}}$}
    \begin{bmatrix} \mathbf{x} \\ \mathbf{s} \end{bmatrix}_k
    \end{gather}
\end{subequations}

or in the form: 

\begin{subequations}
\label{eq:gplfmshort}
    \begin{gather}
    \mathbf{\dot{z}}^{\mathrm{a}}(t) = \mathbf{F}_{\mathrm{c}}^{\mathrm{a}}\mathbf{z}^{\mathrm{a}}(t) + \mathbf{\tilde{w}}^{\mathrm{a}}(t)
    \\
    \mathbf{y}_k = \mathbf{H}^{\mathrm{a}}\mathbf{z}^{\mathrm{a}}_k
    \end{gather}
\end{subequations}

where the spectral density $\mathbf{Q}_{\mathrm{c}}^{\mathrm{a}}$ of the augmented white noise vector $\mathbf{\tilde{w}}^{\mathrm{a}}(t)$ is given as: 

\begin{equation}
    \label{eq:augspectraldensity}
    \mathbf{Q}_{\mathrm{c}}^{\mathrm{a}} = \begin{bmatrix} \mathbf{0} & \mathbf{0} \\
    \mathbf{0} & \mathbf{\tilde{Q}}_{\mathrm{c}} \end{bmatrix}
\end{equation}

Eq. \eqref{eq:gplfmshort} constitutes the continuous-discrete version of the GPLFM, describing the time evolution of both the latent forces and latent system states. The hybrid model adopts the form of a linear SDE driven by a steady-state Gaussian input, and because the solution to the SDE is a linear operation of the input, the resulting system states are also GPs \cite{Sarkka2019}. Therefore, joint posterior inference of the augmented states $\mathbf{z}^{\mathrm{a}}(t)$ reduces to a GP regression problem, where regression of the measured variables provides evidence to simultaneously extrapolate the latent variable response \cite{Hartikainen2011}.
\\
\indent Kalman filtering and RTS smoothing may be implemented to solve GP regression. The model is discretized and noise sources are considered for the purpose of filtering and inference, with $\mathbf{F}^{\mathrm{a}} = \exp(\mathbf{F}_{\mathrm{c}}^{\mathrm{a}} \, \Delta t)$: 

\begin{subequations}
    \begin{gather}
        \label{eq:gplfmshortdiscr1}
        \mathbf{z}^{\mathrm{a}}_k = \mathbf{F}^{\mathrm{a}}\mathbf{z}^{\mathrm{a}}_{k-1} + \mathbf{\tilde{w}}^{\mathrm{a}}_{k-1} 
        \\
        \label{eq:gplfmshortdiscr2}
        \mathbf{y}_k = \mathbf{H}^{\mathrm{a}}\mathbf{z}^{\mathrm{a}}_k + \mathbf{v}_k
    \end{gather}
\end{subequations}

where as in Eq. \eqref{eq:ssm}, $\mathbf{v}_k$ is additive white Gaussian measurement noise with covariance $\mathbf{R}$. Process noise is considered within the term $\mathbf{\tilde{w}}^{\mathrm{a}}_k$, a notion which is discussed further in Section \ref{subsec:noise}. 
\\
\indent We may adopt the same steps as in Section \ref{subsec:gpr_recursive} to derive moments of the LTI SDE solution. In particular, the augmented steady-state covariance $\mathbf{P}_\infty^{\mathrm{a}}$ of the augmented states $\mathbf{z}^{\mathrm{a}}(t)$ is computed from the Lyapunov equation: 

\begin{equation}
    \label{eq:auglyapunov}
    \mathbf{0} = \mathbf{F}^{\mathrm{a}}_{\mathrm{c}}\mathbf{P}^{\mathrm{a}}_\infty + \mathbf{P}^{\mathrm{a}}_\infty(\mathbf{F}^{\mathrm{a}}_{\mathrm{c}})^\textup{T} + \mathbf{Q}^{\mathrm{a}}_{\mathrm{c}}
\end{equation}

and the augmented covariance matrix $\mathbf{Q}^{\mathrm{a}}$ of the discrete noise $\mathbf{w}^{\mathrm{a}}_k$ follows as: 

\begin{equation}
    \label{eq:augcovmatrix}
    \mathbf{Q}^{\mathrm{a}} = \mathbf{P}^{\mathrm{a}}_\infty - \mathbf{F}^{\mathrm{a}}\mathbf{P}^{\mathrm{a}}_\infty(\mathbf{F}^{\mathrm{a}})^\textup{T}
\end{equation}

\subsubsection{Parameter inference for the GP kernel function}
\label{subsec:GPkernels}

Given a GP kernel, the corresponding hyperparameters $\theta$ which govern the length scales of the resulting process may be tuned according to the measurement dataset for a particular system. Nayek et al. \cite{Nayek2019} proposes to use maximum likelihood estimation for parameter inference, in which the optimal hyperparameters are obtained from the least squares solution which minimizes the residual norm between the observed data and model predictions. For Markov processes, maximization of the likelihood function $p(Y|\theta)$ may be reframed as the minimization of the negative log likelihood and computed iteratively using the Kalman filter equations. However, since such linear least squares problems are frequently ill-posed \cite{Kaipio2005}, the negative log likelihood function may not result in real-valued evaluations and its optimization is known to be susceptible to false solutions at local minima \cite{Nayek2019}. A discussion of the computational issues can be found in \cite{Petersen2022}. Rogers et al. \cite{Rogers2020} suggest a random walk Metropolis-Hastings algorithm to enforce a log acceptance ratio and avoid complex values in the objective function. Moreover, they note that initialization of the objective function at the prior $p(\theta)$ produces the maximum a posteriori (MAP) estimate of the hyperparameters.  
\\
\indent As the iterative computation of the objective function using the Kalman filter is computationally intensive, particularly for long time-series data, this work implements a parameter fitting procedure for obtaining optimal values of the hyperparameters $\theta^*$ prior to filtering. Through properties of linear Gaussian systems, the GPLFM inherently places a Gaussian prior over the both latent and observed variables of the system. As prior assumptions are encoded by the kernel function assumed in the GPLFM, the hyperparameters of the Gaussian prior correspond exactly to those of the augmented model. Empirical qualities of the measurement data, such as the empirical mean and covariance, are then leveraged as prior information which is used to tune the kernel hyperparameters. In this approach, model parameters are fit by minimizing the distance $d$ between the empirical distribution of the measurements, $p(Y) \sim \mathcal{N}(\mathbf{0}, \mathbf{P}^{\mathrm{y}}_0)$, and the modeled Gaussian prior on the observed states $p(Y|\theta) \sim \mathcal{N}(\mathbf{0}, \mathbf{\hat{P}}^{\mathrm{y}}_{\infty})$, where by the steady-state condition, the prior covariance is $\mathbf{\hat{P}}^{\mathrm{y}}_{0} = \mathbf{\hat{P}}^{\mathrm{y}}_{\infty}$. For shorthand notation, $p(Y|\theta)$ is referred to by $p^{\theta}(Y)$. The optimal $\theta^*$ is then solved from: 

\begin{equation}
    \label{eq:mle}
    \theta^* = \underset{\theta}{\arg\min} \ d(p, p^\theta)
\end{equation}

In this work, the distance metric $d$ between two probability measures is taken to be the Hellinger distance $H(p, p^\theta)$, which is provided in \ref{sec:app_hellinger}.
\\
\indent As the distributions are taken to be zero-mean, parameter inference reduces to a problem of matching the covariance predicted by the model to the underlying covariance in the data. The empirical measurement covariance $\mathbf{P}^{\mathrm{y}}$ is computed directly from time-series data of the sensor channels. The modeled measurement covariance, $\mathbf{\hat{P}}^{\mathrm{y}}_{\infty}$, may be derived from the augmented steady-state covariance encoded by the GP kernel function, $\mathbf{P}^{\mathrm{a}}_\infty$, by passing it through the measurement model $\mathbf{H}^{\mathrm{a}}$ of the GPLFM with the transformation: 

\begin{equation}
    \label{eq:measurementcov}
    \mathbf{\hat{P}}^{\mathrm{y}}_{\infty} = \mathbf{H}^{\mathrm{a}} \, \mathbf{P}^{\mathrm{a}}_\infty \, (\mathbf{H}^{\mathrm{a}})^\textup{T}
\end{equation}

\indent Rather than computing the likelihood function iteratively over each measurement, this approach takes advantage of the linear Gaussian nature of the GPLFM, such that the likelihood of $\theta$ may be computed efficiently using low-order moments of long time-series data. As the distance between Gaussian distributions in Eq. \eqref{eq:mle} results in a smooth, unimodal optimization surface across a defined search domain, the minimization problem may be solved efficiently using gradient descent methods.

\subsubsection{Treatment of noise in the GPLFM}
\label{subsec:noise}

A feature of the GPLFM which stands in contrast to other Kalman filter-based estimators is that the definition of stochastic process noise on the augmented states, whose covariance $\mathbf{Q}^{\mathrm{a}}$ is solved for by means of the Lyapunov equation in Eqs. \eqref{eq:auglyapunov} and \eqref{eq:augcovmatrix}, is inherent to the formulation of a stationary GP in state-space. That is, the joint noise covariance of inputs and states are parameterized in terms of a handful of GP hyperparameters. This parameterization provides a more complete characterization of the model compared to the parameterization suggested in \cite{Yuen2007} for other estimators, where tunable parameters are limited to diagonal elements of the covariance matrix, or in \cite{Azam2015, Lourens2012b}, where a constant representing the order of magnitude of variance across all states is estimated using the L-curve criterion. Using Eq. \eqref{eq:augcovmatrix}, an appropriate augmented process noise covariance matrix $\mathbf{Q}^{\mathrm{a}}$ may be solved for by a data-driven technique of parameter inference, where the parameters are the GP kernel hyperparameters, rather than by heuristic tuning or ad-hoc methods. The resultant noise covariance matrix captures correlations between the unknown inputs and states which are generally neglected in the AKF and DKF.  
\\
\indent An underlying assumption in the GPLFM of Eq. \eqref{eq:gplfmfull} is that sources of process noise on response states - any dynamical behavior which is unaccounted for in the mechanical model - are treated as additional external input. In this sense, the GPLFM does not distinguish between noise contributions and input excitation, which are characterized together as a GP perturbed by additive white noise. As a consequence, the input estimation is relatively sensitive to large model uncertainty, and the strain estimation may be susceptible to overconfidence. This aspect is generally not a concern in applications where one is less interested in the input estimation for which validation is impractical. However, for applications where an accurate input estimate is necessary, or wider variance bounds on the strain estimate are expected, process noise in the system states $\mathbf{x}$ may be distinguished either by the addition of process noise in the latent system states $[\mathbf{Q}^{\mathrm{x}} \quad \mathbf{0}; \quad  \mathbf{0} \quad \mathbf{0}]$, adding $\mathbf{Q}^{\mathrm{x}} \in \mathbb{R}^{2n_{\mathrm{m}} \times 2n_{\mathrm{m}}}$ to the augmented covariance matrix in Eq. \eqref{eq:augcovmatrix}, as in \cite{Nayek2019, Petersen2022}, or by the incorporation of the spectral density of the process noise $\mathbf{\eta}(t)$ into the vector of hyperparameters $\theta$ to solve for, as in \cite{Rogers2020}. 
\\
\indent The accuracy of the GPLFM is also sensitive to the choice of measurement noise covariance $\mathbf{R}$. Rather than choose a prior estimate of the measurement noise covariance, a method for posterior tuning of $\mathbf{R}$ is adopted from  \cite{Bilbao2021},  where measured data channels are used as an indicator of sensor noise. Upon obtaining a posterior estimate $p(\mathbf{z}^{\mathrm{a}}(t) | Y, \theta) = \mathcal{N}(\hat{\mathbf{z}}^{\mathrm{a}}(t), \mathbf{P}^{\mathrm{a}}(t))$ from Kalman filtering/RTS smoothing, the measurement noise is taken to be the residual $\mathbf{r}_k$ between the observed data $Y$ and prediction on the observed variables $\mathbf{H}^{\mathrm{a}}\hat{\mathbf{z}}^{\mathrm{a}}_k$ as follows: 

\begin{subequations}
    \label{eq:residual}
    \begin{gather}
        \mathbf{y}_k = \mathbf{H}^{\mathrm{a}}\hat{\mathbf{z}}^{\mathrm{a}}_k + \mathbf{r}_k
        \\
        \mathbf{r}_k = \mathbf{y}_k - \mathbf{H}^{\mathrm{a}}\hat{\mathbf{z}}^{\mathrm{a}}_k 
    \end{gather}
\end{subequations}

\begin{figure}[h!]
    \centering
    \includegraphics[width=0.95\linewidth]{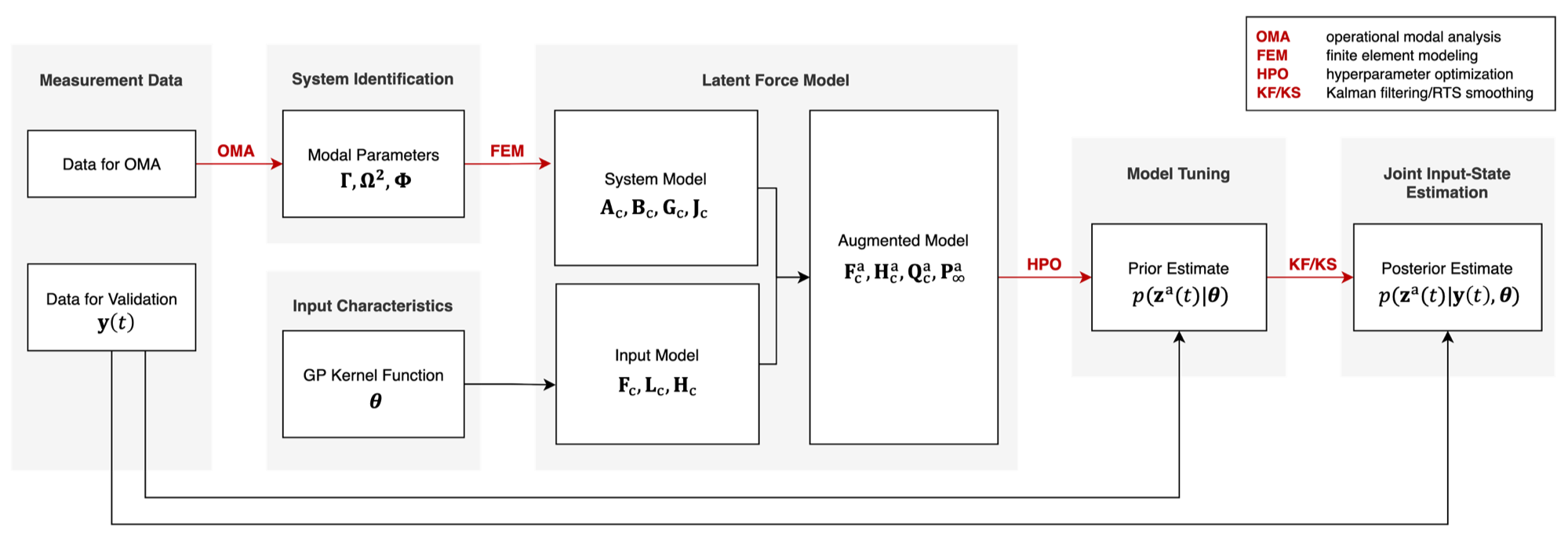}
    \caption{Diagram of GPLFM methodology.}
    \label{fig:flowchart}
\end{figure}

Measurement noise tuning then becomes a problem of minimizing the error between the assumed measurement noise covariance $\mathbf{R}$ in the model and the covariance of the residual time series $\mathbf{\hat{R}} = \mathbb{E}[\mathbf{r}_k\mathbf{r}_k^\textup{T}]$, executed through multiple iterations of posterior estimation using the GPLFM. One may specify termination criteria for arriving at a final estimate of the measurement noise covariance, such as a tolerance between consecutive iterates of $\mathbf{R}$. It is observed that the measurement noise covariance tuning procedure results in greater accuracy of estimates of latent states, as demonstrated in a numerical example in \ref{sec:app_modelerror}.
\\
\indent An overview of the GPLFM methodology is illustrated in Figure \ref{fig:flowchart}.  Source code for this work is publicly available at \href{https://github.com/joannajzou/GPLFM}{https://github.com/joannajzou/GPLFM}.

\subsubsection{Effect of modeling assumptions}
\label{subsec:modelassumptions}

This section discusses the degree of sensitivity of the GPLFM to various sources of model error which may be encountered in virtual sensing applications. In the following, we give emphasis to modeling assumptions chosen for this study and their impact on estimation quality, as well as possible extensions to more general assumptions. To substantiate the discussion on model error, an illustrative example is presented in \ref{sec:app_modelerror} in which the sensitivity of the GPLFM to perturbation in model properties is evaluated on a synthetic model, where the true input and response are known. 
\\
\indent \textit{Gaussianity of the input and structural response.} An underlying assumption in the adoption of the GPLFM for virtual sensing is that the unknown input and response states of the structure are linearly related and Gaussian in distribution. This assumption is deemed valid in structural health monitoring applications which utilize ambient vibrations to determine long-duration changes in structural conditions, a setting where only the linear elastic regime of structural response is engaged. In the case study which follows in Section \ref{sec:casestudy}, it is affirmed that the strain and acceleration data collected from an operating offshore wind turbine have ensemble distributions which are very close to Gaussian in nature. Moreover, an illustrative example of the GPLFM implemented on a synthetic model is presented in \ref{sec:app_modelerror}, where for a structure excited by a non-Gaussian load, the response is still well captured using the GPLFM. Therefore, the assumption of Gaussianity is not expected to introduce significant error in joint input-state estimation results for the present in-situ validation study. While the present work focuses on an approach to fatigue monitoring over long time scales, Rogers et al. \cite{Rogers2020} have explored the coupling of input and state estimation with parameter estimation, to track shifts in modal parameters which can indicate nonlinear damage. 
\\
\indent \textit{Stationarity of the GP.} In the present study, the GP adopted in the formulation of the GPLFM is assumed to be zero-mean stationary. This assumption is hypothesized to be sufficient for the virtual sensing of dynamic strains, in which low-frequency components are not considered, for a structure subject to low-intensity ambient vibration, and is further evaluated in subsequent validation experiments. Assumptions of first-order stationarity allow for the GP to be modeled as a steady-state linear time invariant (LTI) SDE; for such systems, the Lyapunov equation may be invoked as an computationally convenient method to solve for the steady-state probability density over the states, as in Equation \eqref{eq:lyapunov}. The GPLFM may be implemented with assumptions of first-order stationarity removed by computing the time-dependent probability density over states $\mathbf{z}(t)$ via Equation \eqref{eq:meancovsoln} for GPs with time-invariant system matrices, which produces a solution with a non-zero prior mean sequence. A further extension would be to consider time-varying system matrices of the GP, for which the probability density is computed by direct solution to the Fokker-Planck-Kolmogorov equation \cite{Sarkka2012}.
\\
\indent \textit{Choice of GP covariance kernel.} The performance of the GPLFM in joint input-state estimation depends on an appropriate choice of the GP kernel and its hyperparameters. Generally, the kernel is chosen from a standard set of convenient kernels which meet the criteria for conversion to state-space, as discussed in Section \ref{subsec:stationaryGPs}, based on qualitative prior knowledge of the time-varying behavior (e.g. periodicity, smoothness and differentiability) of the stochastic input. One example of a suitable kernel is the Mat\'ern kernel function, which characterizes processes that are finitely differentiable \cite{Hartikainen2010}. In this work, we exclusively make use of the Mat\'ern kernel function, as it is flexible enough to accommodate a broad range of process traits and is commonly employed in engineering applications to represent physically realistic processes. In the case of the Mat\'ern kernel with smoothing parameter $p=2$, covariance relationships are governed by two hyperparameters $\theta = [\alpha, l_s]$, where the vertical length scale $\alpha$ controls the amplitude and the horizontal length scale $l_s$ controls the rate of time variability, and therefore the frequency content, of the resulting function.  For convenience, the state-space model of the Mat\'ern kernel is provided in \ref{sec:app_matern}. Other common kernel choices which are valid for state-space implementation include the white noise (or Brownian motion) kernel function, which represents each time step of the GP as independent and identically distributed; the squared exponential kernel function, which describes infinitely differentiable processes; and the periodic kernel function, which captures strong harmonic components in time series data. Both the squared exponential kernel and periodic kernel require an approximation of the spectral density to be converted into state-space form, the details of which are provided in \cite{Hartikainen2010} and \cite{Solin2014}, respectively. For certain applications, GP regression may be particularly sensitive to the choice of kernel, where the substitution of another kernel function leads to a substantially different posterior prediction. While an evaluation of GP kernel sensitivity is beyond the scope of this work, an optimization-based approach to such a sensitivity study can be found in \cite{Stephenson2021}.
\\
\indent \textit{Identical GP model for all modal force components.} In the virtual sensing case study which follows in Section \ref{sec:casestudy}, it is assumed that all modal forces $f_j(t), j = 1,..., n_\textup{m}$ are endowed with the same GP prior. In other words, in Equation \ref{eq:blockforcemodel}, the modal forces are modeled with state-space matrices $\mathbf{F}_{c,j}, \mathbf{L}_{c,j}, \mathbf{H}_{c,j}, w_j(t)$ which are identical for all $j$ and determined by a common set of GP hyperparameters $\theta$. It is important to note that while these hyperparameters determine the prior variance placed on the companion form of the modal forces $\mathbf{z}_j(t)$, the posterior variance on these states is solved for via a Kalman filter update of the prior variance utilizing information from measurement data. Therefore, the posterior distribution will differ between components of the modal force even if the same GP hyperparameters are used to define the prior distribution. Moreover, in the Bayesian framework to posterior inference, with the availability of a sufficiently large set of measurement data, influence over the posterior is dominated by the likelihood and prior assumptions become negligible \cite{Kaipio2005}. While it is possible to assign unique GP hyperparameters $\theta_j$ to each modal force, it is observed, but not shown in the paper for the sake of brevity, that this approach does not improve accuracy in the posterior estimate of the latent states in the present case study, and comes at the cost of a more expensive high-dimensional optimization problem which increases the complexity of achieving optimal hyperparameter tuning. 
\\
\indent \textcolor{black}{ \textit{Uncorrelated modal forces.} The simplifying assumption is made that all modal forces $f_j(t), j = 1,..., n_\textup{m}$ are uncorrelated with each other, represented by diagonal terms being set to zero in the block-diagonal state-space model in Equation \eqref{eq:blockforcemodel}. In reality, structural systems exhibit coupling between dynamic modes, which must be captured for a complete characterization of the covariance relationship between inputs and states. Neglecting information on the correlation between modal forces leads to reduced estimation accuracy and underestimation of uncertainty. In this work, this assumption is taken as necessary in the absence of prior knowledge of the nature of linear correlation between modal components of the unknown forcing. Nevertheless, our results on joint input-state estimation for a numerical system in \ref{sec:app_modelerror} and the case study wind turbine system in Section \ref{sec:results} provide empirical evidence that this assumption still delivers good estimates. Moreover, the state-space form of the GPLFM provides a useful framework for overcoming this assumption; an initial data-driven approach to modeling the correlation between modal forces is to parameterize the covariance term between $(f_i(t), f_j(t))$ for $i \neq j$ in the appropriate matrix entries of Equation \eqref{eq:blockforcemodel} and append these parameters to the set of kernel hyperparameters jointly inferred in an optimization routine. }

\indent \textit{Fixed modal parameters of the mechanical model.} In practical applications of the GPLFM for virtual sensing, the mechanical model is an idealization of the structural system and cannot be known exactly. In general, the mechanical model is constructed to fit measured properties of the in-situ structure - in particular, modal properties such as the modal damping ratios, natural frequencies, and mode shapes which dictate dynamic response. However, for large scale structures, these modal properties must be inferred rather than measured directly. Modal identification of offshore wind turbines by traditional operational modal analysis (OMA) techniques is generally limited to the prediction of modal properties of the OWT in the idling state, when loading conditions most closely match the white noise assumption for ambient loading in OMA \cite{Peeters1999}. In the run-up and production states, these modal properties shift due to variation in damping sources and the introduction of interaction with harmonic loads from the environment and rotor blade dynamics \cite{vanVondelen2021}. Damping identification presents a particular challenge as the level of contribution of aerodynamic damping, in addition to hydrodynamic damping and soil damping in offshore contexts, is expected to fluctuate widely over time \cite{Devriendt2014, vanVondelen2021}. Therefore, the results of OMA obtained from data from the idling state do not fully reflect the true modal properties of the OWT in other states and are expected to result in an underestimation of modal damping levels, as well as a mismatch in natural frequencies and mode shapes. It is shown in an illustrative example in \ref{sec:app_modelerror} that errors in modal properties of the system affect input estimation by the GPLFM, but as a measure to result in reduced influence on estimates of response states such as displacements and accelerations. It is expected that error in response estimation resulting from error in modal properties will be present but not necessarily significant, due to compensatory effects of the input estimation. 
\\
\indent \textit{Beam theory model for the transformation of displacements to axial strains.} Since the strain response of a structure is required to assess fatigue damage accumulation, the objective of virtual sensing is to estimate the strain response at fatigue-critical locations where direct measurement of strains is infeasible. When representing the structure with a mechanical model of the form in Equation \ref{eq:eom}, dynamic response states of the model correspond to displacements and its higher order derivatives, including velocities and accelerations. Estimates of axial strain response must be computed by an appropriate transformation of the lateral displacement response produced by the model. A simplifying but practically necessary assumption is to perform the transformation using elastic beam theory as presented in \ref{sec:app_disptransf}, which relies on approximations of translational, rotational, and shear deformations described by finite element shape functions. In the present validation study, the shape functions are chosen to be those of Timoshenko beam elements, provided in \cite{Liu2003}. Due to this approximation, it is likely that deterministic conversion of displacements to strains will result in mismatch between the predicted strain response and the true strain response of the structure.

\subsubsection{Practical notes for numerical implementation}
\label{subsec:practical}

This section briefly discusses numerical sensitivities particular to the presented formulation of the GPLFM and approaches to resolve such issues, which are in part sourced from \cite{Grewal2001}.
\\
\indent \textit{High dimensionality.} For physical systems with several degrees of freedom, operations on large matrices in the Kalman filter and RTS smoother is susceptible to roundoff error and challenges with the inversion of sparse matrices. With model order reduction of the mechanical model in Section \ref{sec:mechmodel}, the dimension of the system becomes equal to the $n_\mathrm{m}$ number of modes considered, thus reducing the cost of computation and controlling for stable matrix inversion. Moreover, the reduced order formulation produces an estimate of the modal contribution of any number of forces rather than the exact forces, such that knowledge of the location of the unknown inputs is not necessary \cite{Lourens2017, Lourens2019}. For instance, in the transformation of forces into their modal contributions by $\mathbf{f}(t)= \bar{\Phi}^\textup{T} \mathbf{S}_{\mathrm{p}}\mathbf{p}(t)$, $\mathbf{S}_{\mathrm{p}}$ may be taken to be the identity matrix such that the collective effect of all possible forces $\mathbf{p}(t)$ acting at any degree of freedom is considered and no error is introduced from incorrect assumptions on the force locations.
\\
\indent \textit{Ill-conditioning of covariance matrices.} The Kalman filter algorithm is sensitive to ill-conditioning of the covariance matrices, which more commonly occurs in the companion state-space model where higher derivatives of a process are included in the state vector \cite{Grewal2001}. The ill-conditioning may be resolved through a) reformulating the update equation in the Kalman filter using matrix decomposition methods, such that filtering is performed using stable factors of the covariance matrix $\mathbf{P}_k$, and b) ensuring the addition of non-zero measurement noise $\mathbf{R}$ to stabilize the update equation in events of near-zero change from $\mathbf{P}_{k-1}$ to $\mathbf{P}_k$. 
\\
\indent \textit{Numerical instability from scale difference.} Numerical instability may arise if there exists a large scale difference between components of the state vector $\mathbf{x}$ or measurement vector $\mathbf{y}$, as the filtering of large components may cause cancellation of the significant digits of small components. Scale differences are particularly common when heterogeneous measurement data of different units are used, as is the case with the combination of acceleration and strain measurements. Such numerical errors can be avoided through either a change of units or a rescaling of the vector. In this work, we recommend at a minimum to normalize the measured quantities $\mathbf{y}$ prior to filtering, then applying the reverse operation to the posterior estimate to recover the true signal magnitude.

\section{Offshore wind turbine case study}
\label{sec:casestudy}

In this contribution, validation of the GPLFM for dynamic strain estimation is performed using measurement data from the SWT 3.0-DD offshore wind turbine on a monopile foundation, referred to as the W27 turbine hereon. The W27 turbine is a member of Westermeerwind Park, an operating near-shore wind farm situated in the IJsselmeer Lake in the Netherlands. The instrumentation scheme for the W27 turbine is rare by industry standards, as it contains strain sensors on the monopile foundation at varying depths below the mudline.

\subsection{Measurement setup and data processing}
\label{sec:measurement}

\begin{figure}[hbt!]
    \centering
    \includegraphics[width=0.75\linewidth]{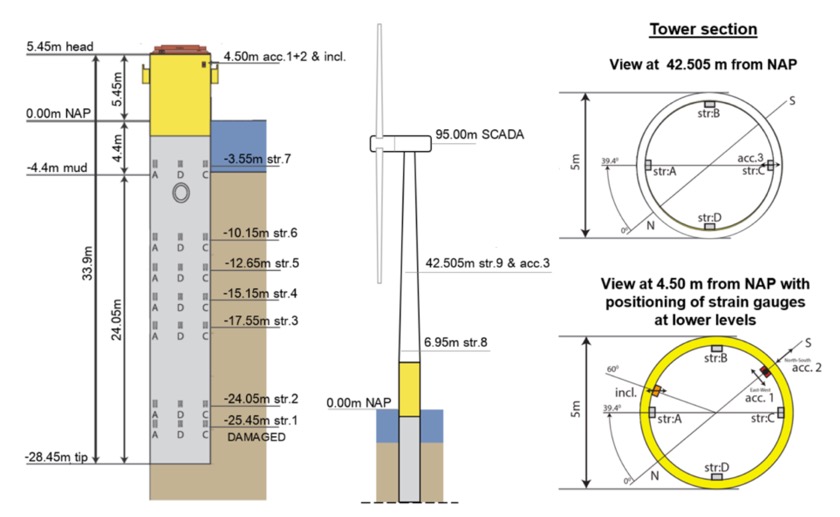}
    \caption{Position of 8 strain gauge rings and 3 accelerometers along the support structure of the W27 wind turbine. Figure adopted from [Versteijlen et al.].}
    \label{fig:sensorsetup}
\end{figure}

The measurement set used for validation consists of the following data, which was gathered from separate data acquisition systems. All positions of sensors are given with respect to the mean water level (NAP) and illustrated in Figure \ref{fig:sensorsetup}.
\begin{itemize}
    \item Rotor-nacelle assembly (RNA) data: data obtained from the SCADA system, where associated sensors are located at the level of the rotor axle at +95.0m NAP. Accelerations in the fore-aft (FA) and side-side (SS) directions of the turbine are recorded at the RNA, where FA references the direction of the rotor axis and SS is the perpendicular direction. In addition, the RNA data consists of measurement channels for wind direction and wind speed, rotor speed, blade pitch angle, power generation level, and yaw angle (rotational position of the nacelle).
    \item Support structure (SUS) data: data from accelerometers and strain gauges installed along the height of the turbine tower and monopile foundation.  Three accelerometers are located above the waterline: Acc. 1, measuring in the east-west direction at +4.5m NAP; Acc. 2, measuring in the north-south direction at +4.5m NAP; and Acc. 3, measuring along the 39.4\textdegree - 219.4\textdegree \ line from North at +42.5m NAP. The SUS data includes nine rings of 4 strain gauges labeled A, B, C, D, measuring axial strains at four faces of the monopile; however, several of the strain gauges at positions C and D, as well as all strain gauges in the first ring level, were damaged during the pile-driving process. Therefore, strain gauge readings at positions A and B in the five rings located below the mudline (Rings 2-6) and three rings located above the mudline (Rings 7-9) are retained for validation. The inclinometer reading at +4.5m NAP is neglected in this study. 

\end{itemize}

\indent The RNA and SUS data are collected continuously and organized into 10-minute increments over a 15-month period between 2016 and 2017. The turbine is monitored over various operational states of the OWT, namely the idling state (minimal rotor motion), production state (full rotor speed), and run-up state (the transitional state). Each of the 10-minute records is categorized by operational state using state curves, such as the wind speed-rotor speed curve in \cite{Sun2019}.
\\
\indent Due to the use of different data acquisition systems, signal processing steps are applied to ensure synchronization between the RNA and SUS data. First, the RNA data is downsampled to meet the 50 Hz sampling rate of the SUS data, which is determined to be sufficient for capturing the frequency resolution of all data channels, as the significant frequency content is expected to fall below 10 Hz. Resampling is followed by the time series synchronization of the data channels based on cross-correlation sequences. Full details on the time series synchronization procedure are provided in \cite{Laanes2021}. Because the primary interest in this study is to estimate dynamic strain response which drives fatigue damage accumulation, quasi-static components are removed via a bandpass Butterworth filter of order 6 with stopband frequencies of 0.12 Hz and 10 Hz. 

\subsection{Modal analysis}
\label{sec:oma}

\begin{figure}[hbt!]
    \centering
    \includegraphics[width=\linewidth]{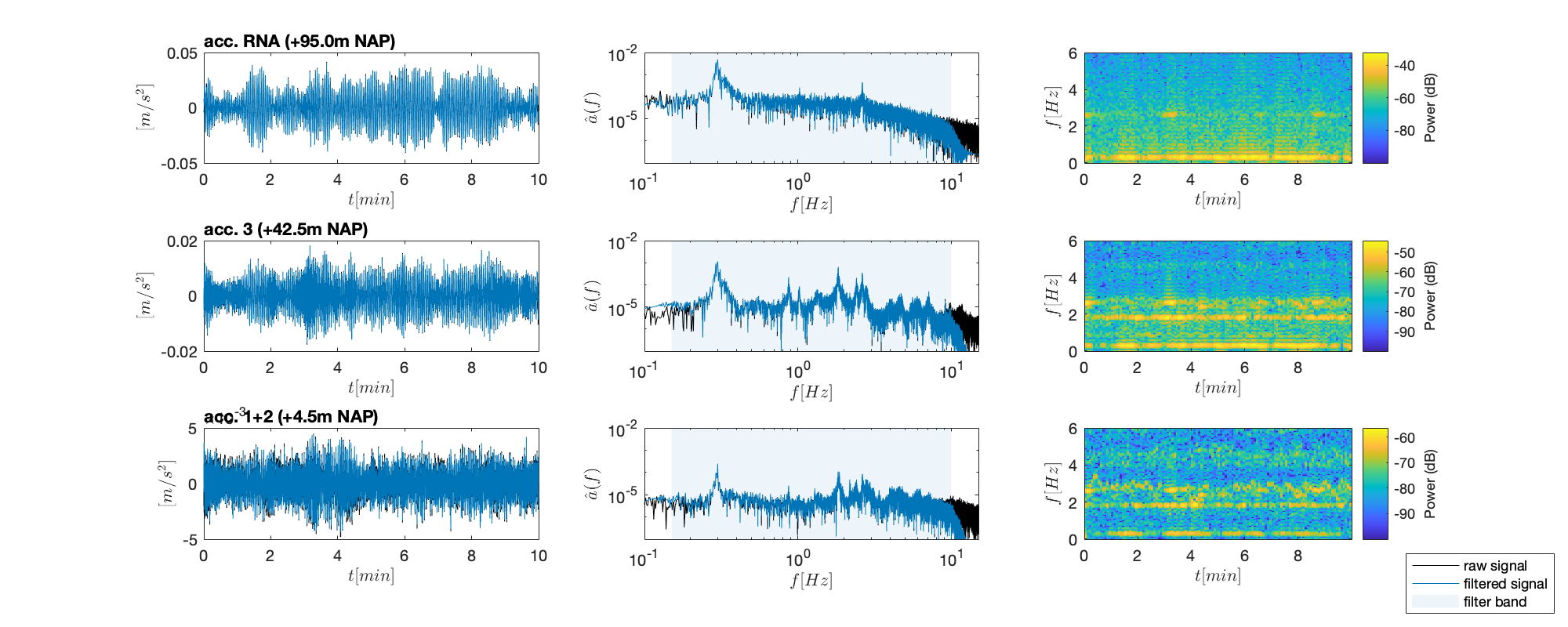}
    \caption{Sample 10-minute record of the three acceleration channels used for modal identification, shown with the filtered time series (left column), discrete Fourier transform (center column), and spectrogram (right column) of the data. The cutoff frequencies of the bandpass filter are indicated in the Fourier spectrum.}
    \label{fig:datasample}
\end{figure}

To identify in-situ modal parameters of the structural system - such as mode shapes, natural frequencies, and damping ratios - operational modal analysis (OMA) is conducted using the covariance-driven stochastic subspace identification (SSI-Cov) technique \cite{Peeters1999}. For the W27 turbine, acceleration measurements at three elevations from Acc. 1 and 2, Acc. 3, and the accelerometer in the RNA are transformed into the local RNA coordinate system to allow for separate identification of the FA and SS modes. From the stabilization diagram produced by the SSI-Cov routine, stable columns corresponding to structural modes are identified using the OPTICS clustering algorithm \cite{Ankerst1999, Boroschek2019}. 
\\
\indent As with the majority of OMA techniques, it is implicitly assumed in SSI-Cov that 1) the system is linear time invariant, meaning the modal parameters remain unchanged over time; 2) the system is driven by unobserved ambient excitation which takes the form of white Gaussian noise; and 3) the system response is not colored by nonstructural damping sources. Therefore, to ensure robust and accurate results, a number of considerations are made in the selection of 10-minute records to use for modal analysis. A) The dataset is filtered for records which correspond to the idling state, as loading conditions in this state most closely match the assumption of white noise excitation in the SSI-Cov method and allow for structural resonance to be discerned from noise \cite{Devriendt2014}. B) The records are further filtered based on the yaw angle of the turbine in order to allow for distinction between the FA and SS modes. The specific sensor configuration of the W27 turbine presents the limitation that Acc. 3 only provides a uniaxial measurement along the 39.4\textdegree - 219.4\textdegree \ line, meaning it cannot measure components of motion perpendicular to its axis. Therefore, the RNA must be oriented parallel to the axis of Acc. 3 for the channel to be utilized in identifying FA modes and perpendicular to its axis for SS modes, such that the measurement corresponds completely to the motion in the respective direction. Records with little to no deviation in yaw angle across the 10-minute period are chosen to reduce conflation with out-of-plane modes. C) Of the remaining records, those which correspond to low wind speeds are prioritized in the selection in order to reduce the effect of aerodynamic damping on the modal estimate. D) Records in the final selection are truncated to sections of the data with relatively consistent time-frequency content which reflects stationary conditions. Following these criteria, the FA and SS modal parameters are each derived from twelve 10-minute records taken across the 15-month duration of the monitoring campaign, with a sample record shown in Figure \ref{fig:datasample}. It should be clarified that these criteria for record selection only applies to the dataset used for modal identification by SSI-Cov; the validation records used in strain estimation are not subject to such constraints, but rather are chosen broadly across operational states and wind speeds. 

\subsection{Finite element model}
\label{sec:fem}

\begin{figure}[hbt!]
    \centering
    \includegraphics[width=0.8\linewidth]{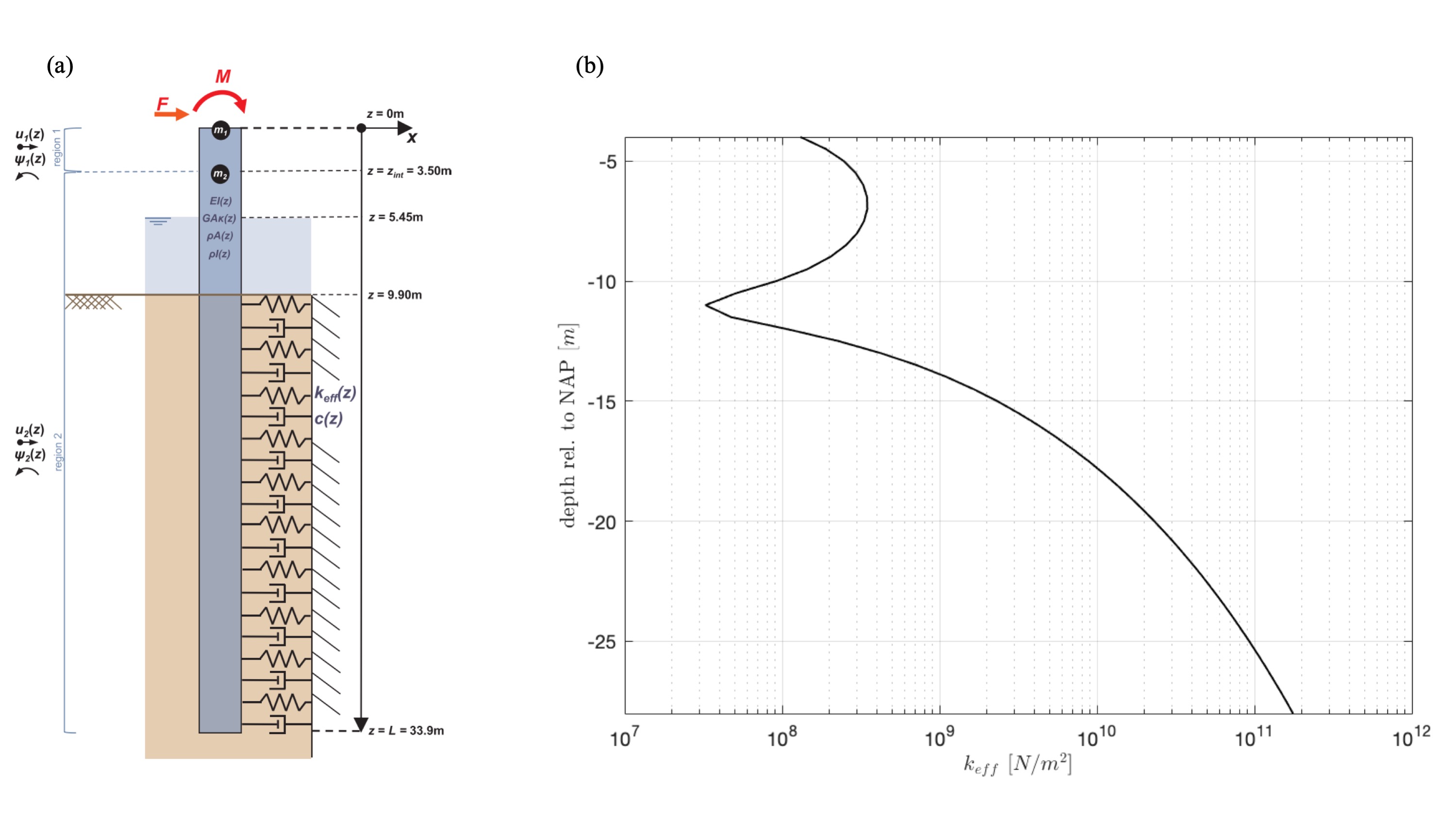}
    \caption{Diagram of Winkler foundation model and soil stiffness profile. Figure adopted from [Versteijlen et al.].}
    \label{fig:foundation}
\end{figure}

\begin{figure}[hbt!]
    \centering
    \includegraphics[width=0.8\linewidth]{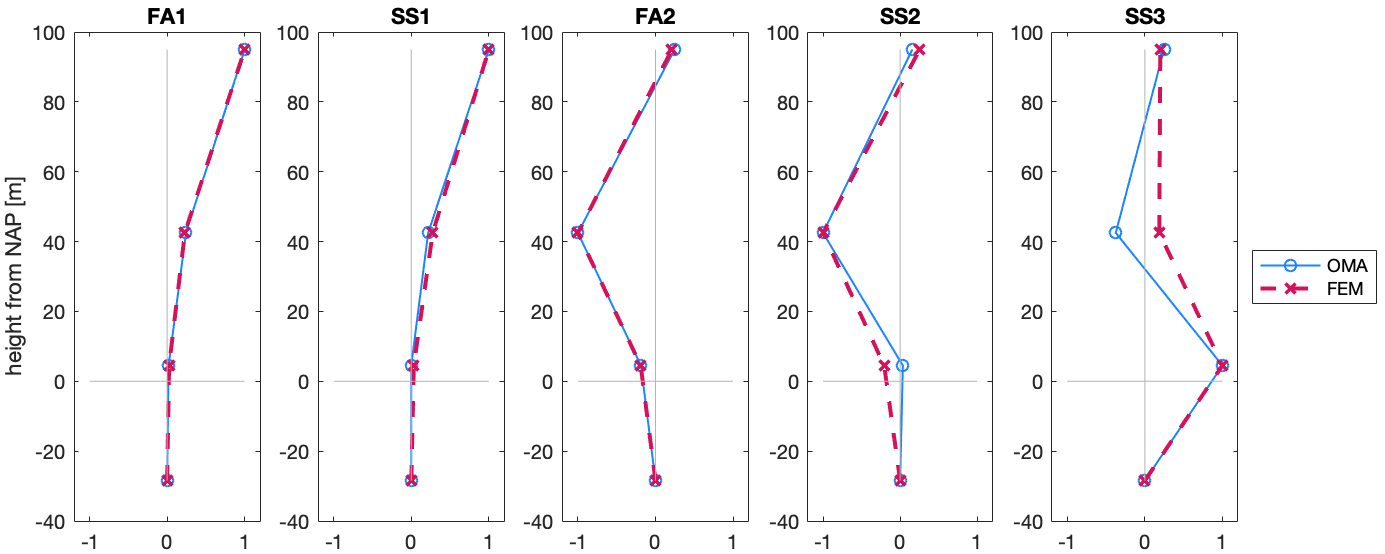}
    \caption{Mode shapes identified by OMA (solid blue) and estimated by the finite element model (dashed red).}
    \label{fig:modalparams}
\end{figure}

A mechanical state-space model of the W27 turbine is derived from a Finite Element (FE) model which is first constructed based on known geometric and material properties, then tuned to match with the modal parameters identified through OMA. A relatively simple FE model, such as a cantilever beam model, is sufficient for capturing the dominant modal properties of the system. The FE model consists of 124 Timoshenko beam elements, each 1.0 m in length, to span across the 29.0 m monopile depth and 95.0 m tower height. The tower does not have a separate transition piece as the monopile and tower have a bolted connection. The tapered profile of the top section of the turbine is modeled using pipe cross sectional elements which diminish in diameter linearly with height. All elements of the turbine are made of structural steel with Young's modulus $E$ = 210 GPa, Poisson's ratio $\nu$ = 0.3, and mass density $\rho$ = 7850 kg/m\textsuperscript{3}. External components such as the RNA, platform, and power unit are represented with concentrated masses at their respective elevations. 
\\
\indent A nontrivial source of model uncertainty is in the subsoil properties of the foundation. Mass and stiffness contributions of the soil are modeled by adding a mass density of 1500 kg/m\textsuperscript{3} and a stiffness profile to the elements located below the mudline. A custom 1D soil stiffness profile for the W27 turbine is taken from Versteijlen et al. \cite{Versteijlen2017a, Versteijlen2017}, where detailed studies of soil characterization and modeling were conducted for the site. In their work, geophysical measurements from Cone Penetration Tests (CPT), Seismic Cone Penetration tests (SCPT), and laboratory testing of borehole samples are used to develop an initial 1D effective Winkler profile $k_{eff}(z)$ which fits elastic properties of the soil-structure system as a function of depth $z$. Then, in-situ shaker tests of the W27 monopile are performed to obtain dynamic strain measurements from Rings 2-7, which are used to solve for an optimal scale factor $\gamma$ which tunes the magnitude of the profile. It is found that while the optimal $\gamma$ varies as a function of frequency, a constant factor $\gamma=0.78$ is most appropriate for the low-frequency regime which governs the response in OWTs. The resultant $\gamma \, k_{eff}(z)$ is illustrated as a distributed springs model in Figure \ref{fig:foundation}. Note that in the absence of high-fidelity soil characterization data, the p-y curve \cite{Versteijlen2017a} is an alternative soil stiffness profile which is commonly adopted in practice.
\\
\indent Table \ref{tbl:modalparams} compares five modes of the final FE model, denoted with subscript $fem$, with those identified by OMA, denoted with subscript $id$. From OMA, natural frequencies are identified with relatively high confidence, reflected by low sample standard deviation, whereas there is greater sample standard deviation in estimates of the modal damping ratios. Between the FE model and OMA results, there is a very good match in the natural frequencies for the first four modes, with the relative error well below 1\%. The high modal assurance criterion (MAC) above 0.9 indicates a strong fit between the mode shapes, as is visible in Figure \ref{fig:modalparams}. In general, the FA and SS modes are close to symmetric; any deviations from total symmetry are likely due to differential local soil damping and stiffness. There is greater difficulty in fitting the third SS mode of the FE model to the OMA results, which is due to the fact that the dynamic response of the OWT is controlled primarily by the first two modes. Therefore, higher modes are identified with significantly greater sample variance and reduced certainty in the OMA procedure; for instance, it was not possible to identify the third FA mode from the dataset. It is expected that the first two modes in each of the FA and SS directions dominate modal contributions to the dynamic response of the OWT across operating states. 

\begin{table}[!htp]
\centering
\fontsize{8}{8}\selectfont
\setlength{\tabcolsep}{8pt} 
\begin{tabular}{llllllllllll}
mode &  & \multicolumn{2}{c}{$\zeta_{id}$} &  & \multicolumn{2}{c}{$\textit{f}_{id}$ \ [Hz]} &  & $\textit{f}_{fem}$ \ [Hz] & $\textit{e}_f$ [\%]      &  & MAC \\ \hline
     &  & mean                  & std.                  &  & mean        & std.        &  &        &         &  &       \\ \cline{3-4} \cline{6-7}
FA1  &  & 1.932                 & 2.051                &  & 0.296       & 0.015       &  & 0.296  & 0.01\%  &  & 0.999 \\
SS1  &  & 3.011                 & 1.703                &  & 0.290       & 0.009       &  & 0.289  & -0.12\% &  & 0.996 \\
FA2  &  & 2.379                 & 2.126                &  & 1.869       & 0.058       &  & 1.874  & 0.23\%  &  & 0.988 \\
SS2  &  & 2.887                 & 1.931                &  & 1.885       & 0.029       &  & 1.883  & -0.09\% &  & 0.939 \\
SS3  &  & 2.604                 & 1.615                &  & 4.563       & 0.252       &  & 5.193  & 13.79\% &  & 0.743 \\ \hline
\end{tabular}
\caption{Mean and standard deviation of modal damping ratios and natural frequencies identified by OMA. Error in natural frequencies ($e_f$) and mode shapes (MAC) of the FE model are computed with respect to the mean modal parameters from OMA.}
\label{tbl:modalparams}
\end{table}

\section{Application to virtual sensing}
\label{sec:results}

\begin{figure}[hbt!]
    \centering
    \includegraphics[width=0.6\linewidth]{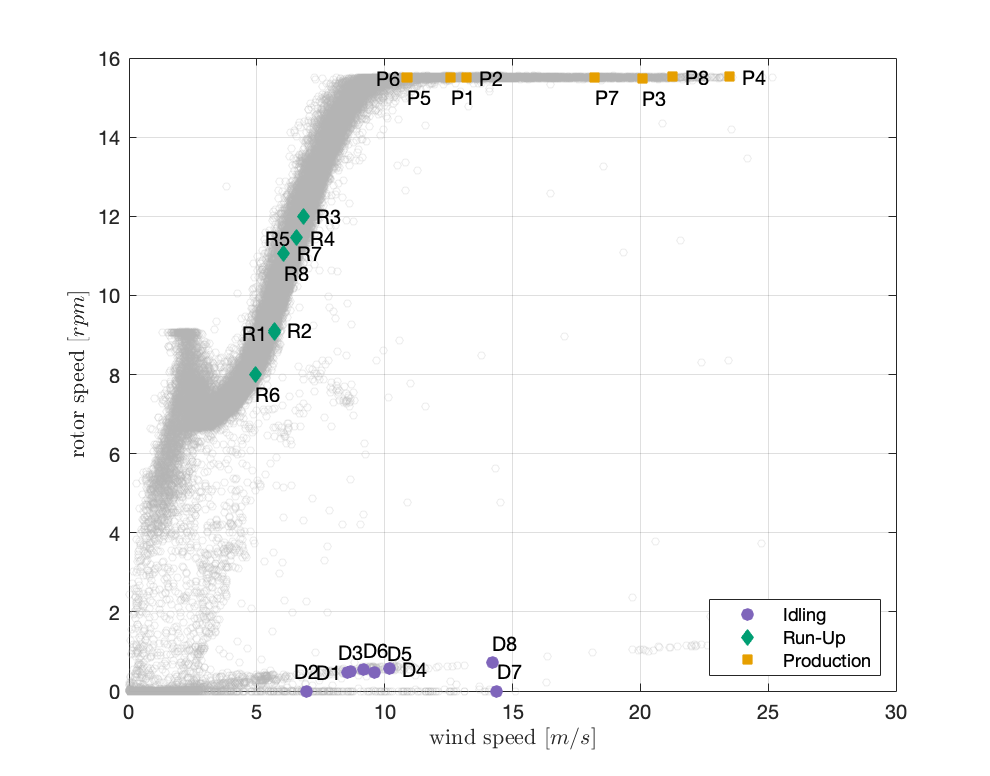}
    \caption{Wind speed-rotor speed state curve of 10-minute records taken over the period of March 2016 to June 2017 for the W27 turbine. The records selected for validation in each of the operating states are labeled. }
    \label{fig:statecurve}
\end{figure}

This section discusses validation of the GPLFM for virtual sensing using operational data from the W27 turbine. First, the fitting of kernel hyperparameters using the approach of Equation \eqref{eq:mle} is presented. The results of dynamic strain and modal force estimation using acceleration-only measurement data are shown for sample records from each of the turbine's operational states. Next, the sensitivity of the GPLFM to the sensor configuration is examined to determine the minimum sufficient measurement setup for characterizing the dynamic strain response of the OWT. Finally, perturbation to the foundation stiffness profile is introduced to evaluate the ability of the GPLFM to accommodate model uncertainty. The performance of the GPLFM is calculated using the correlation coefficient (CC) and mean relative error (MRE) of the predicted strain time series in relation to the validation measurements, as well as by converting the strain estimate into a Damage Equivalent Load (DEL) \cite{Laanes2021, ASTM2017} to measure the impact of estimation error on fatigue assessment. 
\\
\indent In this application of the GPLFM, the finite element model of the W27 turbine is translated into a reduced-order model retaining the first $n_{\mathrm{m}}=3$ modes in each of the FA and SS directions of the structure. The stochastic model of the input assumes the three modal forces are independent and identically characterized by a Mat\'ern kernel with $\nu=5/2$, such that $\kappa_1(\tau) = \kappa_2(\tau) = \kappa_3(\tau)$. The corresponding augmented model is therefore driven by two hyperparameters, $\theta=[\alpha, l_s]$. 
\\
\indent The acceleration-only measurement dataset consists of acceleration time series transformed into FA and SS coordinates at three elevations of the structure; these elevations are referred to as A1 (combination of orthogonal Acc. 1 and Acc. 2 at +4.50m NAP), A2 (uniaxial Acc. 3 at +42.50m NAP), and A3 (accelerometer in the RNA at +95.0m NAP). The validation dataset consists of strain time series recorded at position A for FA motion and position B for SS motion within the rings at six depths below the mudline, referred to by their corresponding ring numbers as Str. 2-7 (refer to Figure \ref{fig:sensorsetup} for an illustration). Validation is shown for eight records from each of the operating states of the turbine, with each record identified by a numeral preceded by the letter 'D' for the idling state, 'R' for the run-up state, and 'P' for the production state. The records are selected with the intention to encompass a wide range of wind and rotor conditions in each operating state, as shown on the wind speed-rotor speed state curve in Figure \ref{fig:statecurve}.

\subsection{Hyperparameter fitting}
\label{sec:prior}

\begin{figure}[hbt!]
    \centering
    \includegraphics[width=0.9\linewidth]{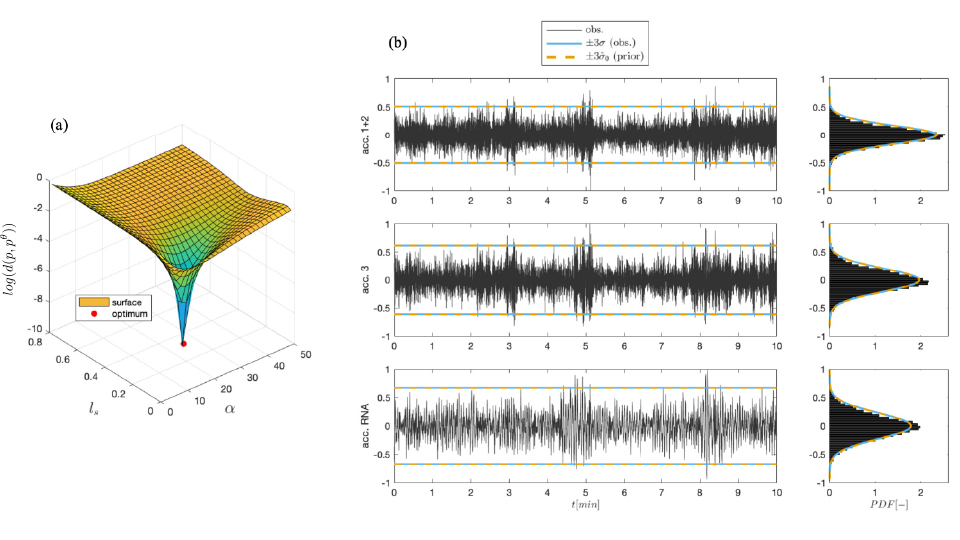}
    \caption{a) The optimization surface across the search domain $\alpha=[0, 50]$ and $l_s=[0, 1]$ for hyperparameters of the Mat\'ern $\nu=5/2$ kernel. b) Measurement time series and histogram of acceleration data at 3 channels for record P3, with the empirical distribution indicated in blue and modeled distribution indicated in orange.}
    \label{fig:HPsurface}
\end{figure}

\begin{figure}[hbt!]
    \centering
    \includegraphics[width=0.95\linewidth]{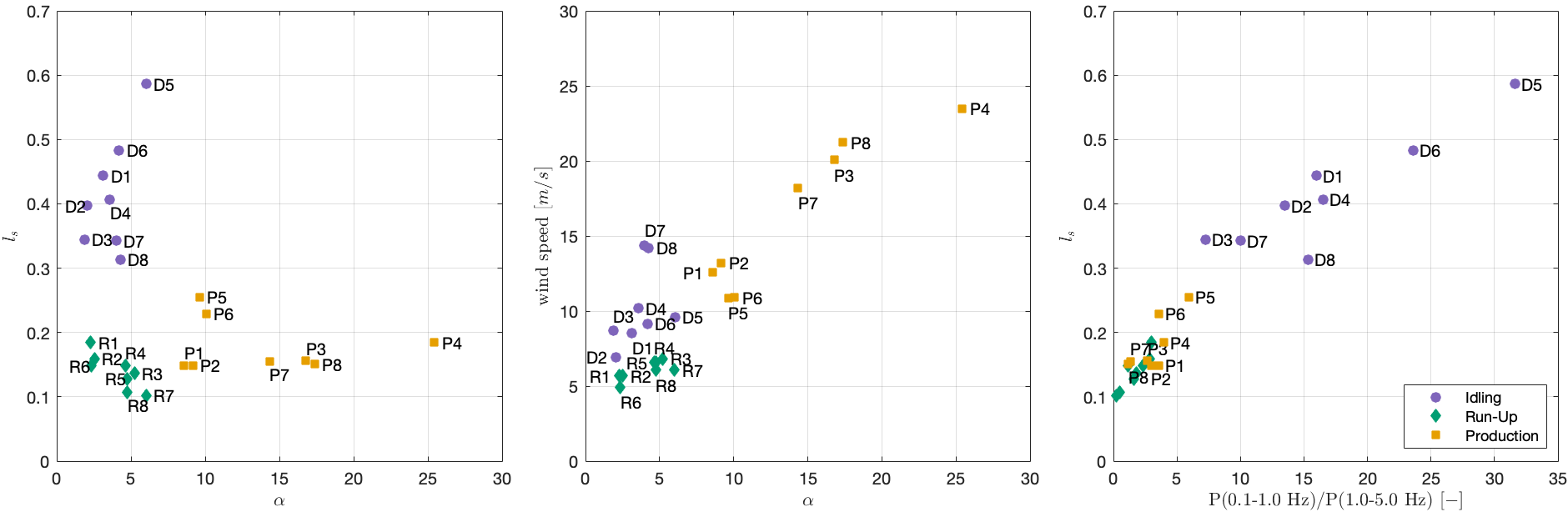}
    \caption{Optimal $l_s$ vs. optimal $\alpha$ (left), wind speed vs. optimal $\alpha$ (center), and optimal $l_s$ vs. relative power in low-frequency range 0.1-1.0 Hz (right) for eight records in each operational state.}
    \label{fig:HPtrends}
\end{figure}

For each record in the validation dataset, the hyperparameters of the GPLFM are fit according to the approach of Equation \eqref{eq:mle}. It is consistently observed that evaluation of the objective function produces a convex surface with a clear optimum $\theta^*$, illustrated in Figure \ref{fig:HPsurface}a for sample record P3 with search domain $\alpha=[0, 50]$ and $l_s=[0, 1]$. To visualize the goodness of fit, the empirical distribution of the measurements (in blue) and the predicted prior distribution corresponding to the optimal model hyperparameters $\theta^*$ (in orange) are overlaid on the time series plots and histograms of the measured variables in Figure \ref{fig:HPsurface}b. The near exact match in the distributions across all three measurement channels indicates an appropriate fit of the model to prior information on the true system dynamics.
\\
\indent The optimal hyperparameters for all validation records are shown in Figure \ref{fig:HPtrends}. In the leftmost plot, the hyperparameter sets $\{\alpha, l_s\}$ for each of the operating states are observed to fall in distinct clusters. In particular, the idling state (records D1, ..., D8) corresponds to large values of $l_s$ and small values of $\alpha$ due to the dominance of low frequency oscillations driven by low-powered ambient excitation of the tower. The production state (records P1, ..., P8) corresponds to small values of $l_s$ and large values of $\alpha$ due to the combination of high wind speeds and higher-frequency motion resulting from interaction with the turbine rotor. From these correlations, we may draw physical interpretation of the controlling hyperparameters of the Mat\'ern kernel as they relate to qualities of the unknown input. To substantiate this reasoning, we refer to the center plot showing a positive correlation between $\alpha$ and the mean wind speed of the record, implying the vertical length scale parameter relates to the amplitude of the input as it varies with the magnitude of wind speed. In the rightmost plot, $l_s$ is compared with a metric of relative power in the low frequency range, which is computed as the ratio of the power (integration of area under the power spectral density function) from 0.1 to 1.0 Hz with the power from 1.0 to 5.0 Hz. Larger values of this ratio indicate greater contributions of low frequencies to the response spectrum. From the positive correlation observed, it can be inferred that the horizontal length scale parameter relates to the dominance of low frequency inputs, which varies with both ambient conditions and rotor-induced loads.
\\
\indent The run-up state (records R1, ..., R8) corresponds to hyperparameters falling between the ranges of the idling and production states, as it marks the transition between these two states. Moreover, lower variance in the hyperparameters is observed compared to the other two states because run-up of the OWT always occurs in predefined wind conditions, whereas the OWT may remain stationed in the idling or production state across a wider range of wind speeds (see Figure \ref{fig:statecurve}).

\subsection{Response estimation}
\label{sec:responseestimation}

\begin{figure}[hbt!]
    \centering
    \includegraphics[width=\linewidth]{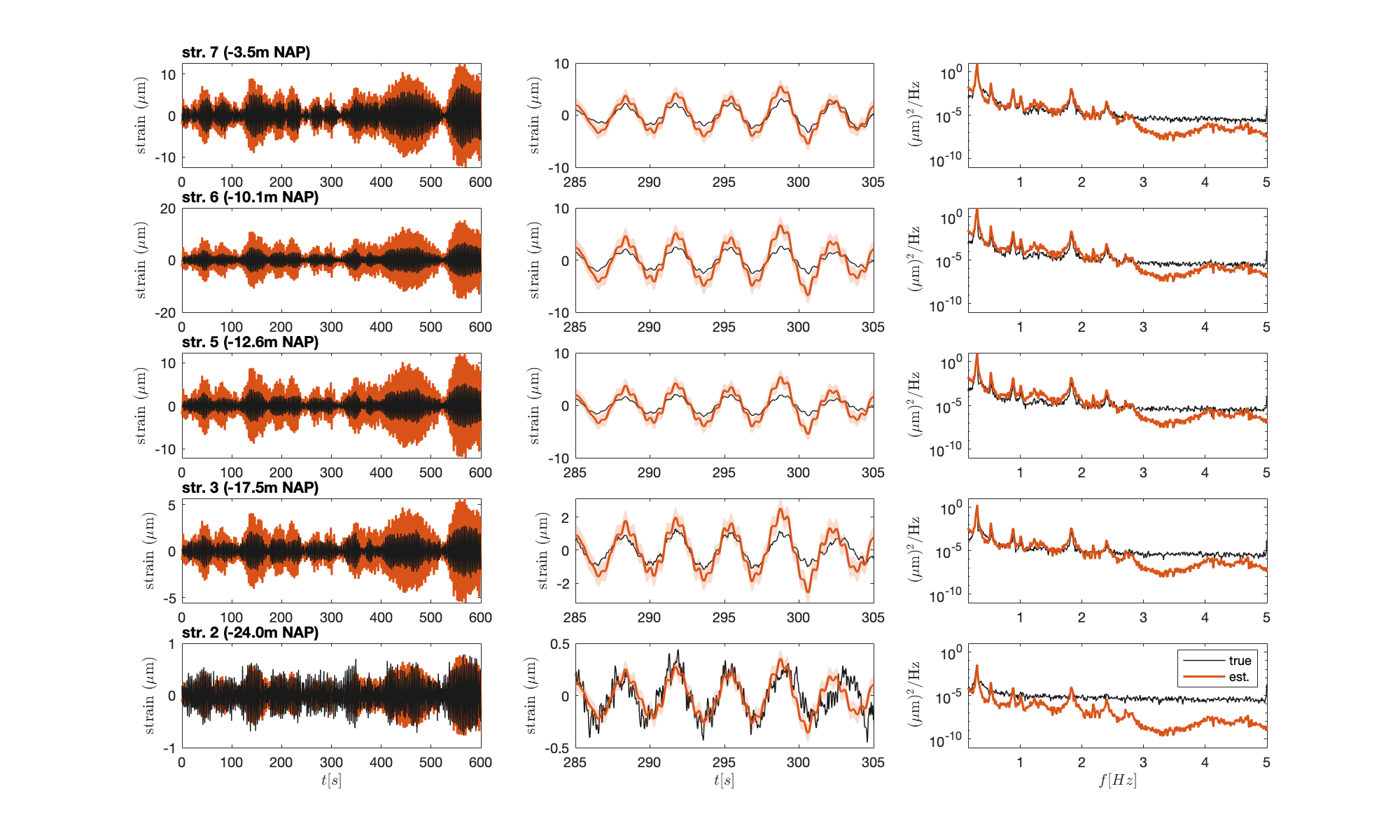}
    \caption{Strain estimation for a record from the idling state (D7), shown with the full time series (left column), zoomed-in view of the time series (center column), and power spectral density (right column) for five strain channels.}
    \label{fig:strain_idling}
\end{figure}

\begin{figure}[hbt!]
    \centering
    \includegraphics[width=\linewidth]{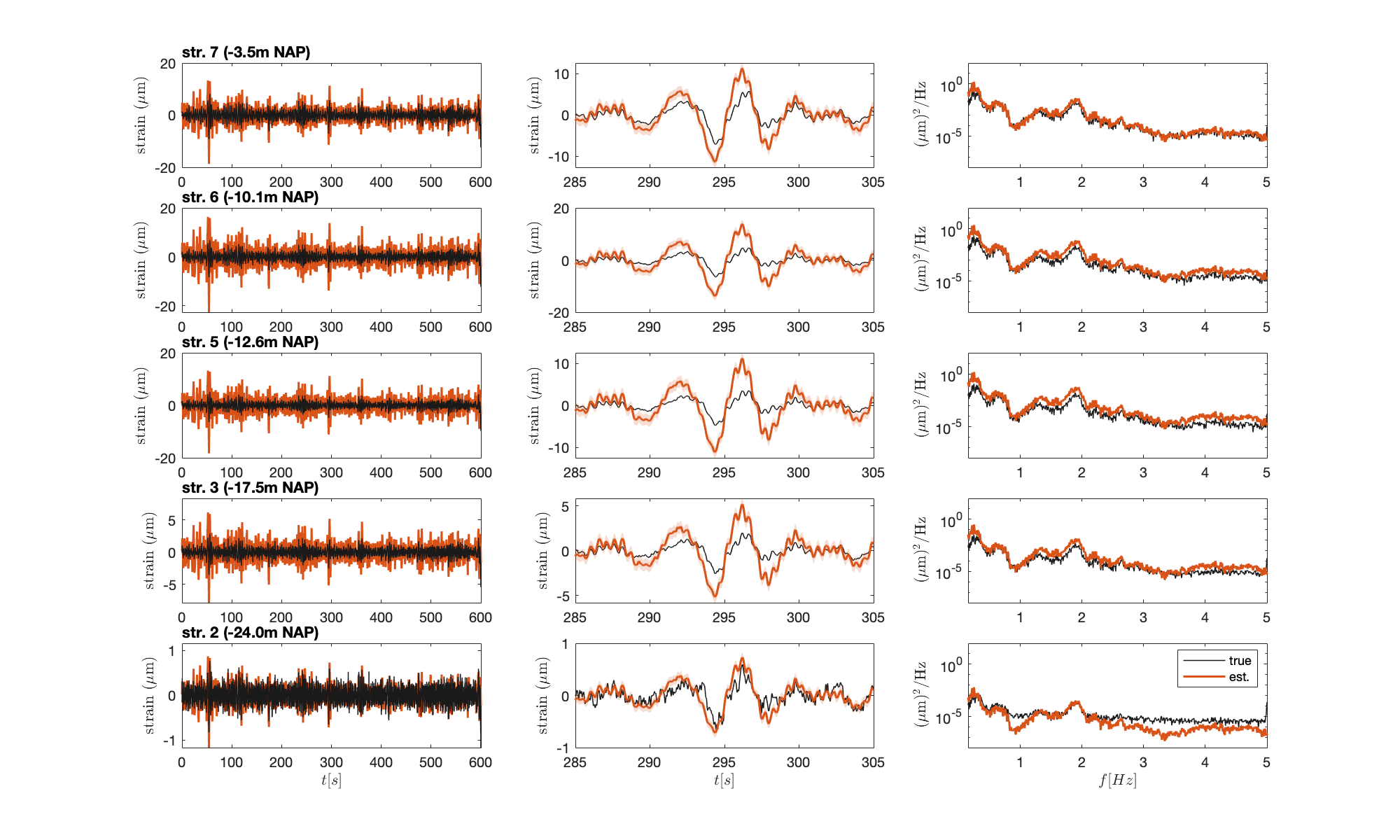}
    \caption{Strain estimation for a record from the run-up state (R5), shown with the full time series (left column), zoomed-in view of the time series (center column), and power spectral density (right column) for five strain channels.}
    \label{fig:strain_runup}
\end{figure}

\begin{figure}[hbt!]
    \centering
    \includegraphics[width=\linewidth]{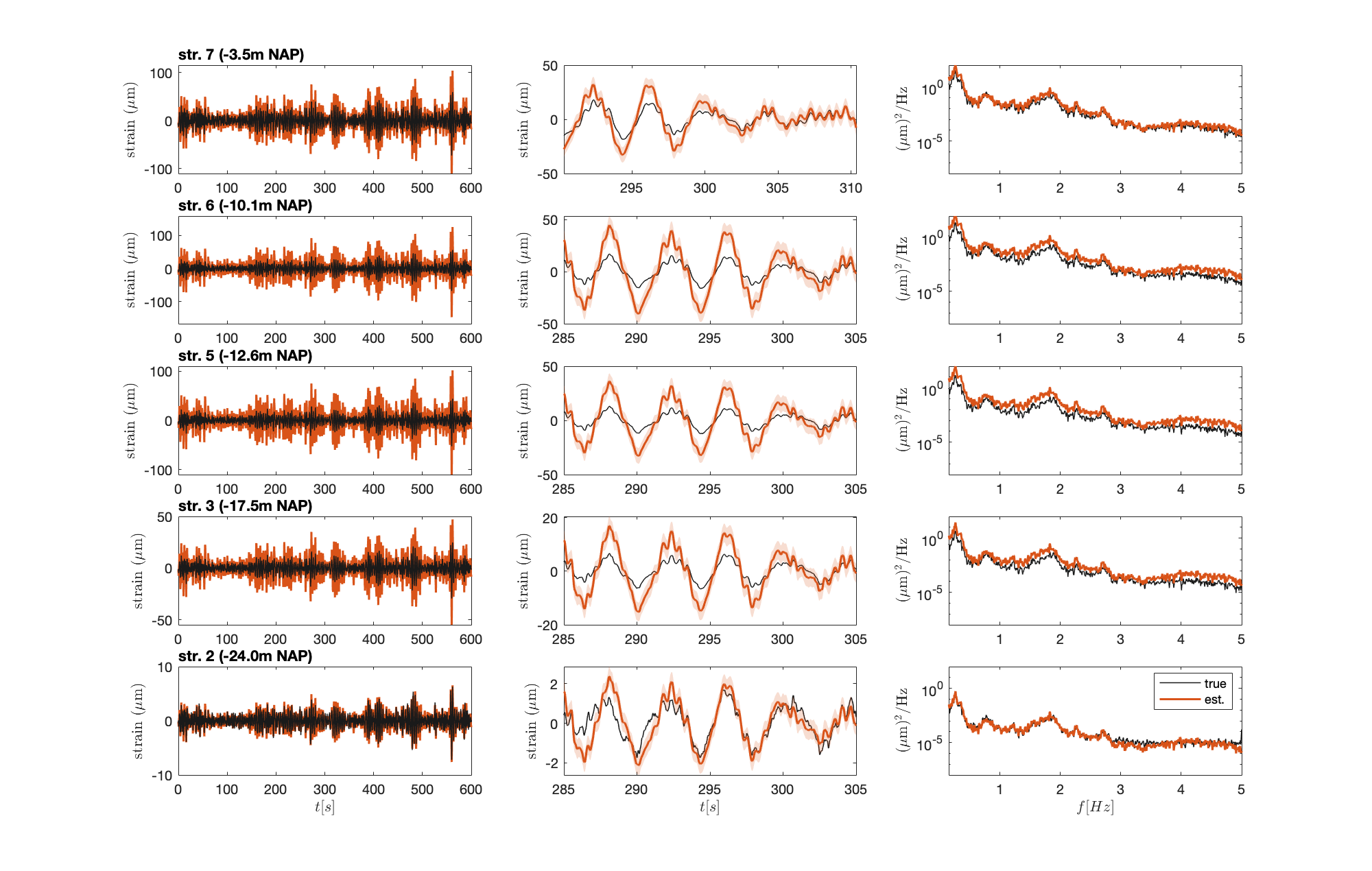}
    \caption{Strain estimation for a record from the production state (P4), shown with the full time series (left column), zoomed-in view of the time series (center column), and power spectral density (right column) for five strain channels.}
    \label{fig:strain_production}
\end{figure}

\begin{figure}[hbt!]
    \centering
    \includegraphics[width=0.9\linewidth]{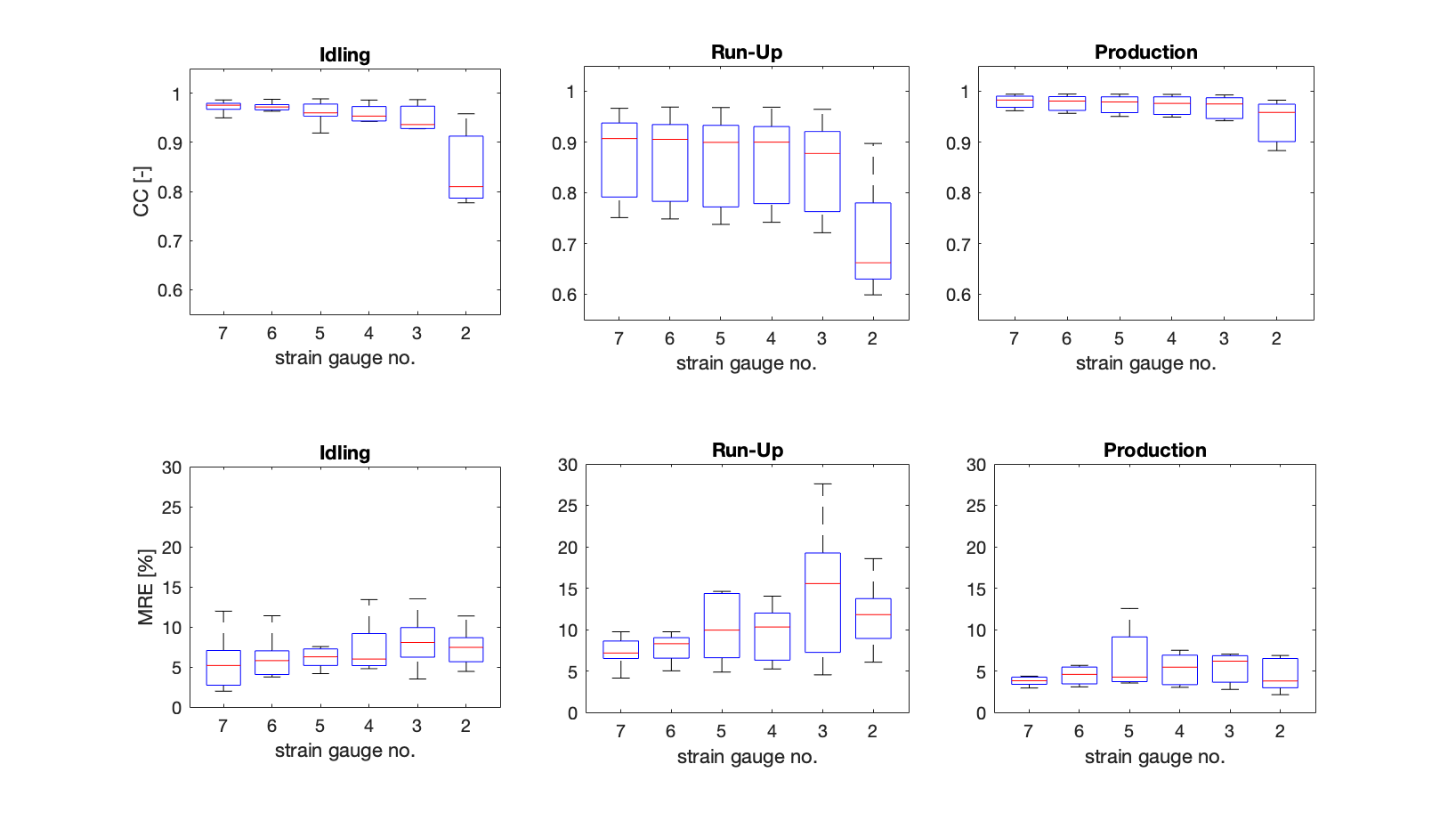}
    \caption{Correlation coefficient (CC) and mean relative error (MRE) across the 6 sub-soil strain channels in the idling (left), run-up (center), and production (right) states.}
    \label{fig:accuracystats}
\end{figure}

The strain estimation results in the idling, run-up, and production states are given in Figures \ref{fig:strain_idling}, \ref{fig:strain_runup}, and \ref{fig:strain_production}, respectively. Overall, the GPLFM attains remarkable accuracy in the estimation of the frequency content of sub-soil strains; discrepancy is observed in the high frequency range in Figure \ref{fig:strain_idling}, where low signal-to-noise ratio in the idling state results in difficulty with capturing the signal where it falls below the level of sensor noise in the validation data. This discrepancy disappears for the run-up and production states, as the response is no longer obscured by the noise amplitude. Despite the good performance in capturing the dominant frequency components, an overestimation of the strain response is evident from both the time- and frequency domain plots. The error in the amplitude is most apparent in the idling state in Figure \ref{fig:strain_idling} as well as at periodic impulses caused by turbine rotor motion in the run-up and production states, shown in the time series of Figures \ref{fig:strain_runup} and \ref{fig:strain_production} respectively. The discrepancy in the predicted signal amplitude is caused by a slight overestimation in the contribution of low frequencies falling below the first natural frequency of the system, which may be attributed to the general challenge of the estimation of low frequency strains. In particular, low frequency noise in measured accelerations is inevitably amplified over double integration of accelerations to obtain strains \cite{Smyth2007}, leading to the observed inflation of low frequency content in the estimated strains. Moreover, effective characterization of the quasi-static behavior of the turbine generally requires a large number of modes to be included in the modal decomposition of the model, a number which exceeds the 2-3 modes which are able to be accurately identified in the present study \cite{Maes2016a}. The inflation of low frequencies may also be indicative of mischaracterization of measurement noise; as shown in the sensitivity study in \ref{sec:app_modelerror}, underestimation of measurement noise variance leads to an overestimation of low frequency content in the displacement response. These challenges are encountered in a similar study in \cite{Maes2016a}, where a high-pass filter which removes frequencies below 0.25 Hz is applied to the predicted strain sequence before drawing comparison to the validation data.
\\
\indent Strain estimation results for all validation records is shown in Figure \ref{fig:accuracystats}, where the boxplots are computed from accuracy metrics of the strain estimate across eight records in each operating state. In general, high accuracy in the frequency domain estimate is reflected by a very high correlation coefficient (CC) near 1, while fair accuracy in the time-domain estimate is reflected by a mean relative error (MRE) generally falling in the range of 5-10\%. There is a reduction in accuracy in the run-up state due to larger variance in the wind and rotor speed (and therefore aerodynamic damping) among the records. Lower accuracy observed at strain gauge 2 may be a result of sensor noise in the particular strain gauge or extrapolation uncertainty, as the sensor is located at the furthest depth in the foundation where uncertainty in the soilbed properties is highest. 

\begin{figure}[hbt!]
    \centering
    \includegraphics[width=0.95\linewidth]{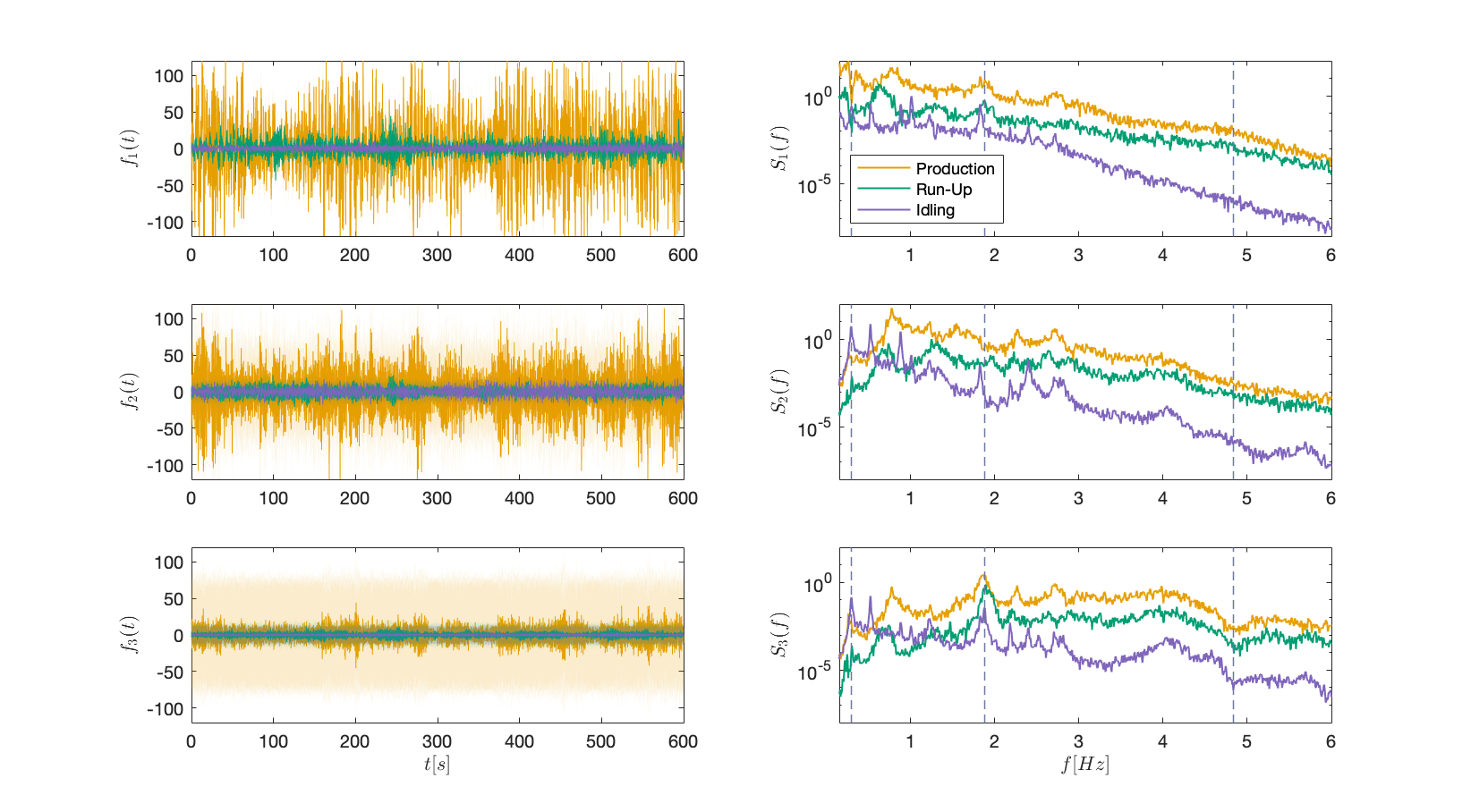}
    \caption{Estimation of the first three modal forces for records in the idling, run-up, and production states, shown with the full time series on the left and power spectral density on the right.}
    \label{fig:modalforce}
\end{figure}

\indent Samples of the modal force estimate from each of the operational states of the OWT are shown in Figure \ref{fig:modalforce}. While it is not possible to obtain direct measurements of the loads for validation of the input estimate, the frequency spectrum of the prediction provides insights to the joint relationship between inputs and states defined in the GPLFM. First, it is clear that the magnitude of the input increases from the idling state to the production state, reflected by the difference in signal amplitude in the time series plots. Higher estimation uncertainty for higher modes is shown by the larger shaded area representing the $\pm3$ standard deviation bounds on the modal force prediction. Of particular interest is the frequency content estimated at the natural frequencies of the system, noted by dashed lines in the PSD plots. For the run-up state and production state, there is a visible dip in the frequency spectrum at the first natural frequency, which is an indicator of underestimation of the damping ratio in the first mode. In particular, when an underdamped model of the structural system predicts response amplitudes which exceed that of the measured acceleration response, the GPLFM compensates for the mismatch by reducing the power at that natural frequency in the frequency spectrum of the modal force.  This effect is also observed in numerical experiments presented in \ref{sec:app_modelerror}, where perturbations to modal damping ratios in a synthetic model causes artificial suppression of frequency content about modes where damping is misidentified. The input estimate is likely also affected by errors in mode shapes and natural frequencies of the system. Since congruence in the mode shapes identified from OMA and those of the mechanical model may only be evaluated at three points where accelerations are recorded, there is a high likelihood for deviation in the section of the mode shape profile falling below the mudline where strains are estimated. Moreover, modal properties in the run-up and production states shift under varying operational and environmental conditions \cite{Devriendt2014}. It is observed in the numerical experiments that while the input estimate is particularly sensitive to errors in the modal properties, the strain estimate is left relatively unperturbed. This effect indicates that in the prediction of latent strain states, the inclusion of input estimation acts to accommodate error in modal properties by attributing its source to stochastic noise, reflected by adjustments to the predicted spectrum of external excitation to the system in order to fit the data provided. However, the input estimate has limited efficacy in accounting for other sources of systematic model error, such as error in the assumed level of measurement noise or mismatch in the relationship defined between displacements and strains. Nevertheless, these results in response estimation provide strong evidence that the GPLFM overcomes several challenges with reconstructing subsoil strain response compared to similar studies conducted using dual-band MD\&E for virtual sensing in \cite{Henkel2019} and the classical Kalman filter in \cite{Maes2016a}, which were not able to achieve the same level of accuracy in the frequency domain estimate and phase synchronization of the time domain estimate.

\subsection{Sensitivity to sensor configuration}
\label{sec:measurevariants}

In this section, the performance of the GPLFM is examined under different  assumptions with respect to the availability and configuration of measurement data. The motivation is twofold: to determine an instrumentation scheme which aids in correcting for the overestimation of strain amplitude, and to determine the minimum instrumentation scheme which provides sufficient evidence to characterize the strain response and inform future measurement campaigns for fatigue monitoring of similarly sized OWTs. Considering the practicality of in-situ implementation, we assume only sensors at accessible locations above the waterline are used for virtual sensing of subsoil response. Strain estimation with the GPLFM is conducted with three sensor configurations, labeled as: 

\begin{itemize}
    \item A3, using only the acceleration channel in the RNA (+95.0m NAP);
    \item A3+A1, using two acceleration channels with one located in the RNA (+95.0m NAP) and the other at +4.50m NAP; and
    \item A3+S8, using the acceleration channel in the RNA (+95.0m NAP) and the strain gauge ring located at +6.95m NAP.
\end{itemize}

\begin{figure}[hbt!]
    \centering
    \includegraphics[width=\linewidth]{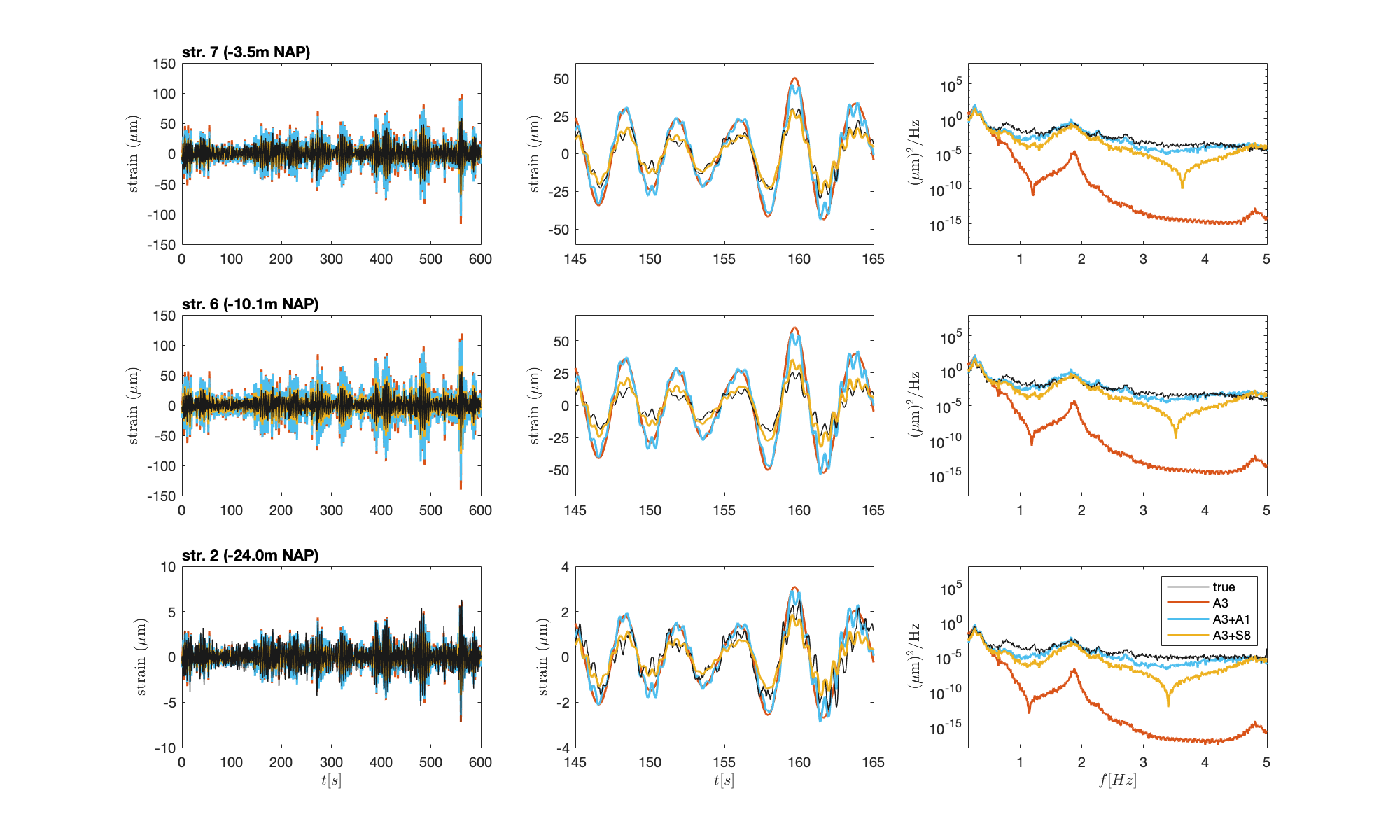}
    \caption{Strain estimation with different sensor configurations for record P3.}
    \label{fig:sensorvar_str}
\end{figure}

\begin{figure}[hbt!]
    \centering
    \includegraphics[width=0.55\linewidth]{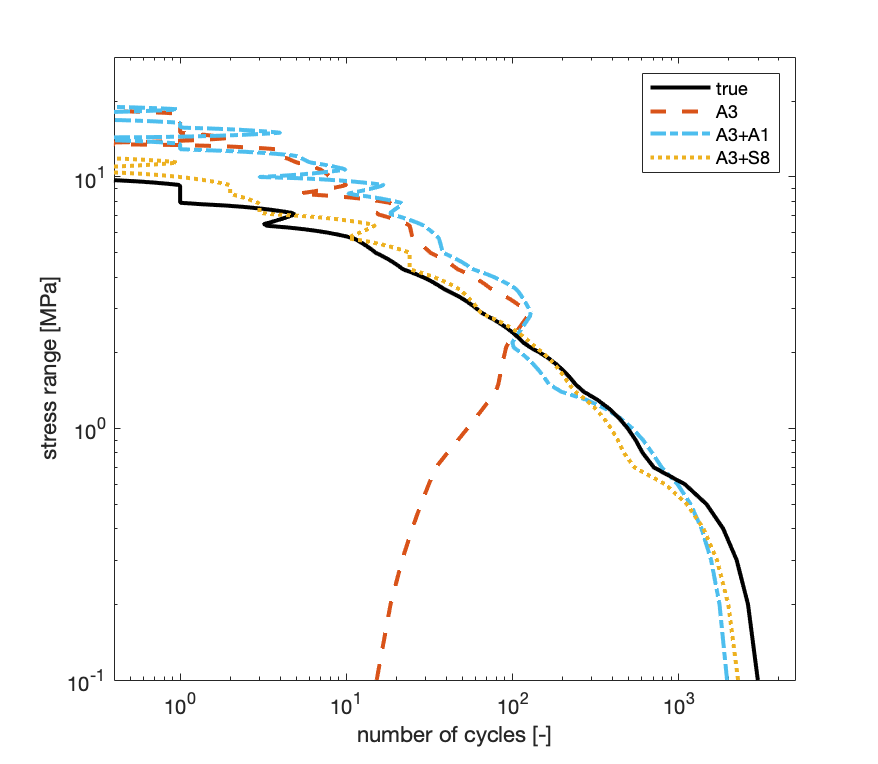}
    \caption{S-N diagram with different sensor configurations across 4 samples in production state.}
    \label{fig:sensorvar_SN}
\end{figure}

While the single sensor set-up of the A3 configuration results in the observability of only the first mode, it is included in the following comparison as a point of reference for the added value of the second sensor in the A3+A1 and A3+S8 configurations. Validation of the strain estimate with Record P3 for the three sensor configurations is given in Figure \ref{fig:sensorvar_str}. For brevity, only strain channels 2, 6, and 7 are shown. It is observed the A3 configuration is only capable of capturing the first modal response, with the accuracy at frequencies above the first natural frequency rapidly declining. The introduction of the second acceleration channel in A3+A1 allows for accurate estimation of higher frequency contributions to the response. However, A3 and A3+A1 lead to a comparable level of overestimation of the signal amplitude due to the inability of the model to overcome challenges with low frequency noise amplification using acceleration-only measurements. In contrast, the inclusion of a strain channel in A3+S8 leads to substantial improvements in capturing the strain amplitude across channels, in exchange for reduced accuracy in the estimation of non-governing frequencies falling between natural frequencies of the system. Compared to A3, the single strain channel in A3+S8 adds information on not only the high frequency response (such as the second modal response) in the tower section closer to the mudline, but also the amplitude of strains, thereby overcoming low frequency error.
\\
\indent The quality of the strain estimate for different sensor configurations may be compared using fatigue assessment measures. In this work, the S-N diagram, also referred to as the Woehler curve, is constructed using a rainflow counting algorithm which translates time series data into the number of cycles counted within defined stress ranges \cite{Laanes2021, ASTM2017}. The S-N diagram of Figure \ref{fig:sensorvar_SN} shows the stress cycle counts of the true signal and the predicted signals at Str. 6, where the highest bending moments in the tower-foundation system are expected to occur, across four records from the production state. From this illustration, it is clear that configurations A3 and A3+A1 lead to an overestimation of the stress magnitude for lower cycle counts, which corresponds to low frequency oscillations. For the A3 case, the contribution of low-magnitude high frequency oscillations to fatigue stress accumulation is completely miscalculated. Figure \ref{fig:sensorvar_SN} affirms the A3+S8 configuration provides the best estimate of the stress-cycle curve. This result is further reflected in the damage equivalent load (DEL) measures of fatigue stress accumulation, where the deviation of the DEL of the prediction from that of the true strain response is 106.89\% for A3,  98.43\% for A3+A1, and 16.04\% for A3+S8. In other words, the fatigue estimate from the acceleration-only configurations A3 and A3+A1 is on the order of double that of the true fatigue damage accumulation.

\subsection{Sensitivity to model uncertainty}
\label{sec:modelvariants}

Previously, in Section \ref{subsec:noise}, we showed that the mathematical construction of the GPLFM attributes noise sources to the unknown input. This section evaluates this effect using in-situ validation data, by imposing controlled model error in the mechanical model and evaluating the consequences on the strain estimation. In this case study, as well as in general for offshore wind turbines, one significant source of uncertainty is in properties of the foundation stiffness, which impacts mode shapes corresponding to the mechanical model. Whereas the stiffness profile $k_{eff}(z)$ is derived from soil characterization data which is customarily collected at wind farm sites, the subsoil strain data used to tune the magnitude of the profile $\gamma$ in \cite{Versteijlen2017a} is generally not available. Without such data, there is the potential for substantial error in the assumed soil stiffness magnitude. We examine both extremes of introducing error in the soil stiffness scaling $\gamma$ by an order of magnitude above and below the baseline stiffness magnitude, with all other FE model parameters remaining constant. The following model variants representing different levels of model fidelity are compared: 

\begin{itemize}
    \item $\gamma = 0.078 $: limiting case of underestimating the soil stiffness scaling by an order of magnitude
    \item $\gamma = 0.78 $: baseline case of soil stiffness scaling to reflect the best fit to true in-situ conditions, determined in \cite{Versteijlen2017a}
    \item $\gamma = 7.8 $: limiting case of overestimating the soil stiffness scaling by an order of magnitude
\end{itemize}

\begin{figure}[hbt!]
    \centering
    \includegraphics[width=0.85\linewidth]{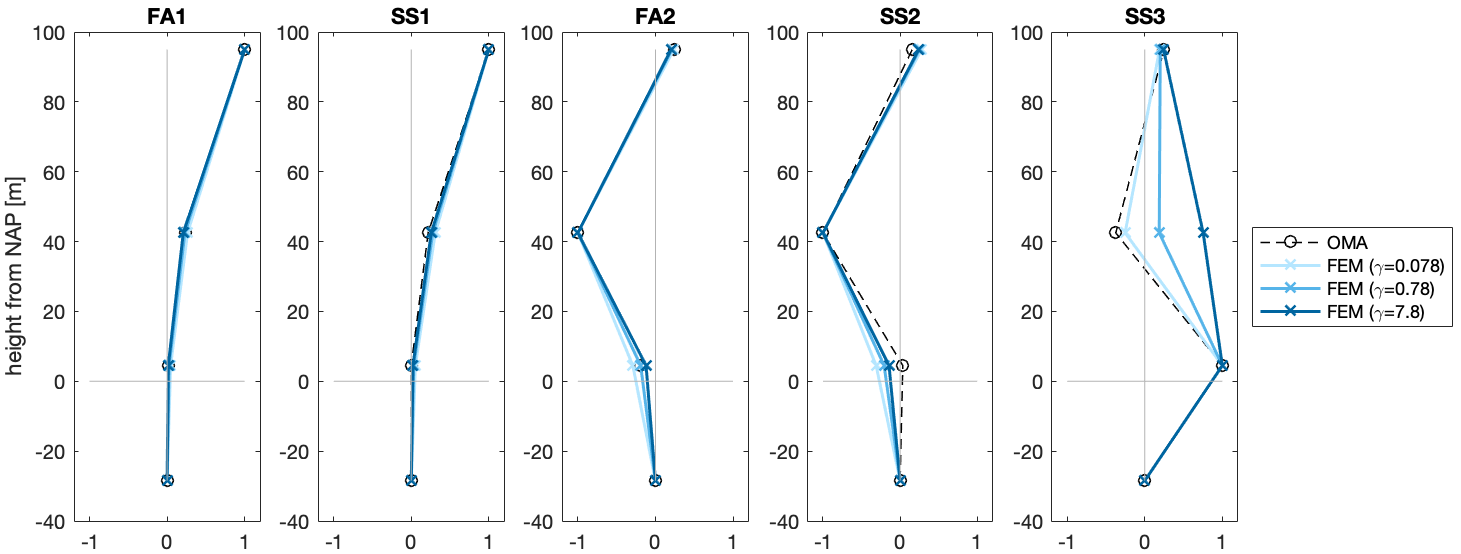}
    \caption{Modal parameters of model variants.}
    \label{fig:modelvariants}
\end{figure}

\begin{figure}[hbt!]
    \centering
    \includegraphics[width=\linewidth]{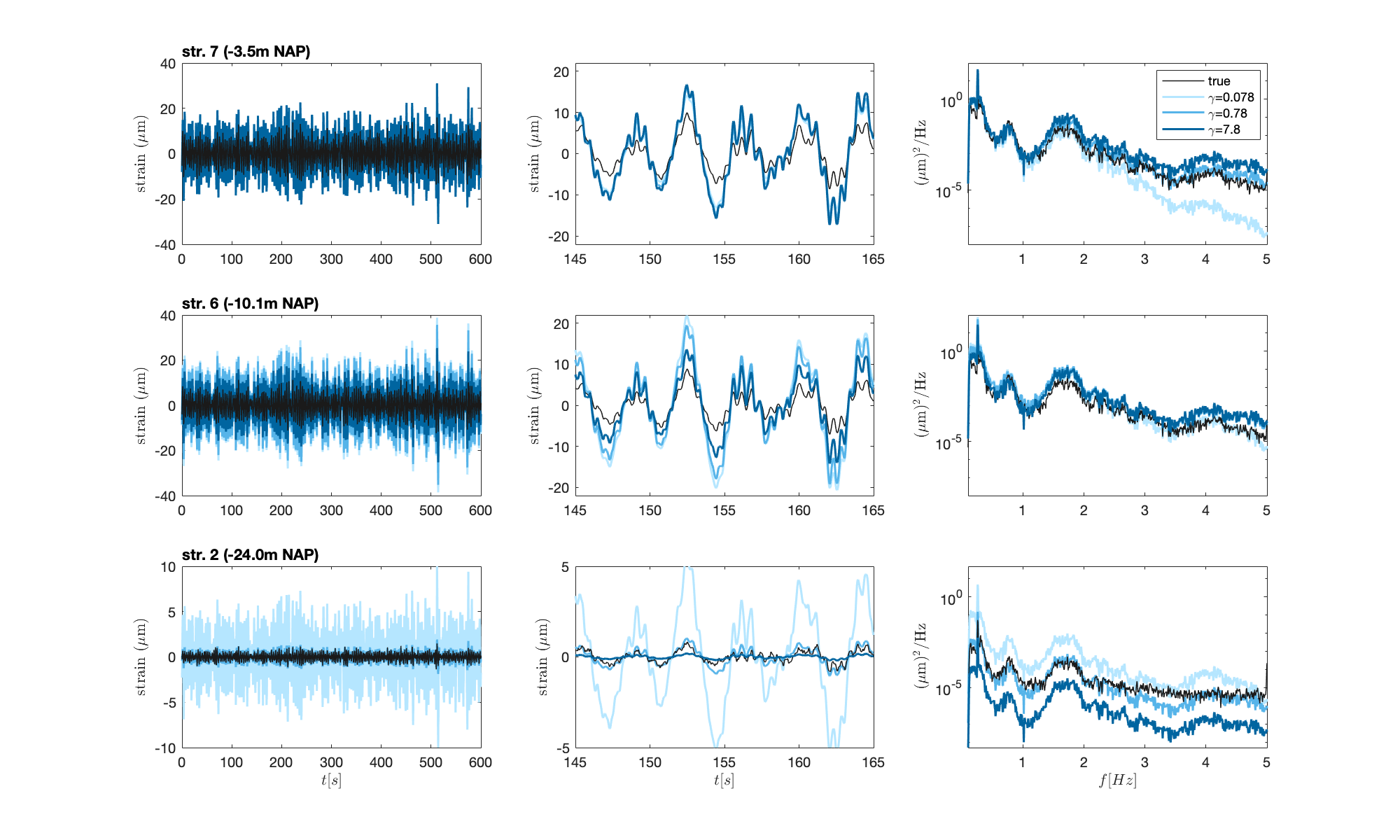}
    \caption{Strain estimation for varying levels of model fidelity using all acceleration measurement channels for record P2.}
    \label{fig:modelvar_str}
\end{figure}

\begin{figure}[hbt!]
    \centering
    \includegraphics[width=0.9\linewidth]{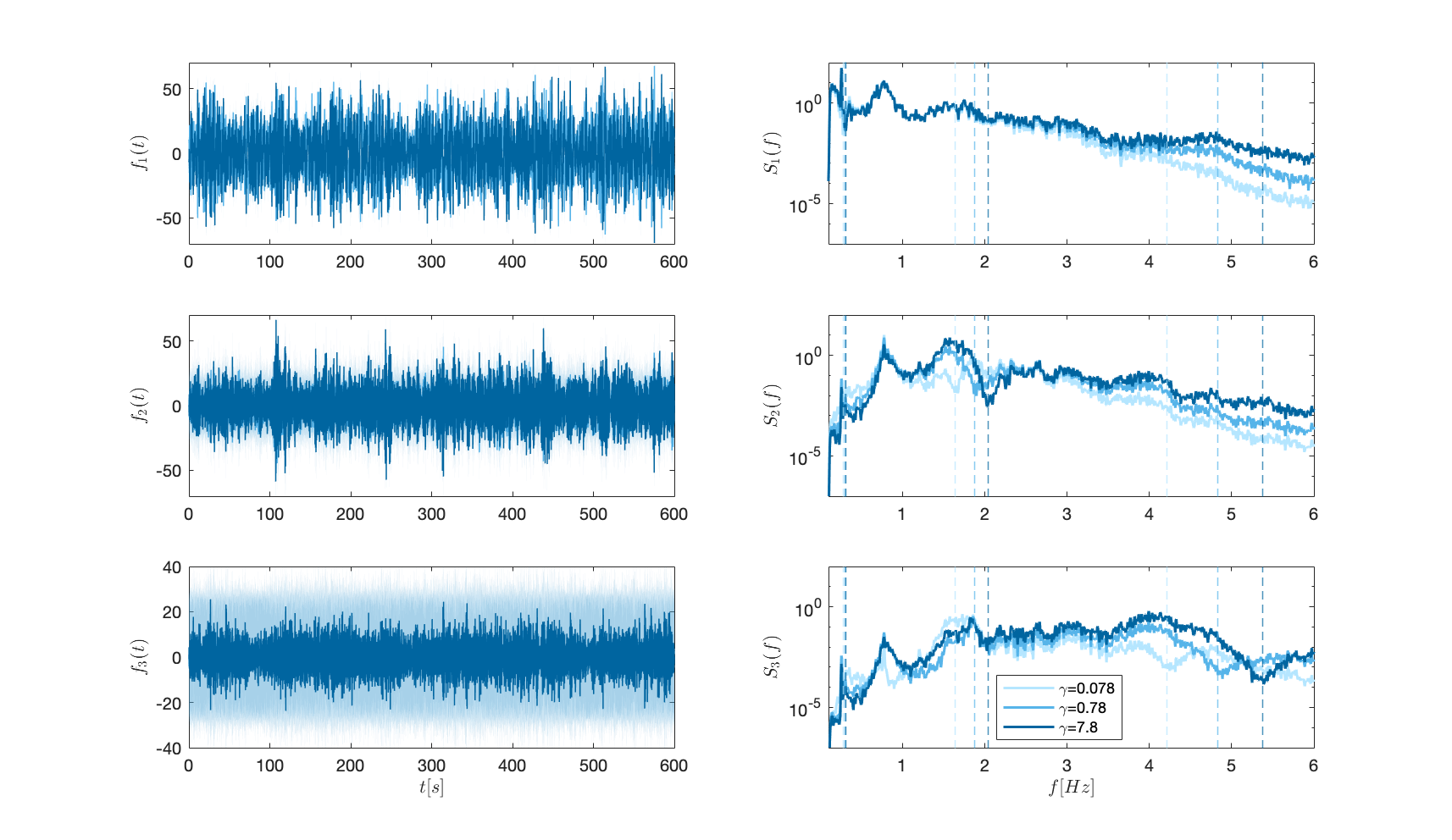}
    \caption{Modal force estimation for varying levels of model fidelity using all acceleration measurement channels for record P2. The natural frequencies of each model variant are noted by dashed lines in the frequency spectrum.}
    \label{fig:modelvar_force}
\end{figure}

\begin{figure}[hbt!]
    \centering
    \includegraphics[width=\linewidth]{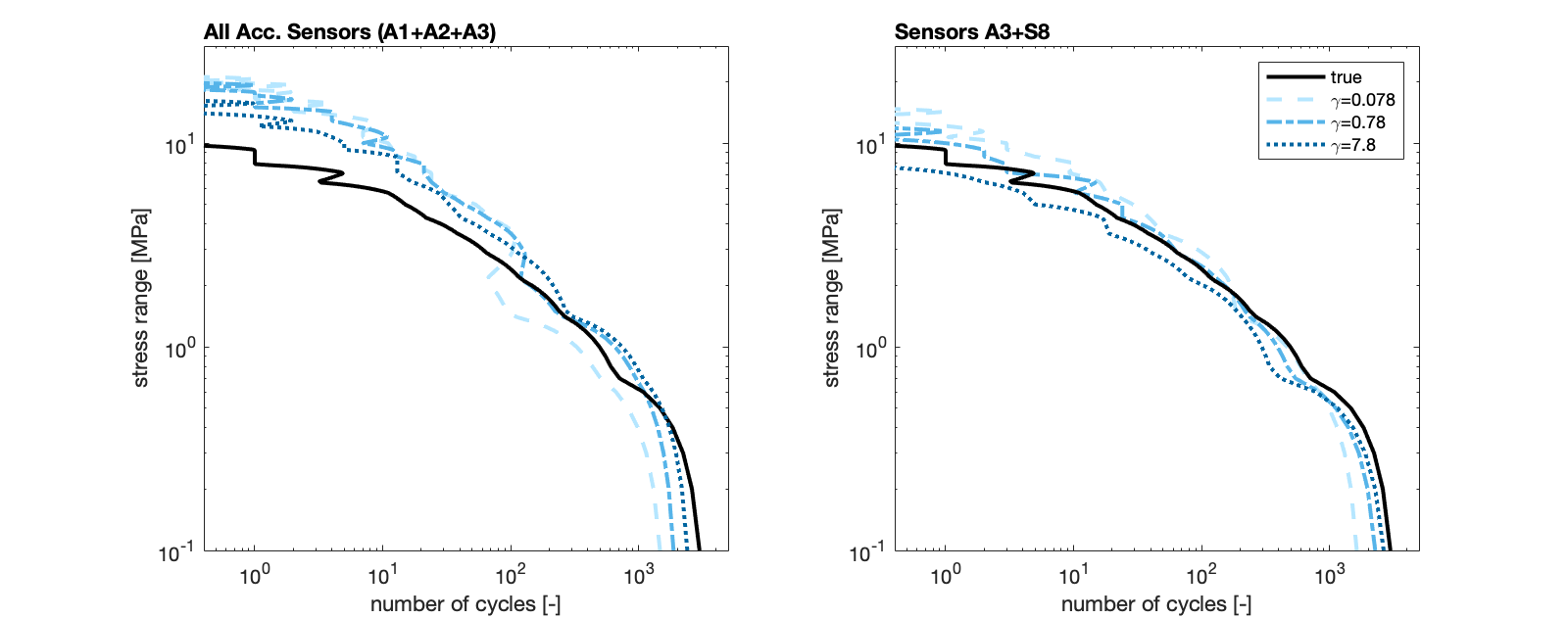}
    \caption{S-N diagram for varying levels of model fidelity for records from the production state, for sensor configuration A1+A2+A3 on the left and sensor configuration A3+S8 on the right.}
    \label{fig:modelvar_SN}
\end{figure}

The modal parameters of the three model variants are shown in Figure \ref{fig:modelvariants} and reported in Table \ref{tbl:modalparams_modelvar}, where the relative frequency error $e_f$ and MAC values are with respect to the modal parameters identified through OMA. It is observed that the lowest two mode shapes in each of the FA and SS directions, which dominate the OWT response, are relatively unaffected by scale differences in the soil stiffness profile. A larger effect is observed in the third SS mode. In terms of both $e_f$ and MAC, the error is greater in the case of underestimation of $\gamma$ than in the case of overestimation.
\\
\indent Figure \ref{fig:modelvar_str} shows the strain estimate obtained with each model variant using the acceleration-only measurement setup. The difference in the accuracy of the strain estimate varies with the depth of the strain location: there is comparatively lower change in the strain estimate at higher elevations (Str. 6 and 7) than at the lowest elevation (Str. 2), where underestimation of stiffness leads to underdamped response and the overestimation of stiffness leads to overdamped response. The difference is non-uniform across channels due to the fact that there is significantly lower stiffness at the upper section of the profile, where $k_{eff}(z)$ is on the order of $10^7 \ N/m^3$, than at lower sections of the profile, where $k_{eff}(z)$ is on the order of $10^{11} \ N/m^3$ (see Figure \ref{fig:foundation}). The scale difference therefore leads to a greater change in subsoil properties at the greatest depths. While the scaling of the predicted signal is affected by the model error, the shape of the distribution of power over frequencies remains relatively unchanged up to the second natural frequency across all channels. 
\\
\indent The difference in prediction by the model variants is more apparent in the modal force estimate, shown in Figure \ref{fig:modelvar_force}. Here, the discrepancy in the natural frequencies between the model variants is illustrated with dashed lines in the frequency spectrum. What is particularly notable is the dip in the frequency spectrum at the $j$th natural frequency for the $j$th modal force, where the natural frequencies differ for each model variant. These dips reflect adjustments to the frequency contributions of the input by the GPLFM in order to compensate for error between the dominant frequencies of the measured response with that of the predicted response by the perturbed model. For instance, where the GPLFM expects greater presence of the perturbed second natural frequency of the model than what exists in the measurement data, the contribution of that frequency in the corresponding input estimate is reduced accordingly. In this manner, error in modal properties is accounted for through the stochastic input estimate, thus retaining the relatively robust accuracy observed in the strain estimate. 
\\
\indent 
To evaluate the effect of model fidelity as a function of the instrumentation scheme, Figure \ref{fig:modelvar_SN} shows the S-N diagram corresponding to four records from the production state, produced from the estimate of Str. 6, for two sensor configurations: the original scheme of using all three acceleration measurement channels (A1+A2+A3) and the best performing scheme from the investigation in Section \ref{sec:measurevariants} (A3+S8). For both instrumentation schemes, the three model variants result in relatively comparable stress-cycle curves, indicating relative invariance to perturbation in the model. As before, A3+S8 provides the most robust reproduction of the true stress-cycle curve while the acceleration-only scheme in A1+A2+A3 universally overestimates the stress magnitude of low-frequency oscillations. The improvement gained from the use of A3+S8 over the acceleration-only scheme is reflected in the DEL values reported in Table \ref{tbl:del}, where A3+S8 results in both lower error and lower variance in the DEL estimate across levels of model perturbation.

\begin{table}[!htp]
    \centering
    \fontsize{8}{8}\selectfont
    \setlength{\tabcolsep}{8pt} % Default value: 6pt
    \begin{tabular}{lllllllllllll}
        ~ & ~ & ~ & $\gamma$ = 0.078 & ~ & ~ & ~ & $\gamma$ = 0.78 & ~ & ~ & ~ & $\gamma$ = 7.8 & ~ \\ \hline
        mode & ~ & $\textit{f}_{fem}$ \ [Hz] & $\textit{e}_f$ [\%] & MAC & ~ & $\textit{f}_{fem}$ \ [Hz] & $\textit{e}_f$ [\%] & MAC & ~ & $\textit{f}_{fem}$ & $\textit{e}_f$ [\%] & MAC \\ \cline{1-1} \cline{3-5} \cline{7-9} \cline{11-13}
        FA1 & ~ & 0.279 & -5.82 & 0.999 & ~ & 0.296 & 0.01 & 0.999 & ~ & 0.307 & 3.55 & 0.999 \\ 
        SS1 & ~ & 0.269 & 6.70 & 0.990 & ~ & 0.289 & -0.12 & 0.996 & ~ & 0.300 & 3.74 & 0.998 \\ 
        FA2 & ~ & 1.636 & -12.47 & 0.991 & ~ & 1.874 & 0.23 & 0.999 & ~ & 2.041 & 9.18 & 0.994 \\ 
        SS2 & ~ & 1.665 & -11.66 & 0.886 & ~ & 1.883 & -0.09 & 0.939 & ~ & 2.025 & 7.44 & 0.964 \\ 
        SS3 & ~ & 4.486 & -1.69 & 0.987 & ~ & 5.193 & 13.79 & 0.743 & ~ & 5.745 & 25.88 & 0.310 \\ \hline
    \end{tabular}
\label{tbl:modalparams_modelvar}
\caption{Modal parameters from each model variant, where accuracy metrics $e_f$ and MAC are computed with respect to the modal parameters identified by OMA.}
\end{table}

\begin{table}[!htp]
\centering
\fontsize{8}{8}\selectfont
\setlength{\tabcolsep}{10pt} % Default value: 6pt
    \begin{tabular}{llllll}
        ~ & \multicolumn{2}{c}{A1+A2+A3} & ~ & \multicolumn{2}{c}{A3+S8} \\ \hline
        ~ & DEL & error [\%] & ~ & DEL & error [\%] \\ \cline{2-3} \cline{5-6}
        true & 2.946 & -- & ~ & 2.946 & ~ \\ 
        gamma = 0.078 & 7.131 & 142.1\% & ~ & 4.295 & 45.8\% \\ 
        gamma = 0.78 & 6.652 & 125.8\% & ~ & 3.418 & 16.0\% \\ 
        gamma = 7.8 & 4.994 & 69.5\% & ~ & 2.227 & -24.4\% \\ \hline
    \end{tabular}
\caption{Damage equivalent load (DEL) values corresponding to Str. 6 of each model variant for two sensor configurations, A1+A2+A3 and A3+S8. The relative error is computed with respect to the true value of DEL.}
\label{tbl:del}
\end{table}

\section{Conclusions}
\label{sec:conclusions}

This work is among the first studies to provide in-situ validation of the GPLFM as a virtual sensing technique. In contrast to other Kalman filter-based techniques to virtual sensing, the GPLFM characterizes the unknown input as a Gaussian process which can describe a greater breadth of time-varying behavior than the standard assumption of white noise input. The joint posterior distribution of latent inputs and states is determined by Gaussian process regression of acceleration measurements, which is performed sequentially in augmented state-space using Kalman filtering/RTS smoothing. Rather than using heuristic or ad hoc tuning procedures, the process noise covariance in the dynamics model is solved for in a data-driven fashion, in which the structure of the GPLFM is utilized to define the joint relationship between input and state variables and the kernel hyperparameters governing the process noise covariance matrix are fit to acceleration data prior to filtering. Compared to other Kalman filter-based techniques for joint input-state estimation, the GPLFM has several notable theoretical advantages, \textcolor{black}{in enforcing a characterization} of covariance between inputs and states consistent with stochastic differential equations theory, as well as practical advantages, in providing an efficient and descriptive parameterization of statistical model parameters.
\\
\indent In this contribution, the GPLFM is applied to the fatigue monitoring of offshore wind turbines. The validation data is taken from a monitoring campaign of an operating nearshore wind turbine from the Westermeerwind Park in the Netherlands, where six strain gauge rings embedded along the profile of the monopile foundation provide subsoil strain measurements for validation. This dataset presents the opportunity to evaluate the GPLFM for its ability to address the challenge in fatigue monitoring of OWTs with high model uncertainty in the soil-structure system of the monopile foundation. First using acceleration-only data from an offshore wind turbine, the GPLFM implemented with a Mat\'ern covariance kernel is able to reconstruct the frequency spectrum of the subsoil strain response with excellent accuracy across operating conditions of the OWT. In the time domain, overestimation of the strain amplitude is attributed to a small level of low frequency noise amplification. In order to test the sensitivity of the GPLFM to the level of model fidelity, perturbation to the scale of the foundation stiffness model is introduced. It is demonstrated that with induced errors in the natural frequencies and mode shapes of the mechanical model, the strain estimate by the GPLFM remains relatively insensitive, particularly at depths of interest near the mudline where the highest bending moments are expected to occur. This result illustrates that robustness in the strain estimate is attained by attributing sources of error in modal properties to the stochastic input estimate. Moreover, it provides evidence that the input estimate may be used as a type of heuristic indicator of error in modal properties, particularly of error in damping levels in the run-up and production states of OWTs. Through a study of the influence of the sensor configuration on the performance of the GPLFM, it is evident that estimation bias may be systematically accounted for through the inclusion of a strain channel in the measurement data. Not only does the strain data aid in correcting the predicted signal amplitude along all depths of the foundation, but it may be used to supply important information on the quasi-static behavior of the strain response which cannot be captured by high-frequency acceleration measurements alone \cite{Noppe2016, Smyth2007}. The best performing instrumentation scheme minimally consists of a single acceleration channel and single strain channel, where the GPLFM achieves a relative error of 16\% in the DEL, mean correlation coefficient of 0.968, and mean relative error of 2.28\% in the estimated subsoil strain time series across records from the production state of the OWT. The use of a heterogeneous dataset consisting of acceleration and strain measurements is deemed sufficient for curtailing model uncertainty, allowing the GPLFM to attain strong robustness and transferrability properties in estimating the subsoil strain response to variable operational and environmental loads.
\\
\indent We believe it is possible to further improve the performance of the GPLFM for subsoil strain estimation by two measures: deliberate estimation of quasi-static response, and explicit consideration of the dynamic evolution of modal properties of the structure. Regarding the former, other than by inclusion of strain data, low frequency error in the predicted strains can be mitigated by estimating measurement noise covariance with greater accuracy or by utilizing accelerometers with a lower noise baseline. Regarding the latter, this work demonstrates that the benefits of the GPLFM in providing a more flexible characterization of the input in the context of filtering are inhibited when this characterization is not extended to the context of modal identification. In other words, traditional OMA techniques, whose formulation relies on the assumption of white noise input, are not appropriate for the modal identification of offshore wind turbines in the run-up and production states, where environmental loads and rotor blade dynamics introduce harmonic components to the excitation \cite{vanVondelen2021}. Therefore, applications of the GPLFM for virtual sensing warrants additional developments to online modal identification techniques which allow for real-time modal parameter estimation accounting for non-white excitation in the system. To this end, the authors recommend further evaluation of proposed adjustments and alternatives to damping identification techniques for LTP systems which are reviewed in \cite{vanVondelen2021}. The reader is also referred to \cite{Rogers2020} where modal identification is treated with a Bayesian approach, in which inference of system parameters such as mass, stiffness, and damping is performed jointly with the estimation of the unknown ambient excitation by means of the GPLFM. This approach enables posterior state inference and modal identification to be performed jointly with a fully cohesive model of the input and may be applied across a greater range of contexts compared to traditional OMA methods.
\\
\indent There is potential for further improvements to the strain estimation through model choices in the GPLFM. In particular,  \textcolor{black}{a major assumption of uncorrelated modal forces may be overcome by explicitly inferring a covariance parameter between modal force sequences in the state-space matrices of Equation \eqref{eq:blockforcemodel}.  Secondly, model uncertainty could be addressed through the definition of a process noise variance parameter on response states (displacements and velocities) in the spectral density of the white noise term of the GPLFM. These additional model parameters, together with the hyperparameters of the GP kernel, constitute an augmented vector of hyperparameters which may be inferred with a maximum likelihood estimation approach. In theory, these improvements would lead to more accurately estimated statistics of the latent states, though at the cost of a higher dimensional optimization problem.} Finally, there are opportunities to investigate the role of the choice of GP covariance kernel in the GPLFM for response data which is driven by inputs of varying degrees of periodicity and temporal correlation. Beyond model choices, the results of this work promote further studies on on efficient and practical sensor placement schemes in monopile-based offshore wind turbines, to select the most informative data for the purpose of virtual sensing-based fatigue monitoring.

\section*{CRediT authorship contribution statement}

\textbf{J. Zou:} Methodology, Data curation, Software, Formal analysis, Investigation, Visualization, Writing - original draft. \textbf{E. Lourens:} Conceptualization, Supervision, Writing - review \& editing. \textbf{A. Cicirello:} Supervision, Writing - review \& editing.

\section*{Acknowledgements}

This research was funded by a Fulbright Student Research Grant awarded by the Stichting Fulbright Commission of the Netherlands. The data used in this paper is obtained from the DISSTINCT project, a collaboration between Siemens Wind Power, Fugro, Delft University of Technology, and DNV-GL which was in part funded by the Dutch government (project number TKIW02001). The authors gratefully acknowledge the people at Siemens-Gamesa Renewable Energy for providing the data, Joaquin Bilbao-Neiva for sharing code for the implementation of the SSI-Cov/OPTICS algorithm, and Mikk Laanes for sharing data preprocessing code.

\appendix

\section{}

%%%%%%%%%%%%%%%%%%%%%%%%%%%%%%%%%%%%%%%%%%%%%%%%%%%%%%%%%%%%%%%%%%%%%%%%%%%%%%%%%%%%%%%%%%%%%%%%%%
\subsection{Effect of model error on an illustrative example}
\label{sec:app_modelerror}

\begin{figure}[hbt!]
    \centering
    \includegraphics[width=0.7
    \linewidth]{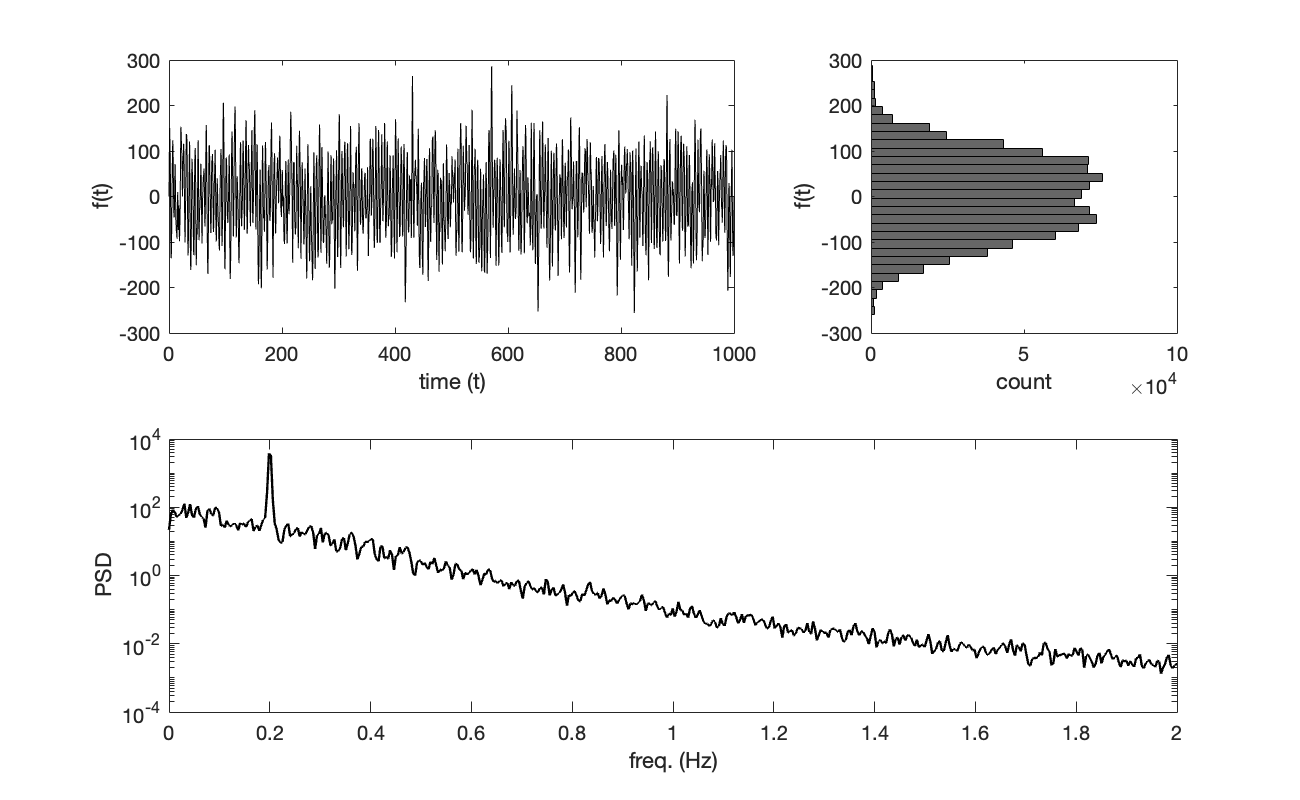}
    \caption{The ground truth input applied to the top level of the 10-dof cantilever structure, shown in terms of the time series (top left), histogram of realized values (top right), and frequency spectrum (bottom). }
    \label{fig:true_force}
\end{figure}

A synthetic model is used to illustrate the quality of estimation by the GPLFM when model error is introduced, as well as to show the efficacy of the measurement noise covariance tuning procedure described in Section \ref{subsec:noise}. The synthetic model is a 10-dof vertical mass-spring system with fixed-free end conditions representing a cantilever structure. The system is defined by ten point masses of $m = 200$ kg and a stiffness of $k = 5 \times 10^3$ N/m. Degrees of freedom are horizontal and aligned with the direction of forcing. Mass proportional damping with modal damping ratios of $\xi_j = 0.02$ for all degrees of freedom is assumed. An artificial input is defined as the combination of a sinusoidal force with an amplitude of 100 N and a frequency of 0.2 Hz, applied at the top (10th) level of the model, and a Gaussian process with a Mat\'ern $\nu=5/2$ kernel with parameters $\alpha=50, l_s=1$, applied to all 10 levels of the model. In this example, the number of degrees of freedom are $n_\textup{u} = 10$ and the number of forces are $n_\textup{p} = 2$, with the locations of the forces denoted by the selection matrix $\mathbf{S}_\textup{p} \in \mathbb{R}^{n_\textup{u} \times n_\textup{p}}$. The synthetic system is constructed to loosely represent environmental and operational load conditions on a wind turbine support structure, where the sinusoidal force represents the harmonic frequency of vibration induced by the rotor motion of the turbine and the GP force represents ambient wind excitation distributed along the height of the tower. Figure \ref{fig:true_force} shows the artificial input acting at the top level of the structure, where the harmonic frequency at 0.2 Hz is clearly visible in the frequency spectrum. The ensemble distribution of the input, shown by the histogram in the top right plot, illustrates how the input is non-Gaussian due to the contribution of the sinusoidal force.

\indent To set up the filtering problem, the ground truth response of the cantilever structure to the artificial load is simulated using the Newmark average acceleration method \cite{Chopra1995} for numerical integration with a time step of 0.001 s. The simulation model used for generating the ground truth is a full-order model of the system, whereas the prediction model is a modally reduced-order formulation of the GPLFM of the form in Equation \eqref{eq:gplfmfull} retaining the first $n_\textup{m} = 3$ modes of the system, such that modal states $\mathbf{x}(t) \in \mathbb{R}^{2n_\textup{m}}$, modal forces $\mathbf{f}(t) \in \mathbb{R}^{n_\textup{m}}$, and its higher derivatives up to order $\beta$ constitute the augmented state vector $\mathbf{z}^a(t)$. For more details on notation, see Section \ref{sec:theory}. The measurement data consists of the acceleration response at all 10 degrees of freedom, observed at a sampling frequency of 100 Hz and colored with white noise with covariance $\mathbf{R} = \sigma_R \mathbf{I}$, where $\sigma_R = 10^{-2}$. It is assumed that no process noise exists in the system, such that response states are known exactly, and that measurement noise is known to be independent across channels of measurement data with variance $\sigma_R$. 
\\
\indent Numerical experiments are conducted to examine the effect of model error. In each experiment, filtering is performed without perturbation, where the prediction model is directly a reduced-order formulation of the simulation model which generated the ground truth. Additionally, filtering is performed where the prediction model is a reduced-order model with perturbation in one of its properties, in order to illustrate the ability of the GPLFM to accommodate for misidentified model parameters in its prediction of latent states. Each version of the prediction model is fit independently using the method of GP hyperparameter inference described in Section \ref{subsec:GPkernels}; in the following studies, we assess the extent to which the GP hyperparameters corresponding to the Mat\'ern $\nu=5/2$ kernel for the prediction model with perturbation differ from that without perturbation. In this synthetic example, the true value of the latent states over time, including displacements, velocities, and modal components of the force, are available for comparison with the prediction by the GPLFM.

\subsection*{\textit{Error in modal damping ratios.}}
\begin{figure}[hbt!]
    \centering
    \includegraphics[width=0.9\linewidth]{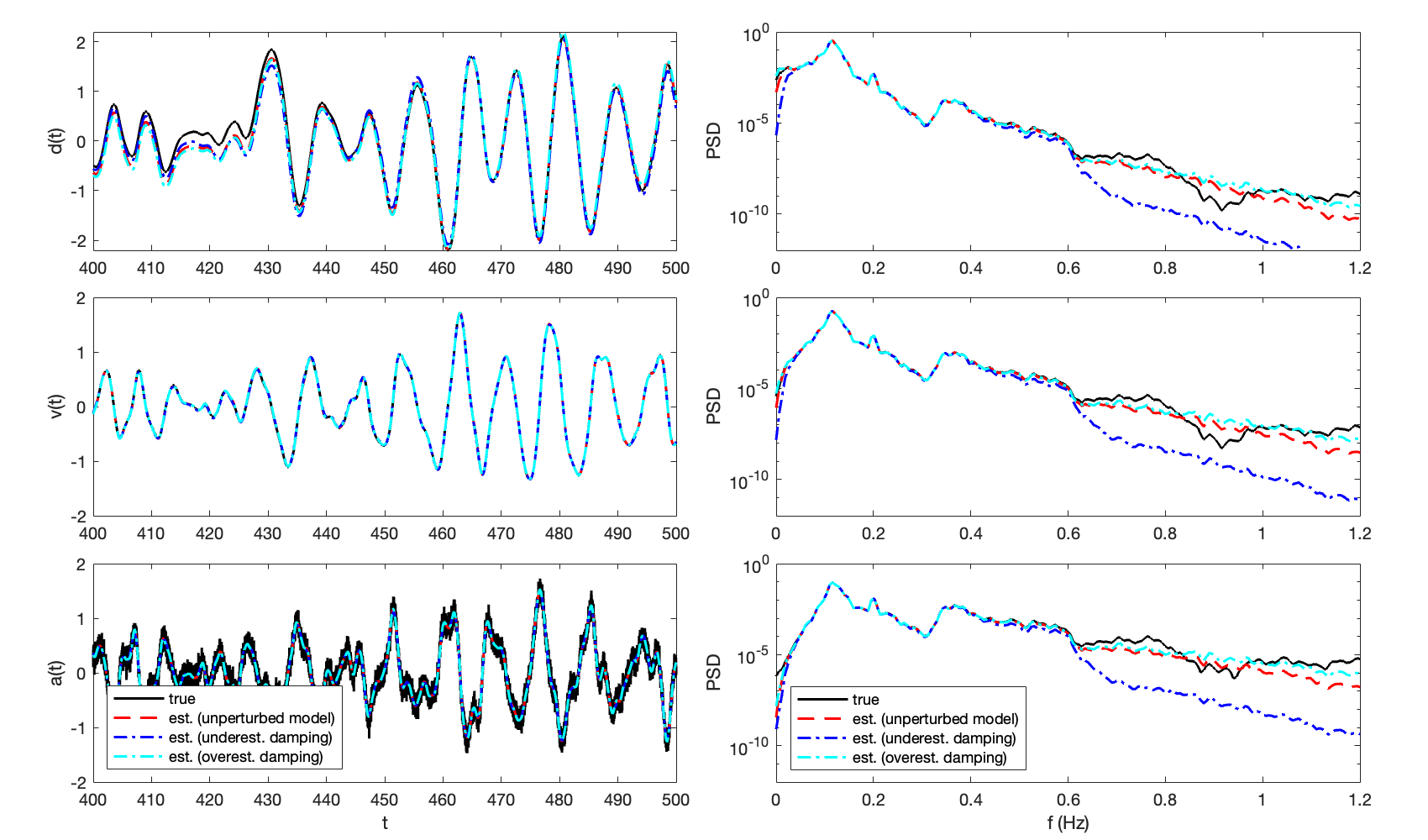}
    \caption{Estimates by the GPLFM of displacement (top row), velocity (center row), and acceleration (bottom row) states at level 5 of the 10-dof cantilever model with no perturbation (red), underestimated modal damping (blue), and overestimated modal damping (cyan), compared to the true response (black).}
    \label{fig:state_damp}
\end{figure}

\begin{figure}[hbt!]
    \centering
    \includegraphics[width=0.9\linewidth]{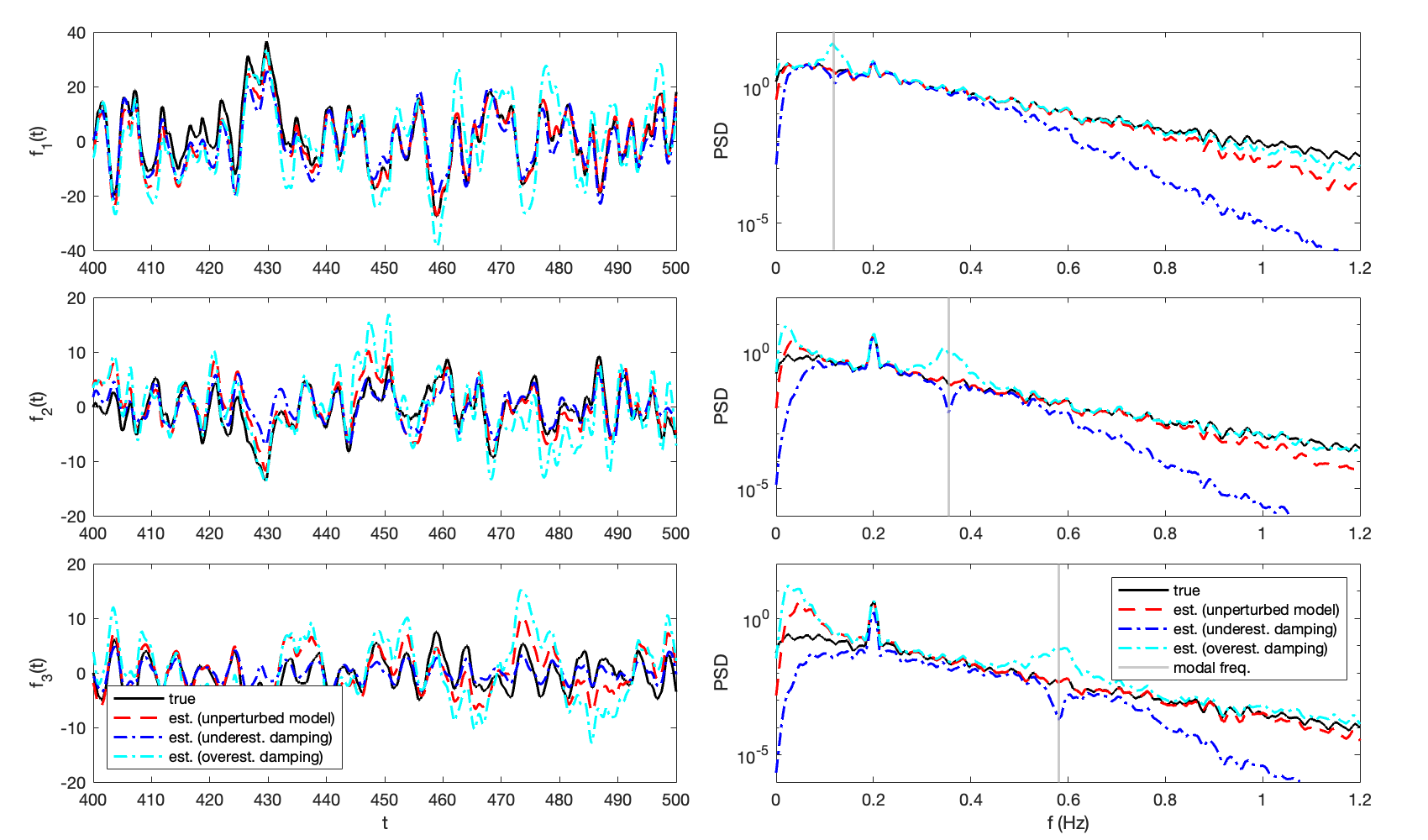}
    \caption{Estimates by the GPLFM of the first three modal forces (top, center, bottom rows respectively) using the 10-dof cantilever model with no perturbation (red), underestimated modal damping (blue), and overestimated modal damping (cyan), compared to the true modal forces (black).}
    \label{fig:force_damp}
\end{figure}

The prediction model without perturbation assumes the modal damping ratios $\xi_j$ are known exactly to be equal to 0.02. Two cases of model perturbation are considered, where a) the first three modal damping ratios are underestimated by an order of magnitude to be $\xi_j = 0.002$ and b) the first three modal damping ratios are overestimated by an order of magnitude to be $\xi_j = 0.2$. 
\\
\indent Figure \ref{fig:state_damp} shows the estimated displacement, velocity, and acceleration response at level 5 of the 10-dof cantilever model for each of the prediction models. We observe good congruence between the true response and predicted response in the time domain for all models, with some error in the predicted displacement time series. Due to the use of model order reduction to 3 modes in the prediction models, the frequency content of the estimate loses accuracy for frequencies above 0.6 Hz, beyond the third natural frequency of the system. In general, there is negligible effect of error in the modal damping ratio on the response states estimated by the GPLFM. This can in part be attributed to compensation in the modal force estimate, shown in Figure \ref{fig:force_damp}, where the estimated first, second, and third modal forces are compared to the modal components of the true force. In the frequency spectrum, a visible drop in the frequency about the $j$th natural frequency of the system is observed in the $j$th modal force estimate by the model with underestimated damping; similarly, a visible inflation of frequency is observed for the model with overestimated damping. The effect illustrates how the input estimate is utilized to account for mismatch between the system model and the measured response: in the case of underdamping, where greater frequency content about the natural frequencies would be expected according to the system model than what is present in the measurement data, the input estimate is adjusted such that the contribution of those particular frequencies from the input are reduced. This experiment lends evidence that Bayesian joint input-state estimators have an advantage over state-only estimators in that the inclusion of input estimation can mitigate certain sources of model error and aid the primary objective of achieving accurate state estimation. It is notable that the presence of the harmonic frequency between the first and second natural frequencies of the structure does not interfere with the accuracy of either the response estimate or the input estimate.

\subsection*{\textit{Error in the measurement noise variance.}}
\begin{figure}[hbt!]
    \centering
    \includegraphics[width=0.9\linewidth]{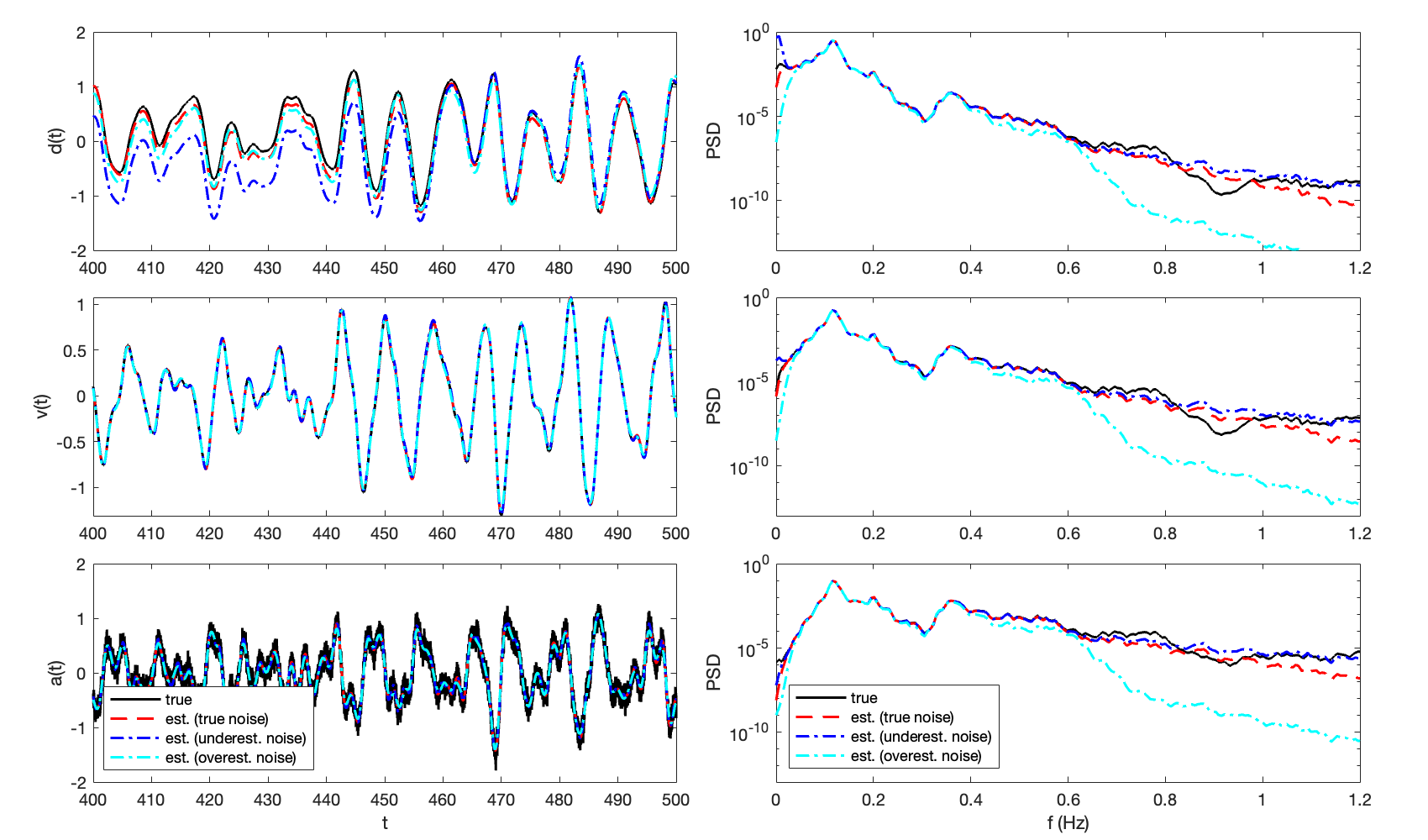}
    \caption{Estimates by the GPLFM of displacement (top row), velocity (center row), and acceleration (bottom row) states at level 5 of the 10-dof cantilever model with known measurement noise variance (red), underestimated measurement noise variance (blue), and overestimated measurement noise variance (cyan), compared to the true response (black).}
    \label{fig:state_noise}
\end{figure}

\begin{figure}[hbt!]
    \centering
    \includegraphics[width=0.9\linewidth]{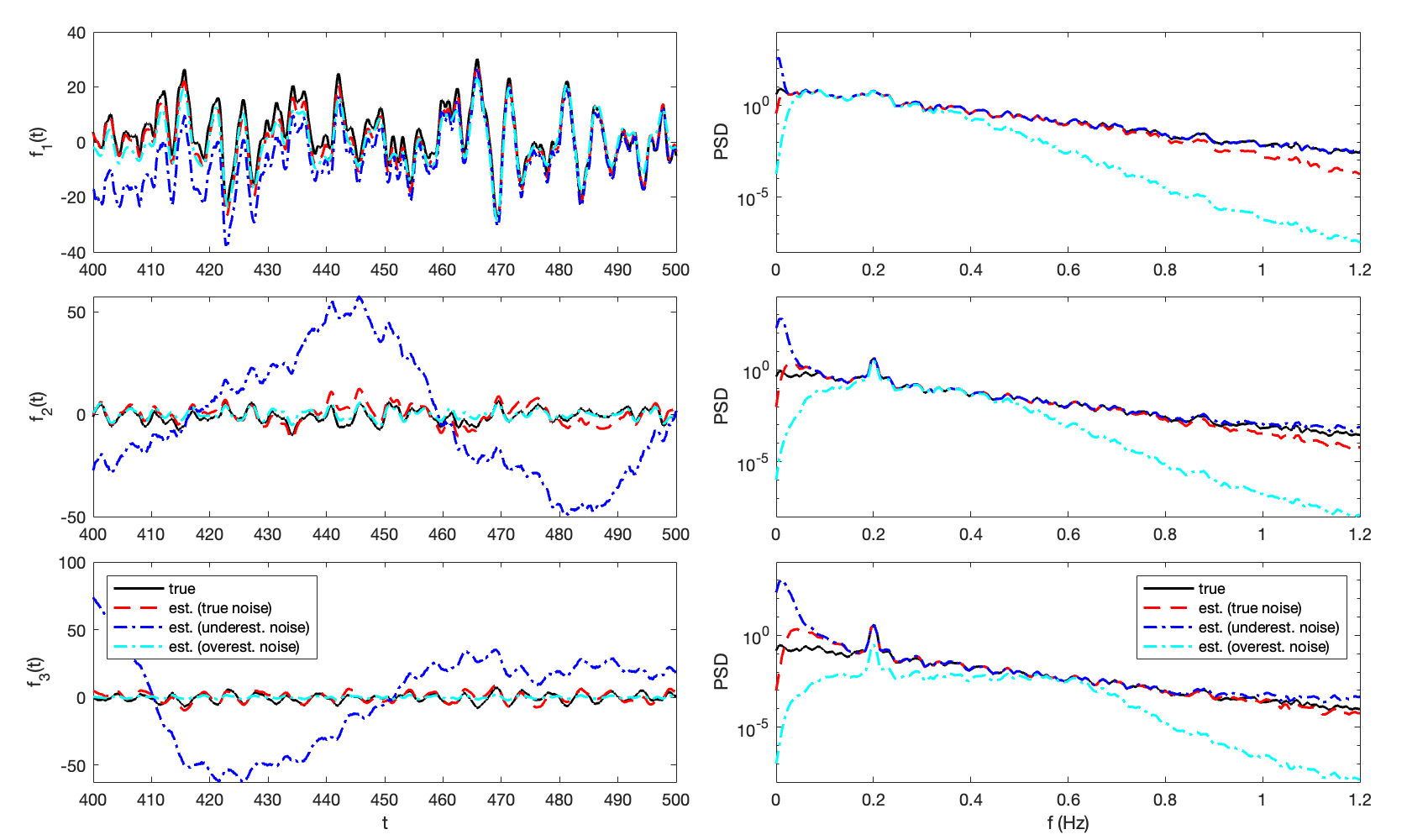}
    \caption{Estimates by the GPLFM of the first three modal forces (top, center, bottom rows respectively) using the 10-dof cantilever model with known measurement noise variance (red), underestimated measurement noise variance (blue), and overestimated measurement noise variance (cyan), compared to the true modal forces (black).}
    \label{fig:force_noise}
\end{figure}

In the prediction model without perturbation, the variance of measurement noise in each channel is known to be $\sigma_R = 10^{-2}$. Two cases of model perturbation are considered, where a) the measurement noise variance is overestimated by 3 orders of magnitude to be $\sigma_R = 10^{1}$ and b) the measurement noise variance is underestimated by 3 orders of magnitude to be $\sigma_R = 10^{-5}$. In this experiment, erroneous values of the measurement noise covariance matrix are provided as input to the Kalman filter with no tuning performed. 
\\
\indent In Figure \ref{fig:state_noise}, it is observed that while high accuracy is retained in the predicted velocity and acceleration response, the introduction of measurement noise results in reduced accuracy in the estimation of low frequency components of the response states, to the greatest degree in the displacement response. This same low frequency drift is observed in the modal force estimate in Figure \ref{fig:force_noise}, with a larger effect in the second and third modal forces. Measurement noise affects the frequency content of the measurement data by introducing greater contributions of higher frequencies. In the case where measurement noise variance is overestimated, the predicted signal has reduced contributions of high frequencies which are expected by the prediction model but not present in the measured signal. In the case where measurement noise variance is underestimated, the predicted signal captures this presence of high frequencies in the measured signal; however, it is countered by inflation of low frequencies which the prediction model expects to observe in the absence of significant measurement noise.

\begin{figure}[hbt!]
    \centering
    \includegraphics[width=0.9\linewidth]{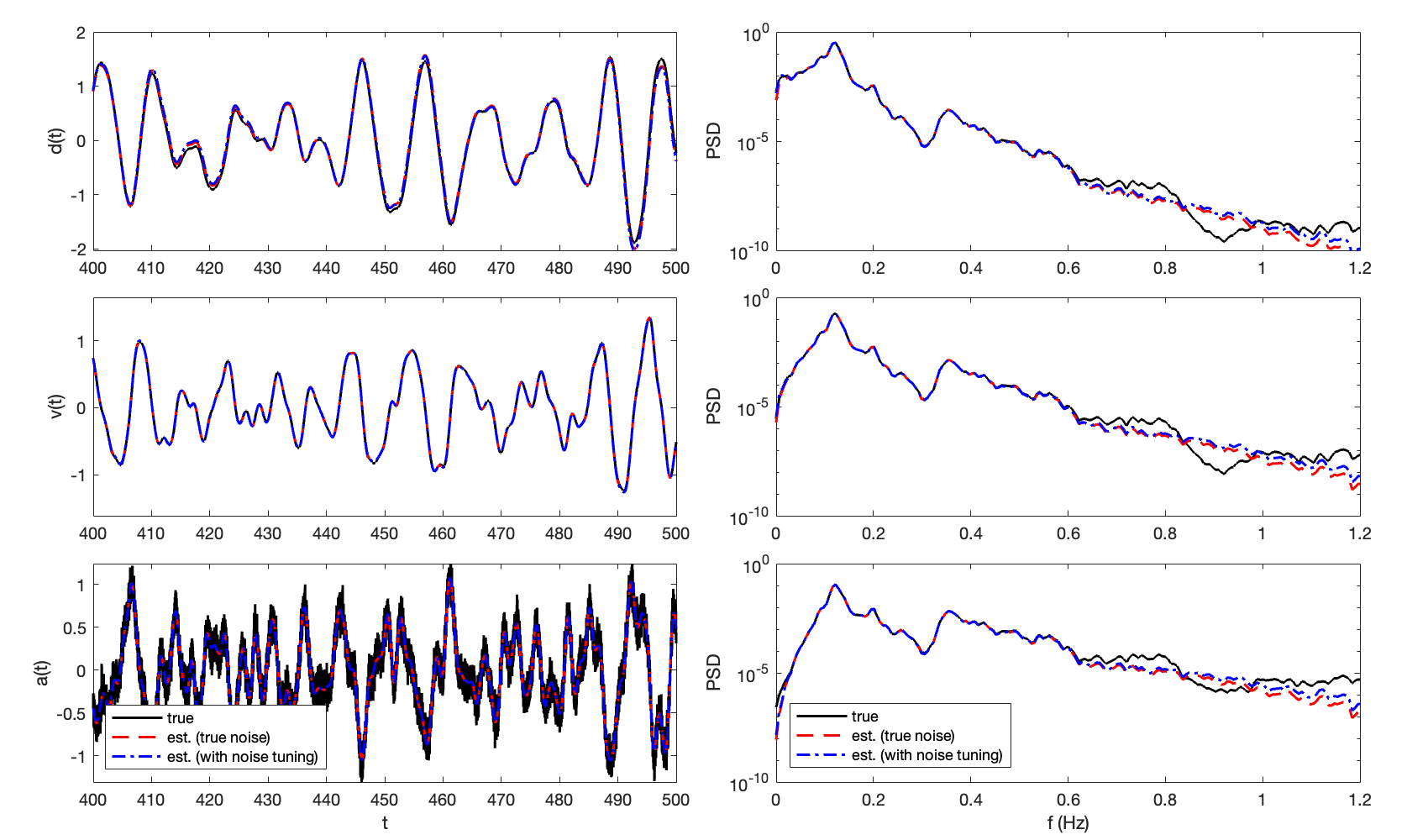}
    \caption{Estimates by the GPLFM of displacement (top row), velocity (center row), and acceleration (bottom row) states at level 5 of the 10-dof cantilever model with known measurement noise variance (red) and with measurement noise covariance tuning using the procedure in Section \ref{subsec:noise} (blue), compared to the true response (black).}
    \label{fig:state_noisetuning}
\end{figure}

\begin{figure}[hbt!]
    \centering
    \includegraphics[width=0.9\linewidth]{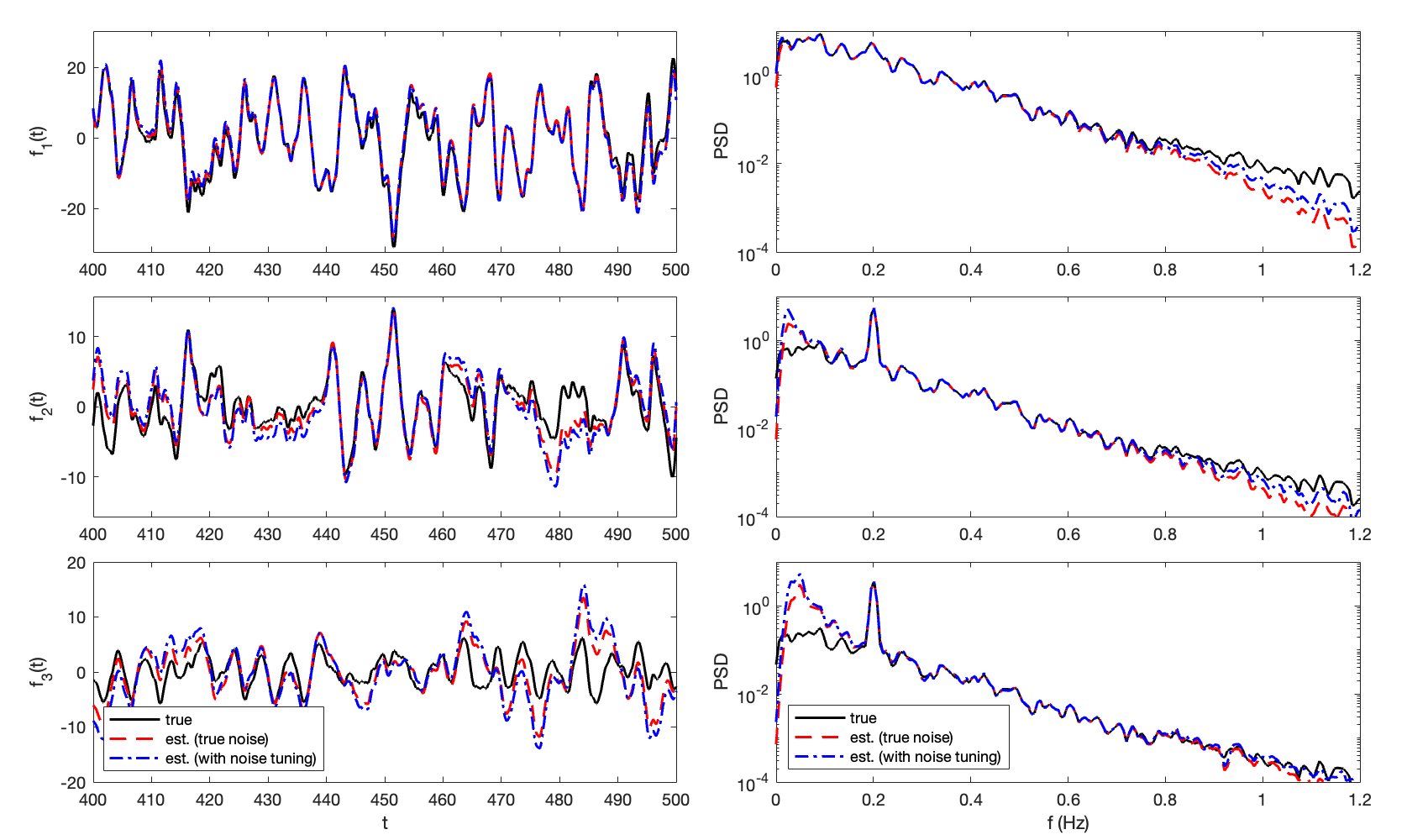}
    \caption{Estimates by the GPLFM of the first three modal forces (top, center, bottom rows respectively) using the 10-dof cantilever model with known measurement noise variance (red) and with measurement noise covariance tuning using the procedure in Section \ref{subsec:noise} (blue), compared to the true response (black).}
    \label{fig:force_noisetuning}
\end{figure}

\indent An additional experiment is conducted to demonstrate the efficacy of the measurement noise covariance tuning procedure in Section \ref{subsec:noise}, which refines the measurement noise covariance matrix $\mathbf{R}$ in the prediction model using the residual sequence between the predicted and measured acceleration response over $k$ iterations. It is observed that the sequence  $\{\mathbf{R}_k \}$ converges to the true value of $\mathbf{R}$ over a number of iterations on the order of $k=3$. While this convergence is observed regardless of the initial value, this experiment presents the case where the measurement noise covariance is initialized at $\mathbf{R}_0 = \sigma_R \mathbf{I}, \sigma_R = 10^{-5}$. Using the tuning procedure, prediction error is effectively eliminated once the order of magnitude of the measurement noise variance is correctly identified, as shown in the response estimation in Figure \ref{fig:state_noisetuning} and modal force estimation in Figure \ref{fig:force_noisetuning}.

\subsection*{\textit{Error in the modal frequencies and mode shapes.}}
\begin{figure}[hbt!]
    \centering
    \includegraphics[width=0.5\linewidth]{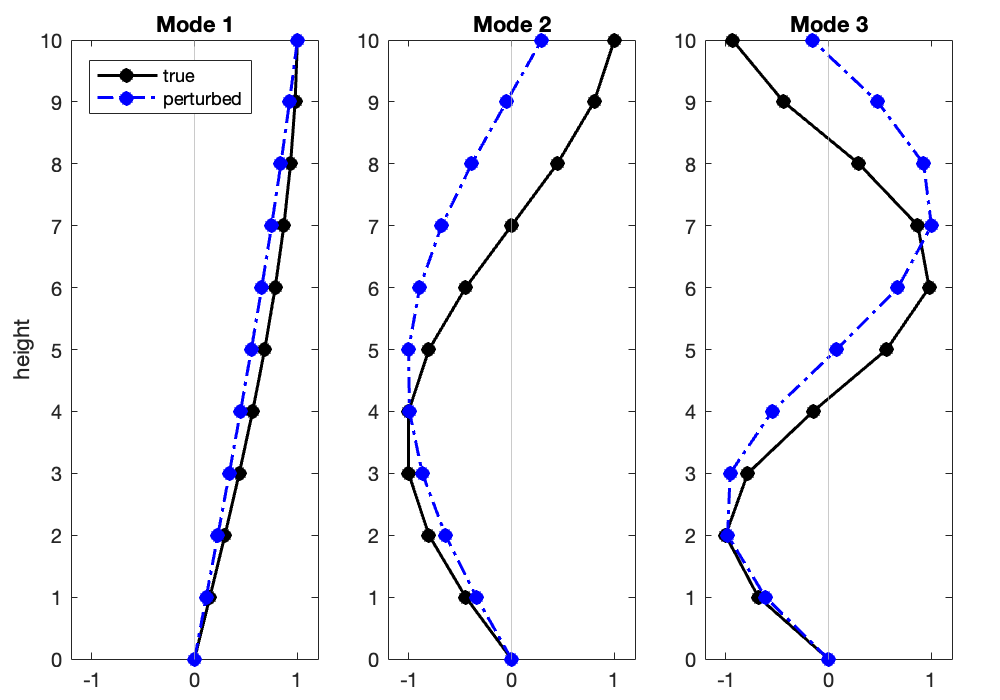}
    \caption{First three mode shapes of the 10-dof cantilever model without perturbation (black) and with perturbation (blue) from the modal properties of the simulation model. }
    \label{fig:perturb_MS}
\end{figure}

\begin{figure}[hbt!]
    \centering
    \includegraphics[width=0.9\linewidth]{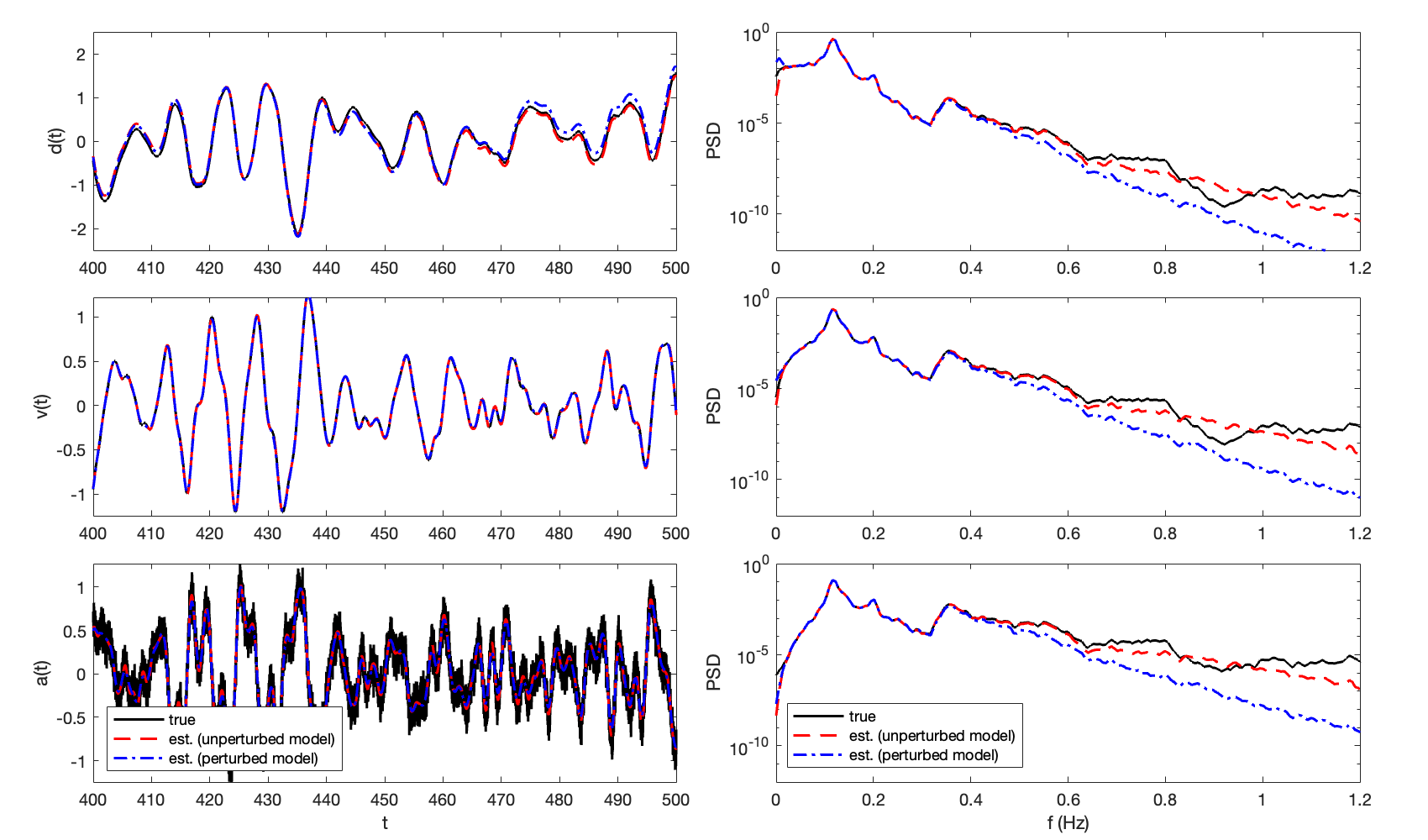}
    \caption{Estimates by the GPLFM of displacement (top row), velocity (center row), and acceleration (bottom row) states at level 5 of the 10-dof cantilever model with no perturbation (red) and with perturbation from the true simulation model (blue), compared to the true response (black).}
    \label{fig:state_MS}
\end{figure}

\begin{figure}[hbt!]
    \centering
    \includegraphics[width=0.9\linewidth]{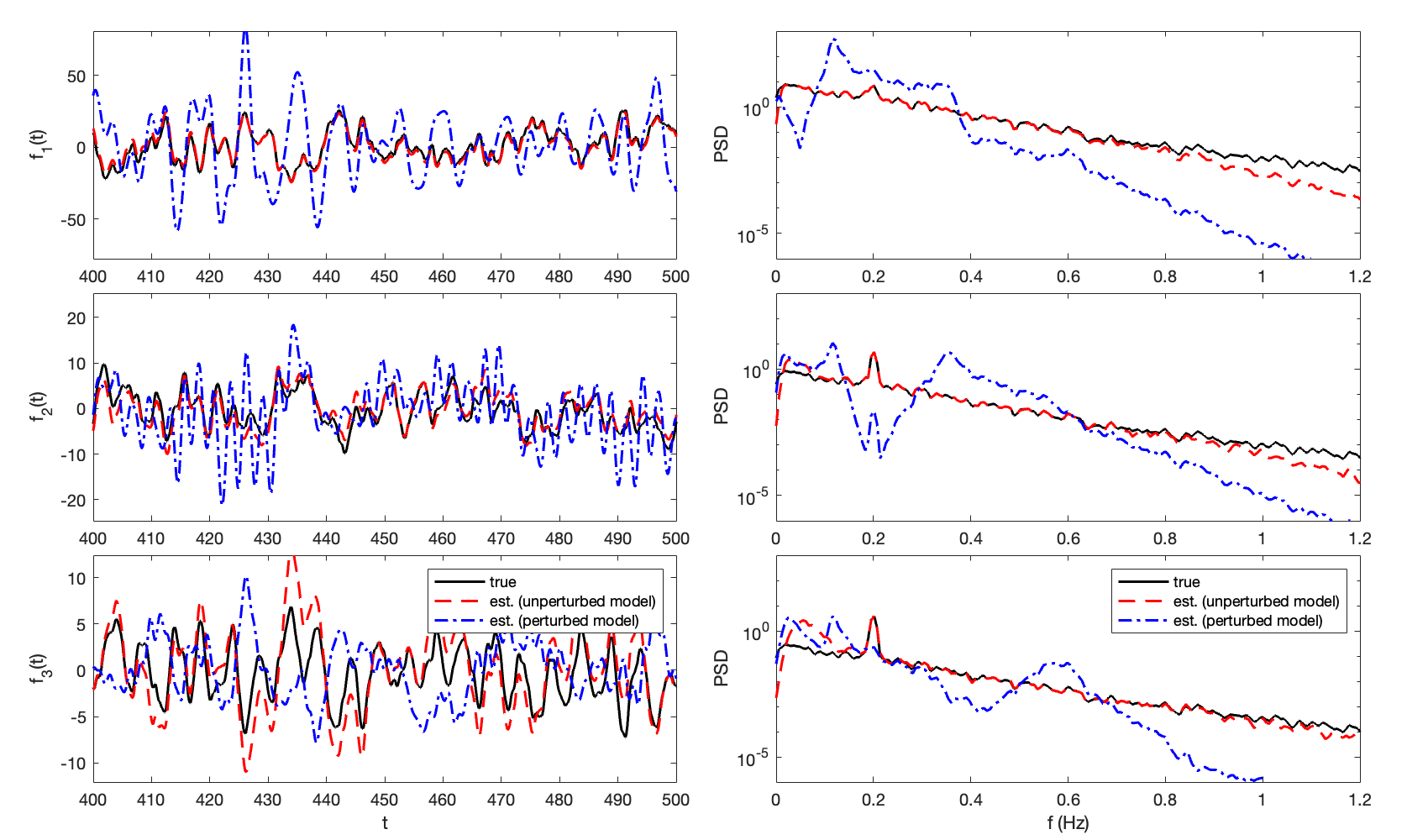}
    \caption{Estimates by the GPLFM of the first three modal forces (top, center, bottom rows respectively) using the 10-dof cantilever model with no perturbation (red) and with perturbation from the true simulation model (blue), compared to the true response (black).}
    \label{fig:force_MS}
\end{figure}

The prediction model without perturbation assumes modal properties are known exactly, where the first three natural frequencies of the system are 0.12 Hz, 0.35 Hz, and 0.58 Hz. One case of model perturbation is considered, where the prediction model has reduced stiffness of $0.5k$ and a concentrated mass of $2000$ kg at the top of the tower, thus having modal properties which deviate from that of the simulation model. The first three natural frequencies of the perturbed model are 0.05 Hz, 0.19 Hz, and 0.36 Hz. The mode shapes of the two prediction models are shown in Figure \ref{fig:perturb_MS}.
\\
\indent In Figure \ref{fig:state_MS}, it is observed that even with error in the modal properties of the prediction model, the estimation of response states is highly accurate with respect to the true response and on par with the estimate using the error-free prediction model. Figure \ref{fig:force_MS} shows that this high accuracy in the response states is attained by a corresponding adjustment to the modal force estimate. In particular, we draw attention to the frequency content of the modal forces about the natural frequencies of both the perturbed and unperturbed prediction models. In the first modal force, the frequency contribution of the first natural frequency of the perturbed model is suppressed, while that of the unperturbed model is inflated. A similar pattern is observed in the second and third modal forces with respect to the second and third natural frequencies, respectively. Even in the second modal force, where the second natural frequency of the unperturbed model nearly coincides with the nonstructural harmonic frequency, the power of frequency in the vicinity of the structural frequency is reduced. This effect illustrates that the input estimate accounts for error in modal properties of the system by identifying frequencies which are dominant in the measurement data as well as frequencies which are less prevalent in the measurement data than what would be expected, such as frequency bands about the natural frequencies of the prediction model.

\subsection*{\textit{Changes in GP kernel hyperparameters in response to model error}}

\begin{figure}[hbt!]
    \centering
    \includegraphics[width=0.9\linewidth]{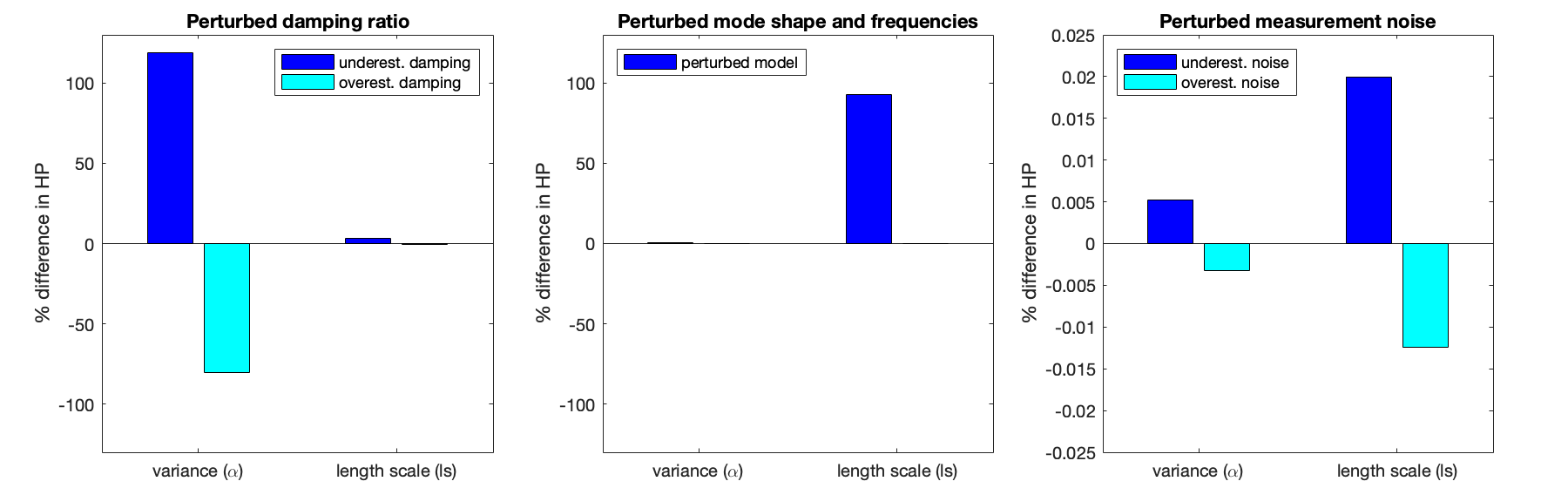}
    \caption{Percent difference in the GPLFM hyperparameters corresponding to a Mat\'ern kernel ($\nu$ = 5/2) with respect to the value of hyperparameters computed with the unperturbed model. The left plot shows percent change in the two hyperparameters from underestimation and overestimation of the first three modal damping ratios. The right plot shows percent change in the two hyperparameters from underestimation and overestimation of the measurement noise variance. }
    \label{fig:perturb_hp}
\end{figure}

While this interchange between the state estimate and input estimate is characteristic of the general class of Bayesian joint input-state estimators, the GPLFM can provide a concise parametric description of how the model adjusts to account for mismatch with measured data. For the prediction model both with and without perturbation, the GPLFM is fit by means of the parameter inference procedure in Section \ref{subsec:GPkernels}, where the parameters in this instance are the GP kernel hyperparameters, $\theta = [\alpha, l_s]$. Figure \ref{fig:perturb_hp} shows the percent change in the parameters $\theta$ fit for the perturbed model with respect to the parameters fit for the error-free model for each of the conducted experiments. When error in the modal damping ratio is introduced, a large change in the variance parameter $\alpha$ is observed - a nearly 100\% increase in $\alpha$ in the case of overestimated damping and a decrease of a similar magnitude in the case of underestimated damping. In contrast, very small change is observed in the length scale parameter $l_s$. The opposite is the case in the experiment where error in the mode shapes and natural frequencies is introduced, where large change in the length scale parameter $l_s$ and negligible change in the variance parameter $\alpha$ are observed. The mechanism by which error in modal properties are compensated for in the GPLFM is reflected in the change in GP kernel hyperparameters, where error in damping levels primarily affects the amplitude of the latent states, measured by $\alpha$, and error in the natural frequencies primarily affects the frequency content of the latent states, measured by $l_s$.  When error in the measurement noise variance is introduced, there is noticeably greater change in $l_s$ compared to $\alpha$, reflecting the influence of measurement noise on frequency content; however, in comparison to the change in parameters observed from errors in the modal properties of the system, the adjustment is considerably smaller or even negligible. This demonstrates that while the GP kernel hyperparameters of the GPLFM are the statistical model parameters which are most sensitive to error in modal properties such as modal damping ratios, natural frequencies, and mode shapes, these parameters are not sensitive to other sources of error, such as error in qualities of the measurement noise. Rather than use the GP kernel hyperparameter tuning process, error from misidentified measurement noise covariance is more effectively addressed by the aforementioned measurement noise covariance tuning process.

%%%%%%%%%%%%%%%%%%%%%%%%%%%%%%%%%%%%%%%%%%%%%%%%%%%%%%%%%%%%%%%%%%%%%%%%%%%%%%%%%%%%%%%%%%%%%%%%%%
\subsection{Transformation of lateral displacements to axial strain}
\label{sec:app_disptransf}

From elastic beam theory, the lateral deformation of a 2D beam element $v$ is represented with a third order polynomial of the form: 

\begin{equation}
    v(\xi) = a_0 + a_1\xi + a_2\xi^2 + a_3\xi^3
\end{equation}

$\xi = \frac{x}{L}$ is the normalized axial distance along length $L$ of the element. The unknown coefficients $a_0, a_1, a_2, a_3$ can be solved for by enforcing boundary conditions which relate the deformation to the vector of degrees of freedom of the element $\mathbf{d}_{\mathrm{e}}$, one translational and one rotational degree of freedom at each end:

\begin{equation}
    \label{eq:strain0}
    v(\xi) = 
    \begin{bmatrix}
    N_1(\xi) & N_2(\xi) & N_3(\xi) & N_4(\xi)
    \end{bmatrix}
    \begin{bmatrix}
    v_1 \\ \theta_1 \\ v_2 \\ \theta_2
    \end{bmatrix} = 
    \mathbf{N}_{\mathrm{e}} \mathbf{d}_{\mathrm{e}}
\end{equation}

The solution provides the element shape functions $\mathbf{N}_{\mathrm{e}}$, which approximate the continuous spatial distribution of deformation between finite element nodes. The derivations of shape functions for an Euler-Bernoulli beam element and Timoshenko beam element are detailed in \cite{Liu2003} and \cite{Bazoune2003}, respectively. 

For small deformations, the axial displacement $u$ may be approximated by the rotational deformation $\theta_z$ and distance from the element centroidal axis $y$ as: 

\begin{equation}
    \label{eq:strain1}
    u = -y\theta_z
\end{equation}

The axial strain $\epsilon_{xx}$ is related to the rate of change of the axial displacement with respect to $x$, which is given as: 

\begin{equation}
    \label{eq:strain2}
    \epsilon_{xx} = \diffp{u}{x} = -y\diffp{\theta_z}{x} = -y\diffp{v}{{x^2}}
\end{equation}

Substituting the expression for $v$ from Equation \eqref{eq:strain0}, Equation \eqref{eq:strain2} becomes:

\begin{equation}
    \label{eq:strain3}
    \epsilon_{xx} = -\frac{y}{L^2}\mathbf{N}_{\mathrm{e}}''(\xi)\mathbf{d}_{\mathrm{e}} = 
    \mathbf{B}_{\mathrm{e}}(\xi)\mathbf{d}_{\mathrm{e}}
\end{equation}

where $\mathbf{B}_{\mathrm{e}}(\xi) = \begin{bmatrix} B_1(\xi) \quad B_2(\xi) \quad B_3(\xi) \quad B_4(\xi) \end{bmatrix}$ is the element strain matrix.
\\
\indent For a finite element model with a chain of $n_{\mathrm{e}}$ identical elements and free end conditions, the global strain vector can be computed as the  matrix product $\mathbf{\epsilon} = \mathbf{Td}$, which is written in expanded form as follows:

\begin{equation}
    \label{eq:strain4}
    \begin{bmatrix}
        \epsilon_{xx}(\xi_1) \\
        \epsilon_{xx}(\xi_2) \\
        \vdots \\
        \epsilon_{xx}(\xi_{n_{\mathrm{e}}})
    \end{bmatrix} = \begin{bmatrix}
        B_1(\xi_1) & B_2(\xi_1) & B_3(\xi_1) & B_4(\xi_1) &  &  &  &  &   \\
        &  & B_1(\xi_2) & B_2(\xi_2) & B_3(\xi_2) & B_4(\xi_2) &  &  &   \\
        &  &  &  &  \ddots &  &  &  & \\
        &  &  &  &  & B_1(\xi_{n_{\mathrm{e}}}) & B_2(\xi_{n_{\mathrm{e}}}) & B_3(\xi_{n_{\mathrm{e}}}) & B_4(\xi_{n_{\mathrm{e}}}) 
    \end{bmatrix}
    \begin{bmatrix}
        v_1 \\ \theta_1 \\
        v_2 \\ \theta_2 \\
        \vdots \\
        v_{n_{\mathrm{e}}} \\ \theta_{n_{\mathrm{e}}} \\
    \end{bmatrix}
\end{equation}

Each entry in the global strain vector provides the strain estimate at a distance $\xi_i$ along the $i$th element, for $i = 1,...,n_{\mathrm{e}}$. If the elements are non-identical, then each row of the global strain matrix $\mathbf{T}$ corresponds to the strain functions of the $i$th element.

%%%%%%%%%%%%%%%%%%%%%%%%%%%%%%%%%%%%%%%%%%%%%%%%%%%%%%%%%%%%%%%%%%%%%%%%%%%%%%%%%%%%%%%%%%%%%%%%%%
\subsection{Kalman filtering and RTS smoothing equations}
\label{sec:app_KFKS}

Given a discrete linear time-invariant state-space model of the form

\begin{subequations}
    \begin{align}
        \mathbf{z}_k = \mathbf{F}\mathbf{z}_{k-1} + \mathbf{w}_{k-1} 
        \\
        \mathbf{y}_k = \mathbf{H}\mathbf{z}_k + \mathbf{v}_k
    \end{align}

where the state vector has a prior distribution $\mathbf{z}_0 \sim N(\hat{\mathbf{z}}_{0|0}, \mathbf{P}_{0|0})$ and the additive white Gaussian noise terms have covariances $\mathbb{E}[\mathbf{w}_k\mathbf{w}_l^\textup{T}] = \mathbf{Q}$, $\mathbb{E}[\mathbf{v}_k\mathbf{v}_l^\textup{T}] =\mathbf{R}$, and $\mathbb{E}[\mathbf{w}_k\mathbf{v}_l^\textup{T}] = \mathbf{0}$. 
\\
\indent Let $\mathbf{z}_{p|q}$ denote the state predicted at time step $p$ given all time steps $1:q$. The Kalman filter of Algorithm \ref{alg:KF} solves for the posterior filtering distribution, where the mean and covariance in $\mathbf{z}_{k|k} \sim N(\hat{\mathbf{z}}_{k|k}, \mathbf{P}_{k|k})$ is predicted from a "forward pass" through time steps $k=1,...,N$ and updated at time steps where measurements are available \cite{Kalman1960}. The RTS smoother of Algorithm \ref{alg:KS} solves for the posterior smoothing distribution, where the predicted mean and covariance from the Kalman filter (denoted with superscript "-" for clarity) is updated with a fixed-interval "backward pass" through time steps $k=N-1,...,1$ \cite{RTS1965}. The final distribution $\mathbf{z}_{k|N} \sim N(\hat{\mathbf{z}}_{k|N}, \mathbf{P}_{k|N})$ constitutes the complete posterior distribution of the state given all measurements in a sequence of $N$ steps to estimate. From this distribution, a posterior estimate of the measurement states may be recovered as $\mathbf{y}_{k|N} \sim N(\hat{\mathbf{y}}_{k|N}, \mathbf{P}_{k|N}^y)$.

\begin{algorithm}[H]
\caption{Kalman Filter}\label{alg:KF}
\begin{algorithmic}
\For{$k = 1,...,N$}
    \State Filter prediction equations:
    \State $\hat{\mathbf{z}}^-_{k|k-1} = \mathbf{F}\hat{\mathbf{z}}_{k-1|k-1}$ 
    \State $\mathbf{P}^-_{k|k-1} = \mathbf{F}\mathbf{P}_{k-1|k-1}\mathbf{F}^\textup{T} + \mathbf{Q}$  
\If{$\mathbf{y}_k$ is observed}
    \State Filter update equations:
    \State $\mathbf{e}_k = \mathbf{y}_k - \mathbf{H}\hat{\mathbf{z}}^-_{k|k-1}$ \Comment prediction error
    \State $\mathbf{S}_k = \mathbf{H}\mathbf{P}^-_{k|k-1}\mathbf{H}^\textup{T} + \mathbf{R}$ 
    \State $\mathbf{K}_k = \mathbf{P}^-_{k|k-1}\mathbf{H}^\textup{T}(\mathbf{S}_k)^{-1}$ \Comment Kalman gain
    \State $\hat{\mathbf{z}}_{k|k} = \hat{\mathbf{z}}^-_{k|k-1} - \mathbf{K}_k\mathbf{e}_k$
    \State $\mathbf{P}_{k|k} = \mathbf{P}^-_{k|k-1} - \mathbf{K}_k\mathbf{H}\mathbf{P}^-_{k|k-1}$
\EndIf
\EndFor
\end{algorithmic}
\end{algorithm}

\begin{algorithm}[H]
\caption{Rauch-Tung-Striebel (RTS) Smoother}\label{alg:KS}
\begin{algorithmic}
\For{$k = N-1,...,1$}
    \State Smoothing update equations:
    \State $\mathbf{G}_k = \mathbf{P}_{k|k}\mathbf{F}^\textup{T} \left( \mathbf{P}^-_{k+1|k} \right) ^{-1}$
    \State $\hat{\mathbf{z}}_{k|N} = \hat{\mathbf{z}}_{k|k} + \mathbf{G}_k \left( \hat{\mathbf{z}}_{k+1|N}-\hat{\mathbf{z}}^-_{k+1|k} \right)$ 
    \State $\mathbf{P}_{k|N} = \mathbf{P}_{k|k} + \mathbf{G}_k \left( \mathbf{P}_{k+1|N} - \mathbf{P}^-_{k+1|k} \right) \mathbf{G}_k^\textup{T}$  
    
    \State Moments of measurement states $\mathbf{y}$:
    \State $\hat{\mathbf{y}}_{k|N} = \mathbf{H}\hat{\mathbf{z}}_{k|N}$ 
    \State $\mathbf{P}_{k|N}^y = \mathbf{H}\mathbf{P}_{k|N}\mathbf{H}^\textup{T} + \mathbf{R}$ 
\EndFor
\end{algorithmic}
\end{algorithm}

\end{subequations}

%%%%%%%%%%%%%%%%%%%%%%%%%%%%%%%%%%%%%%%%%%%%%%%%%%%%%%%%%%%%%%%%%%%%%%%%%%%%%%%%%%%%%%%%%%%%%%%%%%
\subsection{SSM for Mat\'ern kernel}
\label{sec:app_matern}

The Mat\'ern family is a class of covariance functions of the form

\begin{equation}
    \kappa(\tau; \nu, \alpha, l_s) = \alpha^2 \frac{2^{(1-\nu)}}{\Gamma(\nu)} \left( \frac{\sqrt{2\nu}}{l_s} \tau \right)^\nu K_\nu \left( \frac{\sqrt{2\nu}}{l_s} \tau \right)
\end{equation}

where $\Gamma(\nu)$ is the Gamma function and $K_\nu$ is the modified Bessel function of the second kind. The Mat\'ern kernel defines the covariance between two points $t, t'$ of a process to be a function of only the distance $\tau = t' - t$ between them, resulting in stationary stochastic processes. The hyperparameters influence qualities of the covariance relationship derived from the distance: the vertical length scale $\alpha$ controls the magnitude of variance at $\tau=0$, the horizontal length scale $l_s$ controls the distance $\tau$ over which covariance diminishes to zero, and the parameter $\nu$ controls the smoothness of the resulting process. Note that in its limits, a Mat\'ern kernel with $l_s \rightarrow 0$ obtains the Dirac delta function for the covariance, which corresponds to a white noise kernel, and $\nu \rightarrow \infty$ obtains the squared exponential covariance function corresponding to infinitely differentiable processes \cite{Rasmussen2006, Hartikainen2010}. 
\\
\indent This section presents the form of the Mat\'ern kernel with smoothing parameter $\nu=p+1/2=5/2$, where $p=2$, leading to processes which are twice differentiable. The associated hyperparameters are then $\theta=[\alpha, l_s]$. The covariance function becomes:

\begin{equation}
    \kappa_{5/2}(\tau; \alpha, l_s) = \alpha^2 \left( 1 + \frac{\sqrt{5}}{l_s}\tau + \frac{5\tau^2}{3l_s^2} \right) \textup{exp} \left(- \frac{\sqrt{5}}{l_s} \tau \right)
\end{equation}

A Gaussian process $h(t) \sim GP(0, \kappa_{5/2}(\tau))$ may be expressed as a third-order linear stochastic differential equation of the form: 

\begin{equation}
    \label{eq:maternSDE}
    \frac{d^{3}h(t)}{dt^3} +
    3 \lambda \frac{d^{2}h(t)}{dt^{2}} +
    3 \lambda^2 \frac{dh(t)}{dt} +
    \lambda^3 h(t) = w(t)
\end{equation}

where $\lambda = \sqrt{5}/l_s$. The system is driven by white noise with spectral density:

\begin{equation}
     q_w = S_w(\omega) = \frac{400 \sqrt{5} \alpha^2}{3l_s^5}
\end{equation}

Consequently, we can use the form of the Mat\'ern kernel to derive the spectral density of $h(t)$ as $S_h(\omega) = q_w (\lambda^2 + \omega^2)^{-(p+1)}$.
\\
\indent The companion form of \ref{eq:maternSDE} gives rise to the state-space model of process $h(t)$, with the continuous-time matrices \cite{Nayek2019}:

\begin{equation}
    \label{eq:maternforcemodel}
    \mathbf{F}_{c} = \begin{bmatrix} 0  & 1 & 0 \\
    0 & 0 & 1 \\
    -\lambda^3    & -3\lambda^2   & -3\lambda \\ \end{bmatrix}, \quad 
    \mathbf{L}_{c} = \begin{bmatrix} 0 \\ 0 \\ 1 \end{bmatrix}, \quad 
    \mathbf{H}_{c} = \begin{bmatrix} 1 & 0 & 0 \end{bmatrix}
\end{equation}

%%%%%%%%%%%%%%%%%%%%%%%%%%%%%%%%%%%%%%%%%%%%%%%%%%%%%%%%%%%%%%%%%%%%%%%%%%%%%%%%%%%%%%%%%%%%%%%%%%
\subsection{Hellinger distance}
\label{sec:app_hellinger}

The Hellinger distance is a type of f-divergence which integrates the error between two probability densities $p(x)$ and $q(x)$ as follows: 

\begin{equation}
    \label{eq:hellingercont}
    H^2(p, q) = \frac{1}{2} \int \left( \sqrt{p(x)} - \sqrt{q(x)} \right)^2 dx = 1 - \int \sqrt{p(x) \, q(x)} \ dx
\end{equation}

In the discrete case, the Hellinger distance between the probability distributions $p = \{ p_1, ..., p_m \}$ and $q = \{q_1, ..., q_m \}$ may be computed as a Riemann sum over $m$ subdivisions: 

\begin{equation}
    \label{eq:hellingerdiscr}
    H(p, q) = 1 - \sum_{i=1}^{m} \sqrt{p_i \, q_i}
\end{equation}

%%%%%%%%%%%%%%%%%%%%%%%%%%%%%%%%%%%%%%%%%%%%%%%%%%%%%%%%%%%%%%%%%%%%%%%%%%%%%%%%%%%%%%%%%%%%%%%%%%
\subsection{Sensor specifications}
\label{sec:app_sensorspecs}

The instrumentation scheme of W27 wind turbine considered in the present study consists of the following sensor types:

\begin{itemize}
    \item Strain gauge XY101-3/350: cross strain gauge with two grids in the $0\degree$ and $90\degree$ directions, with a resistance of 350 Ohm and active gauge factor of 2.01. From stability tests conducted using a Quantum MX1615B amplifier, the strain gauge is estimated to have an accuracy of approximately 3\%.
    \item Accelerometer AAA323-001: accelerometers with an output of $\pm$5VDC. The bidirectional accelerometers with a grid of $0\degree$ and $90\degree$ have a range of $\pm$2g and the unixirectional accelerometer has a range of $\pm$1g. 
    \item MGCPlus data aquisition system (DAS), which utilizes the HBM Catman data acquisition software to retrieve data at a sampling rate of 50 Hz. A Bessel low-pass filter at 5Hz is automatically applied as an anti-aliasing feature.
\end{itemize}

\bibliographystyle{elsarticle-num} 
\bibliography{mssp_biblio}





% \end{thebibliography}
\end{document}